%% file: arXiv_v2.tex
\documentclass[a4paper,11pt,dvipsnames,usernames]{article}
\pdfoutput=1

\usepackage{jcappub} 
\usepackage{cases}
\usepackage{braket}
\usepackage{aas_macros}
\usepackage{booktabs}   %
\usepackage{rotating}   %

\usepackage{makecell}
\usepackage{subcaption} %
\usepackage{physics}
\usepackage[compat=1.1.0]{tikz-feynman}
\usepackage[english]{babel}
\usepackage{amsmath,amssymb,amsfonts, bm,bbm,slashed, subdepth}
\usepackage{graphicx}
\usepackage[sort&compress]{natbib}
\usepackage{xcolor}
\usepackage[normalem]{ulem}
\usepackage{hyperref}
\usepackage{cleveref}
\definecolor{red}{rgb}{1.0, 0, 0}
\usepackage{enumerate}
\usepackage{setspace}
\usepackage{dsfont}
\usepackage{booktabs, tabularx, multirow}
\usepackage{units} %
\let\oldhat\hat
 \renewcommand{\vec}[1]{\mathbf{#1}}
 \renewcommand{\hat}[1]{\bm\oldhat{\mathbf{#1}}}

\title{Towards a theory of dissipative\\ Dark Matter I: the Born limit}

\author[1,2,3]{Garance Lankester-Broche,}
\affiliation[1]{
Marietta Blau Institute for Particle Physics, Austrian Academy of Sciences, Dominikanerbastei 16, A-1010 Vienna, Austria}
\affiliation[2]{University of Vienna, Faculty of Physics, Boltzmanngasse 5, A-1090 Vienna, Austria}
\affiliation[3]{University of Vienna, Vienna Doctoral School in Physics, Boltzmanngasse 5, 1090 Vienna, Austria}
\author[1,2]{Josef Pradler}
\emailAdd{garance.lankester-broche@univie.ac.at}
\emailAdd{josef.pradler@univie.ac.at}

\begin{document}
\abstract{
We derive the energy-differential cross section and energy loss rate for dissipative self-interacting dark matter (dSIDM) models within the Born regime using perturbative quantum field theory. Six dissipative scenarios are considered, incorporating the emission of particles that may be either massless or possess a kinematically allowed light mass. Both short-range and long-range force-mediated dSIDM interactions are examined. In the non-relativistic regime, we obtain closed-form expressions of the energy-differential cross sections by a controlled expansion in the initial relative dark matter velocity. Up to trivial factors, the leading-order squared emission amplitude is model-independent for massless emissions. Model dependence arises for massive particle emission and at the next-to-leading order. The latter reduces to three distinct cases. The derived analytical expressions exhibit excellent agreement with numerical computations, providing simple, ready-to-use formulas.  Furthermore, we analyze the behavior of these processes in the soft emission limit. Our results show that additional corrections are necessary when applying factorization at the next-to-leading order in a velocity expansion to ensure consistency between the soft energy-differential cross section and the full counterparts across a broad energy range. Finally, we investigate the regime of perturbative validity in terms of the model parameters, identifying the conditions under which our results are applicable.
}

\maketitle

\section{Introduction}

The $\Lambda$CDM paradigm of cold, collisionless Dark Matter (DM) with gravitational only interaction with the Standard Model (SM) provides an excellent fit to observations of the cosmological structure on scales $\gtrsim 1~{\rm Mpc}$. On smaller, galactic and sub-galactic scales, however, several
persistent discrepancies with this simplest picture have been noted over the past decades. The most widely discussed are the core–cusp problem~\cite{Flores:1994gz,Moore:1994yx}, the missing-satellite problem~\cite{Klypin:1999uc}, the too-big-to-fail problem~\cite{Boylan-Kolchin:2011qkt}, and the diversity of dwarf galaxy rotation curves~\cite{Oman:2015xda}; see~\cite{Bullock:2017xww} for a review and references therein. The status of these issues remains a subject of ongoing scientific debate.  Improved observations and simulations with realistic baryonic physics---such as feedback and outflows—can alleviate several of the small-scale discrepancies, including the missing-satellite and too-big-to-fail problems~\cite{Governato:2012fa,Brooks:2012vi,Wetzel:2016wro,Fitts:2016usl}, though reproducing the full diversity of dwarf rotation curves remains challenging~\cite{Oman:2015xda,Santos-Santos:2019vrw}. Irrespective of the concrete status of small-scale structure problems, studying the implications of the departure from the collisionless DM paradigm is an important task.

Dark Matter collisions affect the  small-scale structure of $\Lambda$CDM, once the self-interaction cross section over DM mass, $\sigma/m$, reaches the 1 cm$^2$/g $\sim  $~bn/GeV ballpark~\cite{Spergel:1999mh}; see \cite{arXiv:1705.02358} for details.  Self-interacting dark matter (SIDM) introduces heat transport and effective viscosity to the DM fluid, driving the formation of cores. At late times, halos can undergo eventual gravothermal collapse with the magnitude and timing strongly dependent on the velocity dependence of~$\sigma/m$.
In order to achieve a cross section of the relevant scale, the current SIDM realizations span many models. The simplest arises by the introduction of a singlet-scalar DM candidate \cite{Silveira:1985rk,McDonald:1993ex,hep-ph/0011335} allowing for a quartic self-coupling. This setup naturally yields a constant self-interaction cross section \cite{astro-ph/0003350,hep-ph/0105284,hep-ph/0106249}, which faces tight observational constraints across all scales, see, e.g., \cite{Randall:2008ppe,arXiv:1211.6426,Harvey:2015hha,arXiv:2006.12515,Eckert:2022qia}, and thus renders a reconciliation of the full sub-structure phenomenology difficult. A natural way to relax these bounds and obtain a sufficiently large cross section is well accommodated in sub-GeV DM candidates that interact through the exchange of light mediators \cite{Feng:2009mn,Tulin:2013teo}. This prescription replaces the contact interaction with a Yukawa-type potential that renders the cross section velocity-dependent on smaller scales, while ensuring the right suppression in large-scale structure. Such mediators (vector or scalar) not only extend the scalar setup \cite{1512.05344,1402.6449} but naturally accommodate fermionic DM candidates as well \cite{1507.04931,Kahlhoefer:2017umn}.

Relaxing the assumption of a collisionless DM component also opens the door to additional phenomena, particularly {\it dissipation}, additionally
modifying the thermodynamics of halos: radiative channels shorten the cooling time relative to dynamical and relaxation times, enabling contraction and catalyzing collapse to black holes~\cite{Mohapatra:1999ih,DAmico:2017lqj,Shandera:2018xkn,Choquette:2018lvq,Fernandez:2022zmc,Buckley:2024eoe}, compact substructure~\cite{1412.1839,Buckley:2017ttd,2408.15317}, or the formation of rotationally supported dark disks~\cite{1303.3271,1303.1521,1403.0576,1604.01407}.  Because of constraints on the self-interaction cross sections $\sigma/m$, typically only a fraction of the total DM is assumed to participate in such dissipative sector.
Among other models, a well-studied dissipative realization is Atomic Dark Matter (ADM), which mimics the SM baryonic sector by allowing for the formation of hydrogen-like bound states \cite{Goldberg:1986nk,0909.0753,1209.5752,Cline:2013pca,1312.1336,1705.10341,Ryan:2021dis}.
Although ADM offers a rich phenomenological playground, much of its astrophysical impact stems from the net energy loss within halos, rather than from the specific details of the atomic spectrum. Consequently, some minimal dissipative models 
have been put forward in which the DM is simply composed of at most two DM candidates (fermionic or scalar) associated with either a massless, unbroken dark photon $U(1)_{D}$ \cite{0810.5126,Heikinheimo:2015kra, 1807.04750}, or through a massive dark gauge boson \cite{1812.07000,2405.04575}. These stripped-down setups already offer valuable insight into small-scale structure formation, avoiding the large parameter space inherent to more complex models.  For additional works on dissipative SIDM (dSIDM) see~\cite{Blinnikov:1983gh,Khlopov:1989fj,Cline:2013zca,Foot:2014osa,Petraki:2016cnz,Essig:2018pzq,Huo:2019yhk,Shen:2021frv,Shen:2022opd,Bramante:2023ddr,Gemmell:2023trd} and references therein. 

One key feature of these minimal scenarios is the emission of a boson emerging as a built-in dissipative channel inherent to the elastic self-scattering. Indeed, when the mediator mass is by a velocity-square factor smaller than the DM mass, it may equally well be produced as a final-state in the scattering process. If the mediator is heavier, an analogous cooling mechanism still exists, relying on the emission of an additional lighter particle present in the theory. Either way, one recovers the classic bremsstrahlung framework: a scattering accompanied by the radiation of a quantum that carries away kinetic energy and angular momentum. While other dissipative mechanisms, such as Compton cooling on a dark-radiation bath, are also possible, they typically require additional conditions.

In the SM, the bremsstrahlung process corresponds to the emission of a photon in the scattering of two electrically charged particles, and is one of the most fundamental radiative processes. Two distinct cases arise. First, when the particles have different charge-to-mass ratios, dipole emission occurs. This scenario was first treated classically by Kramers \cite{Kramers:1923}, followed by quantum mechanical (Born) calculations by Oppenheimer \cite{Oppenheimer:1929} and Sugiura \cite{Sugiura:1929}, and later by Sommerfeld who derived the exact non-relativistic solution using confluent hypergeometric functions\footnote{To avoid the cumbersome exact solution, Elwert introduced a multiplicative correction that extends the Born approximation \cite{1939AnP...426..178E}.} \cite{Sommerfeld:1931qaf}, a treatment now standard in advanced textbooks. More recently, dipole emissions have been revisited in \cite{Weinberg:2019mai}.
Second, when the particles share the same mass-to-charge ratio, leading to the vanishing of the dipole piece and the dominance of quadrupole radiation. The corresponding cross section was first derived in a Born calculation by Gould \cite{Gould:1981an, 1990ApJ...362..284G}, and revisited recently in a calculation exact at all orders in the Coulomb interaction of two non-relativistic particles in \cite{Pradler:2020znn}.  Beyond the full energy spectrum, the soft-emission limit is also of particular interest: in this regime, for photon and graviton, the amplitude factorizes into a universal soft factor and a process-dependent hard amplitude \cite{Weinberg:1965nx}. 

Bremsstrahlung-like processes are also found in many astrophysical and beyond the SM contexts. In stellar environments, similar types of radiation can arise from axion emission~\cite{Iwamoto:1984ir,hep-ph/0505090}, neutrino pair production \cite{Cazzola:1971ru,Dicus:1976rj,Raffelt:1993ix}, or scalar emission \cite{2303.00778,2303.03123}.
Unsurprisingly, these multipole emissions also reappear in dissipative dark-sector settings: dipole bremsstrahlung governs cooling in ionized dark-electron–dark-proton plasmas, while quadrupole emission dominates in dark-electron–dark-electron interactions.
However, a comprehensive theoretical framework for bremsstrahlung in generic dark sectors, departing from a simple gauge boson emission that includes dipole and higher multipoles emission, distinguishable and identical scatterers, different emitted quanta, and soft emission, has yet to be fully established.

Motivated by this gap and the diversity of the SIDM proposals, we establish in this work a set of representative minimal dissipative scenarios dominated by bremsstrahlung emission. Our DM candidates include fermions (Dirac or Majorana) and scalars (real or complex), coupled through either scalar or vector mediator. Two distinct species are included in each scenario, allowing the study of both distinguishable and identical scatterers, and both the long-range (light mediator) and short-range (heavy mediator) regimes are covered.
We work throughout in the non-relativistic (NR) limit, appropriate for small-scale structures where the characteristic DM relative velocity is $v_i \ll 1$. Within this framework, analytic formulae are derived for the energy-differential cross section and corresponding energy loss rate in every scenario. The results are compiled into a single catalogue and validated against numerical calculations and existing literature.
Where previous studies relied on process-specific approximations, this catalogue offers a single lookup table covering the parameter space for the most relevant SIDM scenarios, and ready-to-use expressions that can be readily adapted to any combination of couplings, masses, and mediator properties. The present study is restricted to the perturbative Born regime; a detailed investigation of non-perturbative effects will be addressed in a future work.

The paper is organized as follows: In Sec.~\ref {model}, we introduce the set of dissipative self-interacting dark matter (dSIDM) scenarios we consider in this work, summarize the main existing constraints on the corresponding SIDM parameters, and discuss the domains of validity of our results. In Sec.~\ref {dissipativecross-section}, treating separately long-range and short-range mediation, we first identify the dissipative channels that appear in the Born (perturbative) regime and analyse their relevance in the NR limit, making explicit the emergence of a hierarchy of scales among them. Next, we provide a concise textbook review of NR radiation, distinguishing dipole from quadrupole emission. The section concludes with explicit expressions for the squared emission amplitudes and the energy-differential cross sections, both for distinguishable and identical particle scattering.
In Sec.~\ref {soft}, we investigate the low-energy behavior of the different emission amplitudes and energy-differential cross sections obtained. We employ the standard factorisation that expresses the former as an elastic part multiplied by an emission piece, and discuss the limitations of such a method. We conclude this paper with Sec.~\ref{energylossrate}, where we derive energy loss rates. The Appendices provide further details on our calculations, including the case where the emitted particle carries a small, kinematically allowed mass.

By providing a unified analytic backbone for dissipative interactions, we aim to put dSIDM mechanisms on the same quantitative footing as collisionless CDM and elastic SIDM.

\section{Dissipative SIDM models} \label{model}

\subsection{Field content and considered interactions} \label{Lagrangianandmodels}
We consider a set of dissipative minimal models including fermionic and scalar DM candidates, interacting through either vector or scalar mediators. We categorize the field content as follows:  
\begin{align}
    \text{DM candidates:} \quad 
    &\begin{cases}
        \chi_j\, (\tilde\chi_j) & \text{Dirac (Majorana) fermion} \\
        S_j\, (\tilde S_j) & \text{complex (real) scalar}
    \end{cases} \\[8pt]
    \text{Mediators:} \quad 
    &\begin{cases}
       V^{\mu}\, (V'^{\mu}) & \text{light (heavy) vector} \\
         \phi\, (\phi'), & \text{light (heavy) real scalar}
    \end{cases}
\end{align}
An index~\( j = 1,2 \) has been introduced to model the DM self-scattering of 
distinguishable and indistinguishable particles.
As we shall see, the charge-to-mass ratio or, more generally, the ``coupling-to-mass'' ratio controls the leading velocity dependence of cross sections. By introducing the index~$j$, we model cases reminiscent of atomic DM with differing DM masses.
While $V^\mu$ and $\phi$ are the particles that will be emitted in the dissipative  scattering process, the massive mediators~$V'^{\mu}$ and~$\phi'$ are introduced to construct two limiting cases:  
\begin{itemize}
    \item long-range mediation: forces are carried by the light vector boson \( V^{\mu} \) or the light scalar boson \( \phi \) that are also being emitted;
    \item short-range mediation: interactions occur via the heavy vector boson \( V'^{\mu} \), the heavy scalar boson \( \phi' \), or through self-interactions of the scalar DM candidates $ \tilde S_j $ themselves.
\end{itemize}
In this work, we define a ``light'' mediator as one whose mass, \( m_V \) or \( m_\phi \), is smaller than the softest energy scale in the NR regime—the kinetic energy of the relative motion of the colliding DM pair. This ensures that the emission of \( V^\mu \) and \( \phi \) is kinematically allowed while also enabling them to mediate long-range interactions. Conversely, ``heavy'' mediators are those with masses larger than the typical momentum transfer in DM scattering. Their role is to approximate the contact interaction limit.  

The various cases are then constructed through the relevant interactions of mediators and DM particles. We list them in a set of three Lagrangian densities. Gauge interactions are given by the familiar Dirac and scalar QED cases
\begin{equation}  \label{LagrangianVector}
     \mathcal{L}^{V}_{\rm int}= g\sum_{j=1,2} \left\{ -Q_{\chi_j}  \overline{\chi}_{j}\slashed V \chi_{j} + i Q_{S_j} V^{\mu}\left[(\partial_{\mu}S_{j}^{\dagger})S_{j} - S_{j}^{\dagger}\partial_{\mu}S_{j}\right] \right\} + g^2\sum_{j=1,2} Q_{S_j}^2 S_{j}^{\dagger} S_{j} V_\mu V^\mu .
\end{equation}
Here,~$g$ is the gauge coupling and~$ Q_{\chi_j}$ and~$ Q_{S_j}$ are the respective charges of~$\chi_j$ and~$S_{j}$. 
Emission of a scalar mediator is enabled through
\begin{equation} \label{LagrangianScalar}
\mathcal{L}^{\phi}_{\rm int}= \sum_{j=1,2}\left\{  - y_j\overline{\chi}_{j} \chi_{j} \phi  - \frac{1}{2}  \tilde y_j \tilde{\chi}_{j}^{T}\mathcal{C}\tilde{\chi}_{j}  \phi  - A_j |S_{j}|^2\phi 
- \frac{1}{2}\tilde A_j {\tilde S_{j}}^2\phi
-\frac{1}{2}\lambda_{j} |S_{j}|^2\phi^2
-\frac{1}{4} \tilde \lambda_{j}  {\tilde S_{j}}^2\phi^2  \right\}.
\end{equation}
Here,~$y_j$ and~$\tilde y_j$ are Yukawa couplings,~$A_j$ and~$\tilde A_j$ are dimensionful trilinear couplings, and~$\lambda_{j} $ and~$\tilde \lambda_{j} $ are quartic couplings. %
Finally, for scalars, we also have the possibility of self-interactions to play into the phenomenology. Concretely, we consider the following interactions
\begin{equation} \label{angrangianSelfcoupling}
\mathcal{L}^{\rm self}_{\rm int}= - \frac{1}{3!}A_{\phi}\phi^{3} 
- \frac{1}{4!}\lambda_{\phi}\phi^{4}   
-\frac{1}{(1+3\delta_{S_1 S_2})}\lambda_S |S_{1}|^2 |S_{2}|^2 -\frac{1}{4(1+2\delta_{\tilde S_1 \tilde S_2})!}\tilde\lambda_S \tilde S_1^2 \tilde S_2^2
.
\end{equation}
Again, we have a dimensionful trilinear interaction constant $A_{\phi}$ for the real scalar mediator and self-interaction quartic couplings $\lambda_{\phi}$ and $\lambda_S$. We take all couplings as real.%
\footnote{We exclude off-diagonal couplings such as $\lambda_{12}S_{1}S_{2}\phi$, inducing a decay channel $S_{1}\to S_{2}\phi$ (or $S_{2} \to S_{1}\phi$ when $m_{1}>m_{2}+m_{\phi}$), altering the relative DM populations and introducing tangential complications.}

A comment is in order about the Lagrangians~\eqref{LagrangianVector}, \eqref{LagrangianScalar}, and \eqref{angrangianSelfcoupling} when considering indistinguishable particle scattering, for which we then drop any generational index~$j$ on the fields and couplings. In this case, we identify $S_1=S_2=S$ (and similarly for other fields) and it is understood that the sums over~$j$ in \eqref{LagrangianVector} and \eqref{LagrangianScalar} should be \emph{dropped}. In other words, the normalization of coupling constants \emph{remains unchanged} and does not receive a factor of two. 
Finally, the quartic scalar DM interaction term in~\eqref{angrangianSelfcoupling} becomes $-(1/4)\lambda_S |S|^4$ for $S$ and $-(1/4!)\tilde \lambda_S |\Tilde{S}|^4$ for $\tilde{S}$. It is for this reason that we have not included additional quartic $|S_j|^4$ interactions in~\eqref{angrangianSelfcoupling}.

The interaction Lagrangians~\eqref{LagrangianVector}, \eqref{LagrangianScalar}, and \eqref{angrangianSelfcoupling} must, of course, be supplemented by their free-theory counterparts. Particle masses are indicated by field subscripts: scalar and vector mediators have masses $m_{\phi^{(\prime)}}$ and $m_{V^{(\prime)}}$, respectively, while complex scalar DM particles $S_j$ carry masses $m_{S_j}$, and so forth. For convenience, we will soon introduce a shorthand notation $m_j$ for the DM masses when considering concrete scattering processes. The vector mass $m_{V^{(\prime)}}$ may be of St\"uckelberg type, maintaining dark $U(1)$ gauge invariance, or be of Higgsed origin. In the latter case, additional dark Higgs particles become part of the low-energy spectrum and affect the phenomenology; in this work we restrict our attention to a ``hard'' dark vector mass of St\"uckelberg type.

Of course, the possibilities considered above are not exhaustive, but still represent a fair fraction of the dSIDM possibilities. For simplicity, we shall also only consider exclusive cases where DM is of one principal fermionic or bosonic type. Mixed cases are, however, closely related since, as we shall see, the currents between fermions and scalars share many similarities in the NR limit. Notably absent from the above Lagrangians are axial-vector couplings and pseudoscalar couplings; in the latter case, the bremsstrahlung emission has been extensively studied in the context of axion emission from stars~\cite{Iwamoto:1984ir,Raffelt:1993ix,hep-ph/0505090}. 

With the introduced field content, we identify six principal dSIDM combinations. For the processes where the interaction is mediated by a vector boson and involves the emission of a boson of the same type, two types of DM candidates are relevant for the corresponding dissipative processes:~$\chi$ and~$S$. These give rise  to the dSIDM scenarios~$\chi_{1} \chi_{2}\to \chi_{1} \chi_{2} V$, and~$S_{1} S_{2} \to S_{1} S_{2} V$. In contrast, when the interaction is mediated by a scalar boson, and is associated with the emission of a scalar particle, the four DM candidates~$S_j$,~$\tilde S_j$,~$\chi_j$, or~$\tilde\chi_j$ are considered. These lead to the dSIDM scenarios~$\chi_{1} \chi_{2}\to \chi_{1} \chi_{2} \phi$,\,~$\Tilde{\chi}_{1} \Tilde{\chi}_{2}\to \Tilde{\chi}_{1} \Tilde{\chi}_{2} \phi$~and~$S_{1} S_{2} \to S_{1} S_{2} \phi$,\, $\Tilde{S}_{1} \Tilde{S}_{2}\to \Tilde{S}_{1} \Tilde{S} _{2}\phi$.

\subsection{Self-scattering constraints on model parameters} \label{constraints}

Although our focus is on dSIDM, these processes naturally appear as a built-in dissipative channel beyond the elastic SIDM framework. Consequently, many of the constraints derived for the latter remain directly relevant or serve as important benchmarks for assessing the viability of dSIDM scenarios. In the following, we summarize the corresponding principal constraints associated with 2-2 self-scattering processes. For comprehensive reviews, we refer the reader to \cite{arXiv:1705.02358,2207.10638}.

The success of SIDM lies in its ability to remain consistent across a wide range of astrophysical systems. In particular, the associated cross section per unit of DM mass ${\sigma}/{m_{\chi}}$ can be large enough to alter the inner density profiles of dwarf galaxies, yet small enough to avoid constraints on galaxy and galaxy cluster scales. Consequently, the principal bounds and/or regions of interest are typically categorized into three main regimes corresponding to their characteristic initial relative velocity~$v_{i}$:  
 \begin{itemize}
    \item On dwarf galaxy scales ($v_{i}\sim 10$ km/s), a cross section per unit of DM mass in the range $\sigma/m_{\chi} \sim  0.1-10$ cm$^{2}$/g %
    enables efficient heat transfer within the core \cite{arXiv:1201.5892,arXiv:1208.3025,arXiv:1211.6426,arXiv:1412.1477,Fry:2015rta,arXiv:1508.03339}. This mechanism helps alleviate potential mass-deficit problems, i.e., the~\textit{too-big-to-fail} and \textit{core-vs-cusp} problems (see e.g. \cite{Bullock:2017xww}). While larger cross sections can be considered, they eventually drive halos into gravothermal core collapse, yielding overly dense central regions~\cite{arXiv:astro-ph/0110561}. Interestingly, this same process might account for the high central densities observed in certain compact or ultra-faint dwarf galaxies~\cite{1904.07872,1904.10539,2010.02924,2108.03243}. 

\item On galaxy scales ($v_{i}\sim 100$ km/s), the interplay between the DM and baryons becomes non-negligible, rendering the constraints more subject to model dependencies. In Milky Way-like galaxies, very large cross sections per unit of mass can lead to the evaporation of subhalos through frequent scatterings with host-halos DM particles. Such effects are expected to occur once $\sigma/m_{\chi} \gtrsim 5-10$ cm$^{2}$/g \cite{2001.08754,1904.10539,2108.03243}. 
For low surface brightness (LSB) galaxies, which are DM dominated, SIDM can account for the diversity in rotation curve shapes observed for $\sigma/m_{\chi} \gtrsim 3$ cm$^{2}$/g \cite{1611.02716,1808.05695}.
\item On galaxy cluster scales ($v_{i}\sim 1000$ km/s), the measurements of core densities from strong lensing gives the most stringent constraints \cite{arXiv:1508.03339}, typically requiring cross sections per unit of DM mass in the range $\sigma/m_{\chi} \lesssim 0.1-0.35$ cm$^{2}$/g \cite{arXiv:2012.06611,arXiv:2006.12515}. Additional constraints arise from the analyses of spatial offsets between DM and galaxies in merging clusters---most notably the Bullet Cluster---where the observed alignment yields $\sigma/m_{\chi} \leq 1.25 $ cm$^{2}$/g \cite{Randall:2008ppe}. Recent studies have further tightened these constraints to $\sigma/m_{\chi} \lesssim 0.2-0.5$ cm$^{2}$/g  \cite{Harvey:2015hha,1712.06602,1812.06981}.
\end{itemize}

The phenomenology across these different scales has motivated the investigation of SIDM scenarios with a velocity-dependent cross section, arising naturally from the exchange of a light mediator. In this case, the non-relativistic Born transfer cross section~$\sigma_{T}= \int (1 - \cos{\theta^{*}})\, \frac{d\sigma}{d\Omega^{*}}\, d\Omega^{*}$
in the center-of-mass (CM) frame of the distinguishable 2-to-2 DM particle scatterings is given by \cite{arXiv:0911.0422}
\begin{equation} \label{transfertcross-section}
     \sigma_{T} =  \frac{a_{1}^2 a_{2}^2}{8\pi \mu^2 |\vec{v}_{i}|^4 }  \left[ \log{\left( \frac{4\mu^2 |\vec{v}_{i}|^2 + m_{\phi,V}^2}{m_{\phi,V}^2} \right)} - \frac{4\mu^2 |\vec{v}_{i}|^2}{4\mu^2 |\vec{v}_{i}|^2 + m_{\phi,V}^2} \right],
\end{equation}
and applies to $\chi_{1}\chi_{2} \to \chi_{1}\chi_{2}$, $S_{1}S_{2} \to S_{1}S_{2}$, $\tilde{\chi}_{1}\tilde{\chi}_{2} \to \tilde{\chi}_{1}\tilde{\chi}_{2}$, and $\tilde{S}_{1}\tilde{S}_{2} \to \tilde{S}_{1}\tilde{S}_{2}$. Here, 
$\mu = m_{1}m_{2}/(m_{1} + m_{2})$ is the reduced mass, 
$a_{j} = g_{j}, y_{j}, \tilde{y}_{j}, A_{j}/(2m_{j}),$ or $ \tilde{A}_{j}/(2m_{j})$, depending on the DM candidate/mediator, and $|\vec{v}_{i}|=v_i$ is the relative initial velocity.

In the limit where the mediator mass is light compared to the transferred momentum, $m_{\phi,V} \ll \mu v_{i}$,~\eqref{transfertcross-section} reduces to the Rutherford differential cross section and presents a $1/v_{i}^4$ dependence. In the opposite 
limit where the interaction is carried by ``light'' mediators $\phi'$ and $V'^{\mu}$ that transmit a short-range force $m_{\phi',V'} \gg \mu v_{i}$, the corresponding cross section scales as $1/m_{\phi',V'}^4$. Thus, in addition to introducing a velocity-dependent behavior at smaller mediator mass scales, SIDM mediated by a light boson leads to a desirable suppression of the cross section in larger halo structures associated with larger characteristic values of~$v_i$.

\subsection{Validity of the Born regime }

Throughout this work, we adopt a Born-level perturbative framework to describe the different dissipative processes.
We are working in leading order in both the DM self-interaction as well as in the emission of a single quantum. 
In this section, we specify the domain of applicability of our results. 

We may write the dissipative $2\to 3$ differential cross section schematically as the product of the elastic scattering cross section and the probability of emission, $d\sigma_{2\to 3} \simeq \sigma_{2\to 2} dP_{1}$, a form that, for gauge emisson, finds itself well rooted in the soft-photon theorem (see Sec.~\ref{soft}). When $ P_{1} \ll 1$, bremsstrahlung emission can be seen as a perturbative correction to the $2\to 2$ DM elastic scattering process, and the validity of the Born regime is determined by its validity in calculating elastic DM self-scattering.%
\footnote{Exceptions are radiative bound state formation and, in the context of scalar $\phi^3$ theory, a strong infrared sensitivity leading to $  P_{\rm em} = \mathcal{O}(1) $ already in the parameter region of interest; see App.~\ref{VinternalB}.} 
In the NR framework associated with small-scale structure (see Sec.~\ref{constraints}), the scalar and gauge mediating force between two DM candidates can be expressed as a Yukawa potential,
\begin{equation} \label{yukawapotential}
   V(r)= \zeta \frac{a_{1}a_{2}e^{-m_{\phi,V}r}}{4 \pi r},
\end{equation}
where $a_{j}$ denotes the coupling of the mediator to DM candidates, i.e., $g_{j},y_{j}$ 
or $A_{j}/m_{j}$, 
with $\zeta=-1\ (+1)$ 
for a scalar (vector) mediator and $\zeta\operatorname{sgn}{(a_1 a_2)}=-1\ (+1)$ 
being an attractive (repulsive) interaction. 
The Born regime is controlled by a small parameter 
whose form depends on whether the mediating force is short- or long-ranged.

\paragraph{Heavy mediator (short-range; Yukawa)}
For a heavy mediating particle  $m_{\phi^{\prime},V^{\prime}} \gg \mu v_{i}$, the relevant small parameter can be obtained by comparing first and second Born scattering amplitudes, and reads
\begin{equation}
\label{Conditionyukawa}
b=\frac{\left|a_{1}a_{2}\right| \mu }{4\pi m_{\phi^{\prime},V^{\prime}}}  \ll 1 \quad (\text{Born regime}).
\end{equation}
It may be seen parametrically as the ratio of the range of the potential $r_{\phi^{\prime},V^{\prime}} = 1/m_{\phi^{\prime},V^{\prime}}$ to the would-be Coulombic Bohr radius $a_0 = 4\pi/(\mu |a_1 a_2|)$. 
The condition~\eqref{Conditionyukawa} imposes a lower bound on the heavy-mediator mass.%
\footnote{Additional constraints arise from the NR framework we adopt,  further restricting the admissible range of $m_{\phi',V'}$; see Sec.~\ref{NRhierarchyofscales} for details.}

\paragraph{Light mediator (long-range; quasi-Coulomb)}
In the case of a light mediator $m_{\phi,V} \ll \mu v_{i}^2 $,~\eqref{yukawapotential} can be well approximated by a Coulomb form. In this limit, the relevant small parameters are the Sommerfeld parameters for the initial and final states,
\begin{equation} \label{Borncoulomb}
    |\nu_{i,f}| = \frac{\left|a_{1}a_{2}\right|}{4\pi v_{i,f}} \ll 1 \quad (\text{Born regime}).
\end{equation}
Here, $v_{i,f}$ correspond to the initial/final relative velocities of the DM pair. They measure the ratio of potential energy (evaluated at a separation corresponding to the de~Broglie wavelength of the relative motion ${1}/{\mu v_{i,f}}$) to the kinetic energy $\mu v_{i,f}^2$. In the Born approximation, Coulomb wavefunctions of the continuous spectrum can be replaced by plane waves. 
For the $2\to 3$ dissipative processes considered in this work, in which an energy $\omega$ is radiated, energy conservation implies $v_{f}=v_{i}\sqrt{1-x}$ with $x\equiv 2\omega/(\mu v_{i}^2) $. Accordingly, the final state Sommerfeld parameter is given by $\nu_{f}= \nu_{i}/ \sqrt{1-x}$ so that $|\nu_{f}|\geq |\nu_{i}|$. Close to the endpoint energy $x\to 1$, there is inevitably a small energy region where~\eqref{Borncoulomb} is not obeyed, corresponding to the low-velocity emergence of the scattered particle pair.

\paragraph{Multiple emissions} 
 Throughout this work, we restrict our attention to real, single-emission tree-level bremsstrahlung from massive particles. 
In our case, soft enhancements arise solely from an energy integral over the single vector differential emission probability, $\int_E^Q dP_1/d\omega$, where $Q \lesssim \mu v_i^2/2$ denotes the hard scale of the process and $E$ is the IR scale. This integral equals the 
mean number $\lambda$ of unresolved vectors with energies $ E \leq \omega \leq Q $, yielding an approximate Poisson multiplicity,  $P_{n}=(\lambda^n/n!) e^{-\lambda}$.
Using the eikonal factor $\mathcal{E}_{V}(q, \varepsilon^{*})$ for emission with four-momentum $q$ (energy $\omega=q^0$ and three-momentum $\vec q$), polarization vector $\varepsilon^{\mu*}(q)$  from any external leg, Eq.~\eqref{softfactorgauge},
we obtain
\begin{equation} \label{conditionbornsoft}
\lambda=
\int_E^Q d\omega \sum_{\text {pol }}\left|\mathcal{E}_V\right|^2 \frac{d^3 \mathbf{q}}{(2 \pi)^3 2 \omega}
=
\frac{1}{4\pi^2} \ln \frac{Q}{E} \int \frac{d\Omega_{\vec{\hat{q}}}}{4 \pi} \mathcal{F}(\hat{\mathbf{q}}) .
\end{equation}
Here, $\mathcal{F}(\hat{\mathbf{q}}) \equiv \sum_{\mathrm{pol}}\left|\omega \mathcal{E}_V(q, \varepsilon^{*})\right|^2$ is the energy-independent, polarization-summed, angular-only kernel that integrates to $\omega^2$ times the dipole and the quadrupole emission factors. The corresponding angular averages of these emission factors are given in~\eqref{eikonaldipole} and~\eqref{softVquadrupoleheavy}, respectively. 
The Born (single-emission) regime, in which virtual soft corrections can be neglected, and bremsstrahlung is a perturbative correction to the elastic $2\to2$ process corresponds to $\lambda \ll 1$. For $\lambda \gtrsim 1$ multiple emissions are common, and a resummed treatment is required. For scalar emission, the absence of gauge symmetry implies that there is no universal soft theorem. Nonetheless, the leading-order cancellation of soft divergences in real and virtual terms parallels the vector case~\cite{2208.05023}. In Sec.~\ref{soft} and App.~\ref{scalarvsgauge} we discuss the difference with vector emission at the level of the factorization of the amplitude.

The conditions~\eqref{Conditionyukawa},~\eqref{Borncoulomb}, and the mean number of unresolved vectors~\eqref{conditionbornsoft} delineate the parameter space in which our perturbative treatment, and hence all subsequent results, are valid. An exploration of the beyond-Born regime will be presented in a future work.

\section{Energy-differential cross sections} \label{dissipativecross-section}
In order to prepare for the expression of the energy-differential cross section and to set some of the notation, we begin this section by identifying the dissipative channels associated with each dSIDM model and their relevance in the kinematic regimes under consideration, i.e., NR colliding DM candidates with \textit{long-range} and \textit{short-range mediation}. We then introduce the terminology used throughout this work to distinguish between two distinct emission mechanisms: dipole and quadrupole radiation. With these definitions in place, we conclude this section by presenting the expressions for the differential energy cross sections. 

\subsection{Dissipative channels} \label{dissipativechannel}

\begin{figure}[t]
    \begin{subfigure}[b]{0.22\textwidth}
        \centering
        \begin{tikzpicture}
            \begin{feynman}
                \vertex (a1) {\(1\)}; 
                \vertex[right=1.5cm of a1] (a2);
        
                \vertex[right=0.75cm of a2] (a6);
                \vertex[below right =1.5cm of a6] (a7);
                 \vertex[right=0.75cm of a1] (mid1);
                \vertex[right=1.5cm of a2] (a4) {\(3\)};
                \vertex[below=2cm of a2] (a8);
                \vertex[below=2cm of a1] (b1) {\(2\)}; 
                \vertex[right=0.75cm of b1] (mid2);
                \vertex[below=2cm of a4] (b2) {\(4\)};
                  \vertex[left=0.75cm of b2] (mid3);
                \diagram* {
                    (a1) -- [scalar] (a2) -- [scalar] (a4),
                    (b1) -- [scalar] (a8) -- [scalar] (b2),
                    (a6)--[ghost, momentum'={[arrow shorten=0.2,arrow distance=0.18]\(q\)}] (a7),
                    (a8)--[ghost, momentum'={[arrow shorten=0.2,arrow distance=0.18]\(k\)}] (a2),
                };
                 \draw[fill=orange] (mid1) circle (0.05cm);
                 \draw[fill=orange] (mid2) circle (0.05cm);
                 \draw[fill=orange] (mid3) circle (0.05cm);
            \end{feynman}
        \end{tikzpicture}
        \caption{}
        \label{inifini}
    \end{subfigure}
    \hfill
    \begin{subfigure}[b]{0.22\textwidth}
        \centering
        \begin{tikzpicture}
            \begin{feynman}
                \vertex (a1) {\(1\)}; 
                \vertex[right=1.5cm of a1] (a2);
                \vertex[right=0 cm of a2] (a6);
                \vertex[below right =1.5cm of a6] (a7);
                \vertex[right=1.5cm of a2] (a4) {\(3\)};
                \vertex[below=2cm of a2] (a8);
                \vertex[below=2cm of a1] (b1) {\(2\)}; 
                \vertex[below=2cm of a4] (b2) {\(4\)};

                \diagram* {
                    (a1) -- [scalar] (a2) -- [scalar] (a4),
                    (b1) -- [scalar] (a8) -- [scalar] (b2),
                    (a6)--[ghost, momentum={[arrow shorten=0.2,arrow distance=0.18, label distance=-2.5]\(q\)}] (a7),
                    (a8)--[ghost, momentum'={[arrow shorten=0.2, arrow distance=0.18]\(k\)}] (a2),
                };
                \draw[fill=orange] (a8) circle (0.05cm);
            \end{feynman}
        \end{tikzpicture}
        \caption{}
        \label{vertex}
    \end{subfigure}
\hfill
    \begin{subfigure}[b]{0.22\textwidth}
        \centering
\begin{tikzpicture}
    \begin{feynman}
        \vertex (a1); 
        \vertex[above right =1.2cm of a1] (a2) {\(3\)};
        \vertex[above right =0.6cm of a1] (a6);
        \vertex[right =1cm of a6] (a7);
        \vertex[below left =1.2cm of a1] (a3) {\(2\)};
        \vertex[above left  =1.2cm of a1] (a4) {\(1\)};
        \vertex[below right =1.2cm of a1] (a5) {\(4\)};
        \diagram* {
            (a1) -- [scalar] (a2),
            (a3) -- [scalar] (a1),
            (a4) -- [scalar] (a1),
            (a1) -- [scalar] (a5),
            (a6) -- [ghost, momentum'={[arrow shorten=0.1, arrow distance=0.18]\(q\)}] (a7),
        };

        \draw[fill=orange] ($(a3)!0.5!(a1)$) circle (0.05cm); 
        \draw[fill=orange] ($(a4)!0.5!(a1)$) circle (0.05cm);
        \draw[fill=orange] ($(a1)!0.5!(a5)$) circle (0.05cm); 
        
    \end{feynman}
\end{tikzpicture}
        \caption{}
        \label{Contacts}
    \end{subfigure}
    \hfill
    \begin{subfigure}[b]{0.22\textwidth}
        \centering
        \begin{tikzpicture}
            \begin{feynman}
                \vertex (a1) {\(1\)}; 
                \vertex[right=1.5cm of a1] (a2);
                \vertex[right=0 cm of a2] (a6);
                \vertex[below left =1.5cm of a6] (a7);
                \vertex[right=1.5cm of a2] (a4) {\(3\)};
                \vertex[below=1cm of a2] (a9);
                \vertex[right=1.5cm of a9] (a10);
                \vertex[below=2cm of a2] (a8);
                \vertex[below=2cm of a1] (b1) {\(2\)}; 
                \vertex[below=2cm of a4] (b2) {\(4\)};

                \diagram* {
                    (a1) -- [scalar] (a2) -- [scalar] (a4),
                    (b1) -- [scalar] (a8) -- [scalar] (b2),
                    (a9)--[ghost, momentum'={[arrow shorten=0.25,arrow distance=0.18]\(q\)}] (a10),
                    (a9)--[ghost, momentum'={[arrow shorten=0.2, arrow distance=0.3]\(k-q\)}] (a2),
                    (a8)--[ghost, momentum'={[arrow shorten=0.25,arrow distance=0.35]\(k\)}] (a9),
                };
            \end{feynman}
        \end{tikzpicture}
        \caption{}
        \label{VIB}
    \end{subfigure}
    \hfill
    \caption{Emission diagrams for dissipative DM self-scattering. The orange dots show other possible emission points. Depending on the type of diagram, we introduce the following terminology: (a) ``External leg diagrams'' for emission from external legs producing four diagrams for distinguishable and eight diagrams (including crossed channels) for identical particles. These diagrams are relevant for all considered models.
    (b) ``Vertex diagrams'' for scalar DM candidates $S$ and $\tilde S$, where emission occurs at the mediator-exchange vertex. They constitute two diagrams for distinguishable particles and, including crossed channels, four diagrams for identical particles.
  (c) ``Contact diagrams'' due to quartic scalar DM interactions; there are four diagrams of this type. (d) ``Virtual Internal Bremsstrahlung'' (VIB) from a scalar mediator with scalar emission of $\phi$ particles, in principle relevant to all DM candidates; they lead to one (two) diagram(s) in the case of distinguishable (identical) particle scattering. The emitted particle $\phi$ or $V^{\mu}$ carries four-momentum $q$; the four-momentum assignment of propagators is emission-point dependent, and the full kinematic dependency is taken into account when evaluating the various amplitudes.}
    \label{Diagrams}
\end{figure}
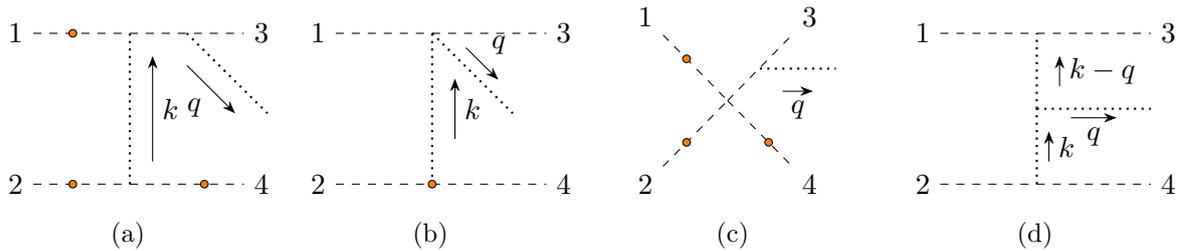

We aim to derive the energy-differential cross section within a perturbative framework. For each dSIDM model, the emission amplitude $\mathcal{M}$ is constructed from Feynman diagrams as shown in Fig.~\ref{Diagrams}.  In the following, we systematically identify the dissipative channels associated with each dSIDM model. 
 
\subsubsection*{Vector emission and mediation}
The DM self-scattering mediated by a vector boson and associated with the emission of the mediator is associated with the dSIDM processes~$\chi_{1} \chi_{2} \to \chi_{1} \chi_{2} V$ and~$S_{1} S_{2} \to S_{1} S_{2} V$. 
Their overall amplitudes are constrained by gauge invariance. 
For~$\chi_{1} \chi_{2} \to \chi_{1} \chi_{2} V$, the emission is exclusively composed of contributions from external leg emission diagrams, Fig.~\ref{inifini}. In contrast, for $S_{1} S_{2} \to S_{1} S_{2} V$, contributions from the vertex emission diagrams, Fig.~\ref{vertex}, must be included. Additionally, the quartic self-interaction~$\lambda_{S}$ contributes to the emission amplitude in this dSIDM scenario, Fig.~\ref{Contacts}. 

We note that, in the particular case where a massless
$V^{\mu}$ mediates the DM interaction and is also the emitted particle, the results are well-established in the literature, corresponding to QED- and scalar QED-like bremsstrahlung processes.

\subsubsection*{Scalar emission and mediation}
The emission of a single scalar boson associated with scalar mediation corresponds to the dSIDM scenarios~$\chi_{1} \chi_{2}\to \chi_{1} \chi_{2} \phi,\ \Tilde{\chi}_{1} \Tilde{\chi}_{2}\to \Tilde{\chi}_{1} \Tilde{\chi}_{2} \phi$, $S_{1} S_{2} \to S_{1} S_{2} \phi$ and $ \Tilde{S}_{1} \Tilde{S}_{2}\to \Tilde{S}_{1} \Tilde{S}_{2} \phi$. As a general remark, unlike gauge vector emission, the dissipative contributions may involve different coupling constants, implying that the relative importance of the various channels for the emission amplitude will depend on their specific values.

For  $\chi_{1} \chi_{2}\to \chi_{1} \chi_{2} \phi$ and $\Tilde{\chi}_{1} \Tilde{\chi}_{2}\to \Tilde{\chi}_{1} \Tilde{\chi}_{2} \phi$, the contributions to the emission amplitude come from the external leg and Virtual Internal Bremsstrahlung (VIB) emission diagrams, Fig.~\ref{inifini} and Fig.~\ref{VIB}, respectively. The latter is induced by the trilinear self-coupling $A_{\phi}$ of the mediator $\phi$. Thus, in the case of distinguishable particle scattering, both dSIDM models share the same dissipative channels. However, in the case of identical particle scattering, $t$-,~$u$- and~$s$-channels have to be considered for \(\Tilde{\chi}\), whereas only the~$t$-~and~$u$-channels are allowed for~\(\chi\).
For $\Tilde{S}_{1} \Tilde{S}_{2}\to \Tilde{S}_{1} \Tilde{S}_{2} \phi$ and $S_{1} S_{2} \to S_{1} S_{2} \phi$, the emission amplitude receives contributions from the external leg, vertex, contact, and VIB emission diagram, Figs.~\ref{inifini}-\ref{VIB}. 
However, for $\Tilde{S} \Tilde{S}\to \Tilde{S} \Tilde{S} \phi$, the absence of any $U(1)$ symmetry allows $\Tilde{S}$ to be the mediator itself in the external leg, the vertex, and VIB emission diagrams, introducing additional contributions compared to when $S$ is DM.

\subsection{Non-relativistic hierarchy of scales  } \label{NRhierarchyofscales}

We consider scattering between DM candidates with a relative velocity that lies in the range \( v_{i}\lesssim 10^{-2} \), corresponding to the low-velocity regimes discussed in Sec.~\ref{constraints}. This NR assumption, when applied in the center-of-mass (CM) frame of the scattering processes, leads to a separation of scales between the relative incoming three-momentum, \(\Vec{p}_{i}\sim v_{i}\mu \), and the available CM-energy for the emission of a particle, \(\omega \sim v_{i}^2 \mu\);~$\mu$ is the reduced mass of the DM scattering pair. 

Moreover, the typical three-momentum transfer in NR elastic DM scattering is $|\vec k| \sim \mu v_i$. If the particle that mediates the elastic scattering is also the particle that is being emitted, this \emph{automatically} places the DM scattering in the regime of \emph{long-range} interaction, since the four-momentum transfer $k$ between DM particles satisfies $-k \simeq  |\vec k| \gg m_{\phi, V}$ where the mediator mass is also the mass of the emitted state. However, when the emitted particle is distinct from the mediator, the mass of the latter is no longer subject to the same kinematic constraint and can instead satisfy \( m_{\phi',V'} \gg |\vec{k}| \). In this case, the DM scattering occurs in the regime of \textit{short-range} interaction.

Before we present the dissipative results for the energy-differential cross sections, we discuss now in greater detail the various behaviors, broken down in long- and short-range mediators for the various models considered.

\subsubsection*{Long-range mediator} \label{longrangeforce}
The separation of scales that arises from the NR assumption naturally establishes a power counting scheme in incoming relative velocity~$v_i$. We hence use the latter to expand the amplitudes for long-range mediated processes of the relevant dissipative processes discussed in Sec.~\ref{dissipativechannel}. As a general result, the dominant scaling of the emission amplitude $\mathcal{M}$ is at most of order~\(\mathcal{O}(v_{i}^{-3})\). If the colliding particle pair has equal ``charge-to-mass'' ratios, the leading order becomes \(\mathcal{O}(v_{i}^{-2})\) instead.

In addition to providing simple analytic expressions, the $v_{i}$-expansion selects a subset of Feynman diagrams that contribute at a given order of $v_i$. 
For all of the dSIDM scenarios considered: 
\begin{itemize}
\item In the case of distinguishable particle scattering, the leading contribution arises from external leg emission diagrams, Fig.~\ref{inifini}, scaling as~$\mathcal{O}(v_i^{-3})$.

\item In the case of identical particle scattering, the $t$-, $u$-, and $s$-channel contributions from the external leg and the vertex emission diagrams, illustrated in Fig.~\ref{vertex}, need to be considered. While the $t$- and $u$-channel amplitudes scale as~$\mathcal{O}(v_i^{-2})$, the $s$-channel involve center-of-mass energy flowing through the propagator $i\Delta(p)$, leading to a scaling of~$\Delta(p_{1}+p_{2})/\mu^2 \sim \mathcal{O}(v_i^0)$. As a result, the $s$-channels can be neglected.

\item The VIB emission diagrams are treated in App.~\ref{diagramcancellation}, and we do not consider their contribution to the emission amplitude in the following.
\end{itemize}
Furthermore, for the dSIDM scenarios involving scalar DM candidates:
\begin{itemize}

\item The vertex emission diagrams contribute only at \(\mathcal{O}(v_i^{-2})\), missing one propagator velocity-enhancement compared to the external leg emission diagrams.

\item The contact emission diagrams, represented in Fig.~\ref{Contacts}, are at most of order \(\mathcal{O}(v_i^{-1})\), and can therefore be neglected in the derivation of the emission amplitudes; see App.~\ref{diagramcancellation} for a detailed discussion. 

\item The exchange, vertex and VIB diagrams mediated by $\Tilde{S}$ have their propagators scaling as $\Delta(p_{2}-p_{4})/m_{S}^2 \sim \mathcal{O}(v_i^0)$, and are therefore suppressed by at least a factor of \(v_i^2\) compared to the exchange of a particle of parametrically smaller mass, as imposed by our kinematic boundary conditions. Consequently, these dissipative channels can be neglected in the computation of the emission amplitudes.
\end{itemize}
In Tab.~\ref{Tablecoupling}, we summarize the different processes considered and their dependencies on the coupling constant and leading power of the initial velocity. We also include the elastic SIDM scattering process for comparison.
\begin{table}[t]
    \centering \small
    \begin{tabular}{r|cccccc}
    \toprule
     \multirow{2}{*}{amplitude $\mathcal{M}$}  & elastic & external leg &  vertex & contact &  internal & annihilation \\
       & --- & Fig.~\ref{inifini} &  Fig.~\ref{vertex} & Fig.~\ref{Contacts} & Fig.~\ref{VIB} & --- \\
          \midrule
       coupling combination & $a^{2}$ & $a^{3}$ &  $g^{3},\, (A/m)\lambda $  & $a\lambda_{S}  $ & $a^2 A_{\phi}$  & $a^{3}$\\
   velocity~scaling &  $ v_{i}^{-2}$ & $ v_{i}^{-3},\ v_{i}^{-2} $ &  $ v_{i}^{-2} $ & $ v_{i}^{-1} $  & $ \ll v_{i}^{-2}$ & $ v_{i}^{-1}$ \\
   \bottomrule
    \end{tabular}
    \caption{Scaling of emission amplitudes $\mathcal{M}$ in the case of long-range mediation for the various scattering channels. The parameter $a = g, y,$ or $A/m$ is the coupling mediating the DM self-scattering process depending on the DM candidate considered. The following columns show associated dissipative processes, their scaling and, in the second row, their leading velocity dependence.
}
    \label{Tablecoupling}
\end{table}

As a general outcome of this discussion, while we initially consider six distinct dSIDM scenarios, the NR framework enhances specific dissipative channels, significantly reducing the number of relevant contributions. This results in only three unique amplitude configurations with a set of coupling constants~\( g_j, y_j, A_j \) together with DM masses~$m_j$, where the specific values depend on the DM candidate. To simplify notation, we make the following identification,
\begin{enumerate}
    \item \textbf{Gauge vector emission}: since the contact diagrams are velocity-suppressed, the leading order contributions for~$\chi_1 \chi_2\to \chi_1 \chi_2 V $ and $S_1 S_2\to S_1 S_2 V $ come from external leg diagrams (Fig.~\ref{inifini}), with the contributions from the vertex diagrams (Fig.~\ref{vertex}) being sub-leading. In the NR limit, spin effects from gauge interactions are suppressed~\cite{Gould:1981an}; see App.~\ref{scalarvsgauge} for a detailed discussion. As a result, the amplitudes we derive are valid for general spin in the case of distinguishable particles. For identical DM particles, the Dirac fermions $\chi$ and the complex scalars $S$ must be treated separately since $t$- and~$u$-channel amplitudes have differing relative sign.
    We define the following general coupling constant for these processes
\begin{align*}
    (g_{j},m_j) &\equiv 
    \begin{cases}
        (Q_{\chi_j} g , m_{\chi_j}) & \text{for gauge emission from Dirac DM } \chi_j, \\
        (Q_{S_j} g , m_{S_j}) & \text{for gauge emission from complex scalar DM } S_j.
    \end{cases} 
    \end{align*}
    \item \textbf{Scalar emission from fermionic DM candidates}: for the two dSIDM scenarios $\chi_{1} \chi_{2}\to \chi_{1} \chi_{2} \phi $ and $ \tilde\chi_{1} \tilde\chi_{2}\to \tilde\chi_{1} \tilde\chi_{2} \phi $, the $v_{i}$-leading and sub-leading order contribution to their emission amplitudes come from external leg diagrams (Fig.\ref{inifini}). This results in these candidates having the same emission amplitude in the case of distinguishable scattering. The same conclusion holds in the case of identical particle scattering because of the suppression of the annihilation channels. The relevance of VIB processes is treated separately in App.~\ref{diagramcancellation}. For the emission amplitudes corresponding to both models, we define
    \begin{align*}
    (y_j,m_j) &\equiv 
    \begin{cases}
        (y_j, m_{\chi_j}) & \text{for scalar emission from Dirac DM } \chi_j,\\
        (\tilde y_j, m_{\tilde\chi_j}) & \text{for scalar emission from Majorana DM } \tilde{\chi}_j.
    \end{cases}
\end{align*}
    \item \textbf{Scalar emission from scalar DM candidates}: for the scenarios $S_{1} S_{2}\to S_{1} S_{2} \phi $ and $\tilde S_{1} \tilde S_{2}\to \tilde S_{1} \tilde S_{2} \phi $, the contact diagrams (Fig.~\ref{Contacts}) and the ones mediated by $\Tilde{S}$ are velocity suppressed. Consequently, the leading contributions to their emission amplitudes come from external leg (Fig.~\ref{inifini}), and  vertex (Fig.~\ref{vertex}) diagrams. The relevance of VIB processes is treated separately in App.~\ref{diagramcancellation}. Therefore, except in specially constructed cases, both types of scalar DM candidates, $\Tilde{S}$ and $S$, yield the same leading and sub-leading analytic expressions in a $v_i$-expansion of the overall emission amplitude. This result suggests a degree of model-independence also in the dissipative dynamics of scalar~DM, allowing us to define
    \begin{align*}
    (A_j, \lambda_{j} , m_j) &\equiv 
    \begin{cases}
        (A_j , \lambda_{j}, m_{S_j} ) & \text{for scalar emission from complex scalar DM } S_j , \\
        (\tilde A_j, \tilde\lambda_{j} , m_{\tilde S_j} ) & \text{for scalar emission from real scalar DM } \tilde S_j.
    \end{cases}
    \end{align*}
    The strength of interaction in the scalar emission scenario is measured by the dimensionless coupling constant $A_j/m_j$. Hence, when we speak of a (vanishing of) ``coupling-to-mass'' ratios to delineate dipole from quadrupole emission below, we mean the combination $A_1/m_1^2 -A_2/m_2^2$.
\end{enumerate}

\subsubsection*{Short-range mediator}\label{shortrangeforce}
In the opposite limit of a short-range mediated force, the emitted particles $(\phi, V^\mu)$ necessarily differ from the exchanged particles $(\phi',V'^\mu,\tilde S_j)$. This implies that, for the Lagrangian interactions discussed in Sec.~\ref{model}, we must introduce a primed version of  associated coupling constants \( g'_{j}\), \(y'_{j}\) and \(A_j'/m_j\). For $S_1 S_2\to S_1 S_2 V $, this leads to the new interaction $g_{j}g_{j}'V_{\mu}V^{\prime\mu}S_{j}^{\dagger}S_{j}$ corresponding to vertex emission (Fig.\ref{vertex}).\footnote{This is a consequence of the covariant derivative being modified to $D_{\mu}=i\partial_{\mu}+igV_{\mu} + ig'V'_{\mu}$.} Similarly,  for scalar emission with scalar DM candidates introduced in~\eqref{LagrangianScalar}, we account for the interactions \(\lambda'_{j}\phi \phi'|S_{j}|^2 \) and \(\Tilde{\lambda}'_{j}/2\,\phi \phi'\Tilde{S}_{j}^2 \). Finally, we consider a direct coupling between the light and heavy mediators only in the scalar-mediated, scalar-emission case, which gives rise to VIB processes (Fig.~\ref{VIB}). Since this contribution is significantly more suppressed than the other dissipative channels, we defer its discussion to App.~\ref{VinternalB}.

In this scenario, we identify another small parameter, the ratio of typical momentum transfer in the DM scattering to mediator mass,
\begin{align}
    r \equiv \frac{\mu v_{i}}{ m_{\phi',V'}} .
\end{align}
The condition $r \ll 1 $ then ensures that we are in the contact limit---our nomenclature of a short-range mediator. This allows for an additional expansion of results in $r$, noting that  \(\vec{p}_{i}/m_{\phi',V'} \sim \mathcal O(r)$, and $ \mu \omega/m_{\phi',V'}^2 \sim \mathcal{O}(r^2)\).

From a phenomenological perspective, however, we note that {\it long-range} mediated cases likely continue to be relevant to the emission process. The reason is simple: even in the presence of another heavy mediator, the coupling that enables the emission from the dark matter leg also continues to be responsible for mediating a long-range force. For fermionic DM candidates, the parametric dependencies of amplitudes are such that the condition for short-range processes to dominate over the long-range irreducible ones reads 
\begin{align}\label{shortrangedominance}
\text{short-range dominance: }\quad 
\frac{g_1 g_2}{g_1' g_2'},\ \frac{y_1 y_2}{y_1' y_2'} \lesssim r^2 \quad \text{for fermionic DM}.
\end{align}
For scalar DM candidates, as detailed below, additional contributions in the emission amplitude arise from the contact emission diagrams, modifying this condition to
\begin{equation}
\label{shortrangedominance2}
\text{short-range dominance: }\quad 
 \frac{g_1 g_2}{\left(g_1' g_2' + \lambda_{S}\frac{m_{V'}^2}{4 m_{1}m_{2}}\right) },\ \frac{A_1 A_2}{\left(A_1' A_2' - \lambda_{S}{m_{\phi'}^2}\right)} \lesssim r^2\quad \text{for scalar DM.}
\end{equation}
Whereas the requirement for the presence of a contact interaction is mild, $m_{\phi',V'} \gtrsim v_i \mu \sim 10^{-3} \mu$, the typical coupling of $\phi'$ or $V'^{\mu}$ to DM must be larger than the coupling of $\phi$ or $V^{\mu}$ by a factor of $1/v_i \sim 10^3$. This demands a clear hierarchy of couplings in the dark sector.
Such a scenario is then best conceivable in gauged models of SIDM where the DM self-interaction is due to (non-Abelian) strong interactions in the hidden sector.

The conditions~\eqref{shortrangedominance} and~\eqref{shortrangedominance2} also guide us to focus on mediator masses that satisfy,~$m_{\phi',V'}< \sqrt{s}$ where $s =(p_1+p_2)^2$ is the squared CM energy. Heavy mediators $m_{\phi',V'} > \sqrt{s}$ are, first, less prospective in successfully realizing SIDM scenarios with sufficiently large self-scattering cross section---unless strong interactions are invoked---and second, typically do not dominate the dissipative process, thereby violating~\eqref{shortrangedominance}--\eqref{shortrangedominance2}. Hence, by the standards of $\sqrt{s}\simeq m_1 + m_2$ we continue to consider ``light'' mediators $\phi'$ and $V'^{\mu}$ as the prototype case, but such that they transmit a short-range force in the context of NR DM self-scattering. In this regime, we can then safely neglect any $s$-channel self-scattering diagrams that would otherwise contribute to the emission process. By the same token, we are also allowed to disregard contributions in scalar cases where the exchanged mediator is the DM itself. 

Under these conditions, again only a subset of diagrams yield the dominant contributions in the joint expansion in $v_i$ and~$r$. For fermionic DM candidates, the same hierarchy between the dissipative channels arises as for long-range mediations. In contrast, for scalar DM candidates, the dissipative processes from the external leg, vertex, and contact emission diagrams, Figs.~\ref{inifini}-\ref{Contacts}, respectively, are now velocity-enhanced in a similar manner.  Among these, the external leg and contact emission contributions dominate at leading order in the $r$-expansion, while the vertex diagrams are relevant only at sub-leading order.

It is clear, that in the heavy mediator limit $r\ll1$ the external leg amplitudes Fig.~\ref{inifini} and contact amplitudes Fig.~\ref{Contacts} for scalar DM scenarios will share similar analytic structure, apart from coupling-constant prefactors.
Consequently, the contributions from these different dissipative channels unify, for~$S_{1}S_{2} \to S_{1}S_{2}\,\phi, \Tilde{S}_{1}\Tilde{S}_{2} \to \Tilde{S}_{1}\Tilde{S}_{2}\,\phi$,  to
\begin{align} \label{shortrangecontact}
     \mathcal{M}_{\text{external leg}} +  \mathcal{M}_{\text{contact}} & = \mathcal{F}
     (\lambda_{S},A_{j},A_{j}',m_{j}) \mathcal{T}(r,\omega,m_{\phi})
     + \mathcal{O}(r), 
\end{align}
and, for $S_{1}S_{2}\to S_{1}S_{2}V$, to
\begin{align}
     \mathcal{M}_{\text{external leg}} + \mathcal{M}_{\text{vertex}} + \mathcal{M}_{\text{contact}} & = \mathcal{F}(\lambda_{S},g_{j},g_{j}',m_{j}) \mathcal{T}(r,\omega,m_{V})+ \mathcal{O}(r).
\end{align} 
Here $\mathcal{F} $ are the model-dependent prefactors containing masses and coupling constants, and $\mathcal{T}$ the overall amplitudes for either dipole or quadrupole emission (see following sections). 

Finally, for the scenarios $S_1 S_2 \to S_1 S_2 \phi$ and $\tilde{S}_1 \tilde{S}_2 \to \tilde{S}_1 \tilde{S}_2 \phi$, the three dissipative channels contributing to the sub-leading order of the $r$-expansion of the emission amplitude involve different coupling constants. We distinguish two cases: (i) when the external leg and contact emission diagrams parameters dominate the emission amplitude, and (ii) when the coupling constant from the vertex emission diagrams becomes large enough to provide the leading contribution. For convenience, we refer to the first case as the ``external leg-dominated emission'' and the second as the ``vertex-dominated emission.'' The full amplitude is given in App.~\ref{scalarscalarquad}.

\subsection{Dipole vs.~quadrupole emission} \label{dipolevsquadrupole}
In this section, we shall clarify the terminology of dipole and quadrupole emission used throughout this work, before presenting the formula for the energy-differential cross section.

We start by considering the initial (final) DM states $|\varphi_{i(f)}\rangle$  of the dissipative process to satisfy the eigenvalue equation $H_{0}\ket{\varphi_{i (f)}}=E_{i(f)}\ket{\varphi_{i(f)}}$, where $H_0$ is  Hamiltonian of the non-relativistic colliding particle-pair in the presence of a vector-mediated force,
\begin{equation} \label{hamiltonien}
    H_{0}= \frac{\Vec{p}_{1}^2}{2m_{1}} + \frac{\Vec{p}_{2}^2}{2m_{2}} + \frac{g_{1}g_{2} e^{-m_{V} |\Vec{x}_{1}-\Vec{x}_{2}|  }}{4 \pi|\Vec{x}_{1}-\Vec{x}_{2}|^2}. 
\end{equation} 
 If the force is mediated by the scalar, the substitutions $g_{1}g_{2}\rightarrow -y_{1}y_{2}$ and $m_{V}\rightarrow m_{\phi}$ apply in~\eqref{hamiltonien}. The transition matrix element for the emission of a 
vector $V^{\mu}$ or scalar $\phi$ of energy $\omega$, three-momentum $\vec{q}$  (and of helicity $\lambda$ for $V^\mu$) is given by
\begin{equation} \label{Smatrix}
\mathcal{T}_{fi}=-2\pi i \delta(E_{i}-E_{f}-\omega)\bra{\varphi_{f}} a(\Vec{q},\lambda){V}_{\text{int}}\ket{\varphi_{i}},
\end{equation}
where $a(\Vec{q},\lambda)$ is the raising operator for the energy state of the emitted particle. The interaction potential ${V}_{\text{int}}$ relevant to the emission process reads
\begin{subnumcases}{V_{\text{int}}=}
 - \sum_{n}{g_{n}}/{m_{n}} \vec{A}(\Vec{x}_{n}) \cdot \vec{p}_{n} &\text{for vector emission}, \label{potential1}\\
    \sum_{n} y_{n} \phi(\vec{x}_{n})&\text{for scalar emission}, \label{potential2}
\end{subnumcases}
where $\Vec{A}(\Vec{x}_{n})$ and $\phi(\vec{x}_{n})$ are the field operators in the interaction picture, with the time-dependence already factored out; $\vec x_n$ ($\vec p_n$) are the particle coordinates (three-momenta) participating in the emission process. 
One evaluates~\eqref{Smatrix} in the center-of-mass (CM) frame. We define
$\Vec{P}=\Vec{p}_{1}+\Vec{p}_{2}=0$ as the CM momentum, $\vec{p}_{i}=\mu (\vec{p}_{1}/m_{1}-\vec{p}_{2}/m_{2})$, $\vec{p}_{f}=\mu (\vec{p}_{3}/m_{1}-\vec{p}_{4}/m_{2})$, as the DM particles’ respective initial and final relative momenta,  $\mu = m_{1}m_{2}/(m_{1} + m_{2})$ as the reduced mass, and \( \vec{r}=\vec{x}_{1}-\vec{x}_{2}, \, \vec{X}= \left({\vec{x}_{1} m_{1}}/{\mu} + {\vec{x}_{2} m_{2}}/{\mu} \right) \) as the CM coordinates. In the CM frame, the transition matrix element~\eqref{Smatrix} acquires an overall phase \( e^{-i \vec{q} \cdot \vec{X}} \), which can be absorbed into the final state as $ 
    \bra{\varphi_ {\bar f}} = \bra{\varphi_f} e^{-i \vec{q} \cdot \vec{X}},$
corresponding to a final state with energy \( E_{\bar{f}} = E_f - \omega^2 / (2\mu) \), that is, to $E_{f}$ minus the recoil kinetic energy transferred to the system by the emitted particle. While this recoil energy is negligible in most of NR system, such that \( E_{\bar{f}} - E_i \simeq -\omega \), we retain the full expression for generality.

\subsubsection*{Vector emission}

In the case of vector emission, we work in the Coulomb gauge, where the polarization vector $\varepsilon^{\mu*}(q)=(0,\Vec{e}^{*})$ reduces to its spatial component \(\vec{e}^{*}\), which is transverse to the emitted vector momentum. Using the definition of the interaction potential given in~\eqref{potential1}, the $\mathcal{T}$- matrix element, up to the energy conserving delta function and the factor $-2\pi i$, takes the form
\begin{equation} \label{Mvectorbefore}
   \mathcal{T}_{fi} \propto -\frac{g_{1}}{\sqrt{2 \omega}}\bra{\varphi_{\bar f}}\frac{\Vec{p}_{i}\cdot \Vec{e}^{*}}{m_{1}}e^{-{i \mu \Vec{r} \cdot \Vec{q} }/{m_{1}}}\ket{\varphi_{i}} + \frac{g_{2}}{\sqrt{2 \omega}}\bra{\varphi_{\bar f}} \frac{\Vec{p}_{i}\cdot \Vec{e}^{*}}{m_{2}} e^{ {i \mu \Vec{r} \cdot \Vec{q} }/{m_2}}\ket{\varphi_{i}}.
\end{equation}
This result can be further simplified when the wavelength of the emitted particle is larger than the typical separation $|\vec{r}|$ of particles from the center of mass, i.e., $\nu \equiv \Vec{r}\cdot\Vec{q} \ll 1$. Expanding the exponential in~\eqref{Mvectorbefore} in $\nu$, and integrating over $d^3\vec X$ yields
 $\mathcal T_{fi} = -i(2\pi)^4 \delta(E_{i}-E_{f}-\omega)\delta^{(3)}(\Vec{p}_{1}+\Vec{p}_{2}-\Vec{p}_{3}-\Vec{p}_{4}-\Vec{q}) \mathcal{M}$ where
\begin{equation} \label{Mvector}
    \mathcal{M} \simeq -\frac{1}{\sqrt{2 \omega}}\left(\frac{g_{1}}{m_{1}}- \frac{g_{2}}{m_{2}}\right) \bra{f} \Vec{p}_{i}\cdot \vec{e}^{*}\ket{i} +i\frac{1}{\sqrt{2 \omega}}\left(\frac{g_{1}}{m_{1}^2}+ \frac{g_{2}}{m_{2}^2}\right) \mu \bra{f}(\vec{p}_{i} \cdot \vec{e}^{*} )(\vec{r}\cdot \Vec{q})\ket{i} 
    .
\end{equation}
Here, $\ket{i,f}$ are the initial and final scattering states of the DM pair in the CM frame.\footnote{The matrix elements derived in the subsequent sections  are obtained in relativistic QFT normalization; their translation to the non-relativistic QM normalization of the current section reads $|\mathcal{M}_{\text{QFT}}|^2 = 2\omega (2 m_{1})^2 (2m_{2})^2|\mathcal{M}_{\text{QM}}|^2.$}
The first term in~\eqref{Mvector} describes {\it electric dipole} radiation and cancels in the case of equal charge-to-mass ratio. 

In our setting, i.e., for a two-particles system without explicit magnetic moment interaction, for massless vector particles the second term in~\eqref{Mvector} corresponds to \textit{electric quadrupole radiation}.\footnote{If spin effects are included, the magnetic moment yields an additional interaction potential~$V_{\text{int}}^{\prime}=-\sum_{n} \boldsymbol{\mu}_{n} \cdot \left( \text{curl} \, \vec{A}(\Vec{x}_{n}) \right)$, where $|\boldsymbol\mu_{n}|=g_{n}/(2m_{n})$ for a $1/2$-spin particle. In a NR system, however, the latter can be neglected compared to electric ones both at dipole and quadrupole order~\cite{Gould:1981an}.} 
Using transversality of the polarization state and the canonical commutation relations, the operator in the bracket can be reduced to its symmetric part,
\begin{equation} \label{symmetryquadope}
e^{* l}q^{j}\bra{f}p_{i}^{l}r^{j}\ket{i}= \frac{e^{*l}q^{j}}{2}\bra{f}{\{p_{i}^{l},r^{j}\}}\ket{i}.
\end{equation}
Squaring the corresponding matrix element, summing over the transverse polarization states and averaging over the emitted particle direction $\hat{q}$ yields
\begin{equation} \label{demo}
\langle |\mathcal{M}_{\mathcal{Q}}|^2 \rangle_{\hat{q}}= \frac{ \mu^2 \omega }{8} \left(\frac{g_{1}}{m_{1}^2}+ \frac{g_{2}}{m_{2}^2}\right)^2 \Big\langle \sum_{\lambda}{\oldhat{q}}^{j}{\oldhat{q}}^{k}{e}_{(\lambda)}^{*l}{e}_{(\lambda)}^{m} \Big\rangle_{\hat{q}} \bra{f}\{p_{i}^{l},r^{j}\}\ket{i} \bra{f}\{p_{i}^{m},r^{k}\}\ket{i}^{*},
\end{equation}
where the subscript $\mathcal{Q}$ labels the quadrupole nature of the emission. The angular average decomposes into symmetric and anti-symmetric parts, 
\begin{equation}\label{averagingquadrupole1}
\begin{aligned}
\Big\langle \sum_{\lambda}{\oldhat{q}}^{j}{\oldhat{q}}^{k}{e}_{(\lambda)}^{*l}{e}_{(\lambda)}^{m} \Big\rangle_{\hat{q}}
&=
\frac{1}{15}\left[ \frac{5}{2}\left(\delta^{jk}\,\delta^{lm}-\delta^{jm}\,\delta^{kl}\right)
+\frac{3}{2}\left(-\frac{2}{3}\,\delta^{jl}\,\delta^{km}
+\delta^{jm}\,\delta^{kl}
+\delta^{jk}\,\delta^{lm}\right) \right],
\end{aligned}
\end{equation}
respectively.
The first, antisymmetric part cancels in the contraction~\eqref{demo}, yielding the familiar, trace-free Cartesian quadrupole form,
\begin{equation} \label{operatorquadrupoleV}
    \langle |\mathcal{M_{Q}}|^2\rangle = \frac{\mu^2 \omega}{120} \left(\frac{g_{1}}{m_{1}^2}+ \frac{g_{2}}{m_{2}^2}\right)^2 \left[ 3|\bra{f} \{ r^{k},p_{i}^{l} \}
    \ket{i}|^2 - |\bra{f} \{r^{l},p_{i}^{l} \}\ket{i}|^2 \right]. 
\end{equation}
We note in passing, that the absence of \textit{magnetic dipole} contributions is also in agreement with the classical expectation of vanishing magnetic dipole radiation for a two-particle system~\cite{Landau:1975pou}.

\subsubsection*{Scalar emission}
We now consider scalar emission. Using~\eqref{potential2}, the corresponding $\mathcal{T}$-matrix element decomposes analogously to the previous case,
\begin{equation} \label{Mscalarbefore}  
  \mathcal{T}_{fi} \propto y_{1}\bra{\varphi_{\bar f}}e^{-{i \mu \Vec{r} \cdot \Vec{q} }/{m_{1}}}\ket{\varphi_{i}} + y_{2}\bra{\varphi_{\bar f}}e^{ {i \mu \Vec{r} \cdot \Vec{q} }/{m_2}}\ket{\varphi_{i}}.
\end{equation}
By assuming that $\nu \ll 1$, the expansion of the exponentials leads to
\begin{equation} \label{Mscalar}
  \mathcal{M} \simeq  -\frac{i}{\sqrt{2\omega}}\left(\frac{y_{1}}{m_{1}}- \frac{y_{2}}{m_{2}}\right)\mu \bra{ f} \Vec{r}\cdot \Vec{ q }\ket{i}
    -\frac{1}{2\sqrt{2\omega}}\left(\frac{y_{1}}{m_{1}^2}+ \frac{y_{2}}{m_{2}^2}\right)\mu^2 \bra{f}(\Vec{r}\cdot \Vec{q} )^2\ket{i}.
\end{equation}
In contrast to the vector emission, the $\mathcal{O}(\nu^0)$ order vanishes, and~\eqref{Mscalar} is composed of second- and third-order-in-$\nu$ terms. We, of course, encounter this structure again in the Feynman-diagrammatic calculations; see App.~\ref{scalarvsgauge}. In analogy to the case of vector emission, the first term in~\eqref{Mscalar}, which cancels in the case of equal coupling-to-mass ratio, is identified as a dipole transition. 

By squaring and averaging the second term over the direction of the emitted particle, one obtains a contraction with
\begin{equation} \label{averagingquadrupole2}
    \left\langle {\oldhat{q}}^{j}{\oldhat{q}}^{k}{\oldhat{q}}^{l}{\oldhat{q}}^{m} \right\rangle_{\hat{q}} = \frac{1}{15}\left( \delta^{jm} \delta^{kl} + \delta^{jk} \delta^{lm} +\delta^{jl}\delta^{km} \right). 
\end{equation}
The second term of~\eqref{Mscalar} then yields the contribution
\begin{equation} 
\label{quadrupolescalaruniversal}
\langle |\mathcal{M_Q}|^2 \rangle_{\hat{q}}= \left(\frac{y_{1}}{m_{1}^2}+ \frac{y_{2}}{m_{2}^2}\right)^2 \frac{\mu^4 \omega^3 }{120}  \left[  2|\bra{f} r^{k}r^{l} \ket{i}|^2 + |\bra{f} r^{l}r^{l} \ket{i}|^2  \right],
\end{equation}
which, in contrast to~\eqref{operatorquadrupoleV}, includes a trace component. Throughout the paper, we still refer to such scalar emission as being ``quadrupole'', having in mind the relative {\it velocity ordering} rather than a specific tensorial structure.
Indeed, as can be seen from~\eqref{Mvector} and~\eqref{Mscalar}, quadrupole radiation is of higher order in velocity when compared to the dipole case: in the amplitudes by a factor of $\vec{r}\cdot \vec{q} \sim v_{i} $, and in terms of emission rates, by a factor of $v_i^2$. Therefore, dipole radiation dominates, unless it vanishes because of an identical charge-to-mass ratio of the colliding particle pair. In turn, this means that for identical particle scattering, the quadrupole order is the dominant emission mode.
At this point, we note that the Hamiltonian in~\eqref{potential2}---on which~\eqref{quadrupolescalaruniversal} is based on---receives model-dependent corrections at quadrupole order; the expressions derived below take this into account.

Furthermore, in analogy to the derivation of the matrix elements presented within this section, we always first expand the dissipative amplitudes in $v_{i}$ or $r$ before squaring them in order to avoid any interferences between the dipole and quadrupole contributions to the emission amplitude. 
The non-relativistic emission amplitudes of the $2\rightarrow3$ processes involving scalar candidates $S (\Tilde{S})$ are obtained through a long but straightforward calculation. For dissipative channels involving fermionic DM candidates $\chi(\Tilde{\chi})$, there is also a $v_{i}$-expansion of spinors involved, involving hence extra steps; see App.~\ref{Powercounting} for details on the methodology. The derivation yields insights into how the different dissipative models are related in detail; see App.~\ref{scalarvsgauge}.

\subsubsection*{Energy-differential cross section}

Having clarified the principal leading analytical structure of emission amplitudes, we close this section by quoting the NR formula for the energy-differential cross section in the CM frame
\begin{equation} \label{crosssection}
   \frac{d\sigma }{d\omega} = \frac{  S_{f} \mu}{2^7 \pi^3m_{1}^2m_{2}^2|\Vec{p}_{i}|}\int_{0}^{\infty} d|\Vec{p}_{f}| \, |\Vec{p}_{f}|^2 \int_{-1}^{1} d \cos{\theta_{if}} \, \omega \sqrt{1-\frac{m_{\phi,V}^2}{\omega^2}} \frac{ \langle \, |\mathcal{M}|^2 \rangle_{\Hat{q}}}{d_1 d_2} \, \delta{\left(\frac{|\vec{p}_{i}|^2}{2\mu} - \frac{|\vec{p}_{f}|^2}{2\mu} -\omega \right)} . 
\end{equation}
Here, $\langle \, |\mathcal{M}|^2 \rangle_{\Hat{q}}$ is the squared matrix element, summed over \emph{both}, initial and final DM spins~$s_j$, with spin multiplicity $d_j=2s_j+1$, and averaged over the directions $\Hat q$ of the emitted particle three-momentum $\vec q$; $S_{f}$ is the symmetry factor of the particles in the final state ($1/2$ for identical particle and $1$ otherwise) and  $\theta_{if}$ is the scattering angle between the relative initial and final DM momenta, $\vec p_i$ and $\vec p_f$, respectively. The emitted particle is on-shell and has an energy   
\begin{equation}
\omega = \sqrt{|\Vec{q}|^2+ m_{\phi,V}^2} =
\frac{|\Vec{p}_{i}|^2}{2 \mu}-\frac{|\Vec{p}_{f}|^2}{2 \mu} .
\end{equation}

\subsection{Dipole transitions} \label{Leading order}
We now present the leading expressions of the amplitude in the expansion of initial relative DM velocity~$v_{i}$ and of the contact parameter $r$. These expressions are non-vanishing when the charge-to-mass ratios of incoming DM particles differ, generically the case of $m_{1} \neq m_{2}$ or opposite sign charges.

\subsubsection*{Long-range mediated force} \label{sectionlabellongdipole}
First, we consider the case of a long-range mediated force.  After squaring the amplitude, averaging over the initial spins, summing over the final ones, and integrating over the direction of the emitted particle, we obtain
\begin{equation} \label{generaldipole}
  \frac{1}{d_{1} d_{2}}  \langle |\mathcal{M_{D}} |^2 \rangle_{\Hat{q}}=  \frac{\mathcal{F}_{\mathcal D}^2 m_{{1}}^2m_{{2}}^2 }{\mu^2} \frac{ 16 f_{\phi,V} }{3 \omega^2 |\Vec{v}_{i}-\Vec{v}_{f}|^2 } .
\end{equation} 
Here, $ \vec{v}_{i,f} = \vec p_{i,f}/\mu$ are the relative velocities of the DM pairs in the initial ($i$) and final ($f$) state; $d_{1},d_{2}$ are the number of spin states in the initial state and
the dipole prefactors $\mathcal{F_D}$ read
\begin{equation} \label{prefactordipole}
\begin{aligned}
 \mathcal{F_D} = 
 \begin{cases}
     \displaystyle{ g_{1}g_{2} \left( \frac{g_{1}}{m_{{1}}} -  \frac{g_{{2}}}{m_{{2}}}  \right) } 
     & \text{for } \ \chi_1 \chi_2\to \chi_1 \chi_2 V,\ S_1 S_2\to S_1 S_2 V, \\[10pt]
     \displaystyle{ y_{{1}}y_{{2}} \left( \frac{y_{1}}{m_{{1}}} -  \frac{y_{2}}{m_{{2}}}  \right) } 
     & \text{for } \ \chi_1 \chi_2\to \chi_1 \chi_2 \phi,\   \tilde \chi_1 \tilde\chi_2\to \tilde\chi_1 \tilde\chi_2 \phi , \\[10pt]
     \displaystyle{ \frac{1}{8} \frac{A_{1}}{m_{{1}}} \frac{A_{ {2}} }{m_{{2}}}
     \left(\frac{A_{1}}{m_{1}^2}-\frac{A_{2}}{m_{2}^2} \right)  }  
     & \text{for } \ S_1 S_2\to S_1 S_2 \phi , \ \tilde S_1 \tilde S_2\to \tilde S_1 \tilde S_2 \phi .
 \end{cases}
\end{aligned}
\end{equation}

The factors $f_{\phi}=1$ for scalar emission and $f_{V}=2$ for vector emission reflect the difference that arises in the angular average of the squared matrix elements,
\begin{align} \label{averageqrelations}
    \left\langle \oldhat q^i \oldhat q^j \right\rangle_{\Hat q} = \frac{1}{3} \delta^{ij} & \quad \text{for scalar emission}, \\
    \left\langle \sum_{\lambda} e_{(\lambda)}^i  e_{(\lambda)}^{*j} \right\rangle_{\Hat q} = \frac{2}{3}\delta^{ij}& \quad \text{for vector emission}. \label{averageqrelationsvector}
\end{align}
 Here, $ \oldhat q^i$ and $ e^{*j}$ are the Cartesian components of the  emission direction $\Hat q = \vec q/|\vec q|$ and of the transverse polarization vectors $\vec e_{(\lambda)}^*$ in Coulomb gauge;  the polarization sum only needs to run over the two transverse polarizations, as the emission of the longitudinal mode of $V$ is suppressed as $m_V^2/\omega^2$ (see below.)

The model-independence of~\eqref{generaldipole} arises from the dominance of external-leg emission diagrams relative to others, making it the exclusive contribution to the emission amplitude at leading order in~$v_i$. Within the amplitude of these dissipative channels, the emitted momentum scales as \( |\vec{q}|/\mu \sim \mathcal{O}(v_i^2) \), causing all $\vec{q}$-dependent contributions in the emission current to become negligible at leading order; see~\eqref{withoutemissionYuka} and~\eqref{withoutemissionQED}. Once these contributions are neglected, the amplitude reduces to a universal form, where the nature of the emitted particle and of the DM candidates only enters as corrections in the quadrupole emission amplitude. We refer the reader to App.~\ref{scalarvsgauge} for a detailed discussion. We note that, in the time-dependent perturbation theory framework of Sec.~\ref{dipolevsquadrupole}, this universality arises from neglecting the recoil energy \cite{Weinberg_2015}---i.e., neglecting the $\vec{q}$-corrections---which leads the operators of the first term of the matrix elements~\eqref{Mvector} and~\eqref{Mscalar} to become equivalent\footnote{This equivalence is obtained by using the commutator relation \( [H_0, \vec{r}]\ket{\varphi_{i}} = -i \vec{p}_i / \mu \ket{\varphi_{i}} \),
\begin{equation}
    \bra{\varphi_{\bar f}} \vec{p}_{i} \ket{\varphi_{i}} = i\mu(E_{\bar f} - E_{i}) \bra{\varphi_{\bar f}} \vec{r} \ket{\varphi_{i}} \simeq -i\mu \omega \bra{\varphi_{\bar f}} \vec{r} \ket{\varphi_{i}}.
\end{equation}} up to a factor~$i\mu\omega$.

This model-independent behavior of~\eqref{generaldipole} for scalar and vector emission from Dirac fermions was already noticed in \cite{2303.00778}. In addition to agreeing with their work, we have added the case of scalar emission from scalar DM candidates~$S_j$ and~$\tilde S_j$. 
Allowing the emitted particle to carry a kinematically allowed mass, we may, in fact go further and obtain the mass-dependent corrections
of the emission amplitudes,
\begin{equation} \label{dipolescalar}
    \langle |\mathcal{M_{D}}|^2 \rangle_{\hat{q}} = \left( 1 - \frac{m_{\phi}^2}{\omega^2} \right) \langle |\mathcal{M_{D}}|^2 \rangle_{\hat{q},{m_{\phi}=0}}\quad \text{for scalar emission}.
\end{equation}
Below, the factor in parentheses is denoted by $C_{\phi}$. 
This result is in agreement in the case of emission from Dirac fermions~$\chi$ with \cite{2303.00778,2303.03123}, %
 where the dipole amplitude was derived for electron-nucleon scalar bremsstrahlung in the $m_{e} \gg m_{p}$ limit; $m_{e(p)}$ is the electron (proton) mass. For massive gauge vector emission, taking into account the longitudinal part of the polarization vector $V^{\mu}$, the NR leading order emission amplitude becomes
\begin{equation} \label{dipolegauge}
    \langle |\mathcal{M_{D}}|^2 \rangle_{\hat{q}} = \left( 1 + \frac{m_{V}^2}{2\omega^2} \right)  \langle |\mathcal{M_{D}}|^2 \rangle_{\hat{q},{m_{V}=0}} \quad \text{for vector emission}.
\end{equation}
Hereafter we refer to the factor in parentheses as~$C_{V}$. 

We point out that the $m_{\phi,V}$-dependent terms entering~\eqref{dipolescalar} and~\eqref{dipolegauge}, prior to squaring, averaging over the emitted particle direction, and---in the latter---summing over polarizations, possess the same analytic structure for $\phi$-emission and for the longitudinal mode of $V$-emission. This results from the combined effects of  Ward identity and NR reduction of emission amplitudes; see App.~\ref{Massiveemission} for more details. 
The differences in the energy-differential cross sections between scalar and vector emission are then captured in overall factors $C_\phi$ and $C_V$, respectively. 
Given a fraction $x\equiv 2 \omega / \mu v_{i}^2 \leq 1$  of CM energy carried away by the emitted particle, the energy-differential cross section for dipole emission then reads, 
\begin{equation} \label{dipolelongrange}
\centering
     \omega \frac{d \sigma}{d \omega} = C_{\phi,V} f_{\phi,V} \mathcal{F}_{\mathcal D}^2  \sqrt{1-\frac{m_{\phi,V}^2}{\omega^2}} \frac{ 1   }{24 \pi^3  v_{i}^2}\ln{\left(\frac{1 + \sqrt{1-x}}{1-\sqrt{1-x}}\right)} . 
\end{equation}

\begin{figure}[h!]
    \centering
\includegraphics[width=0.49\linewidth]{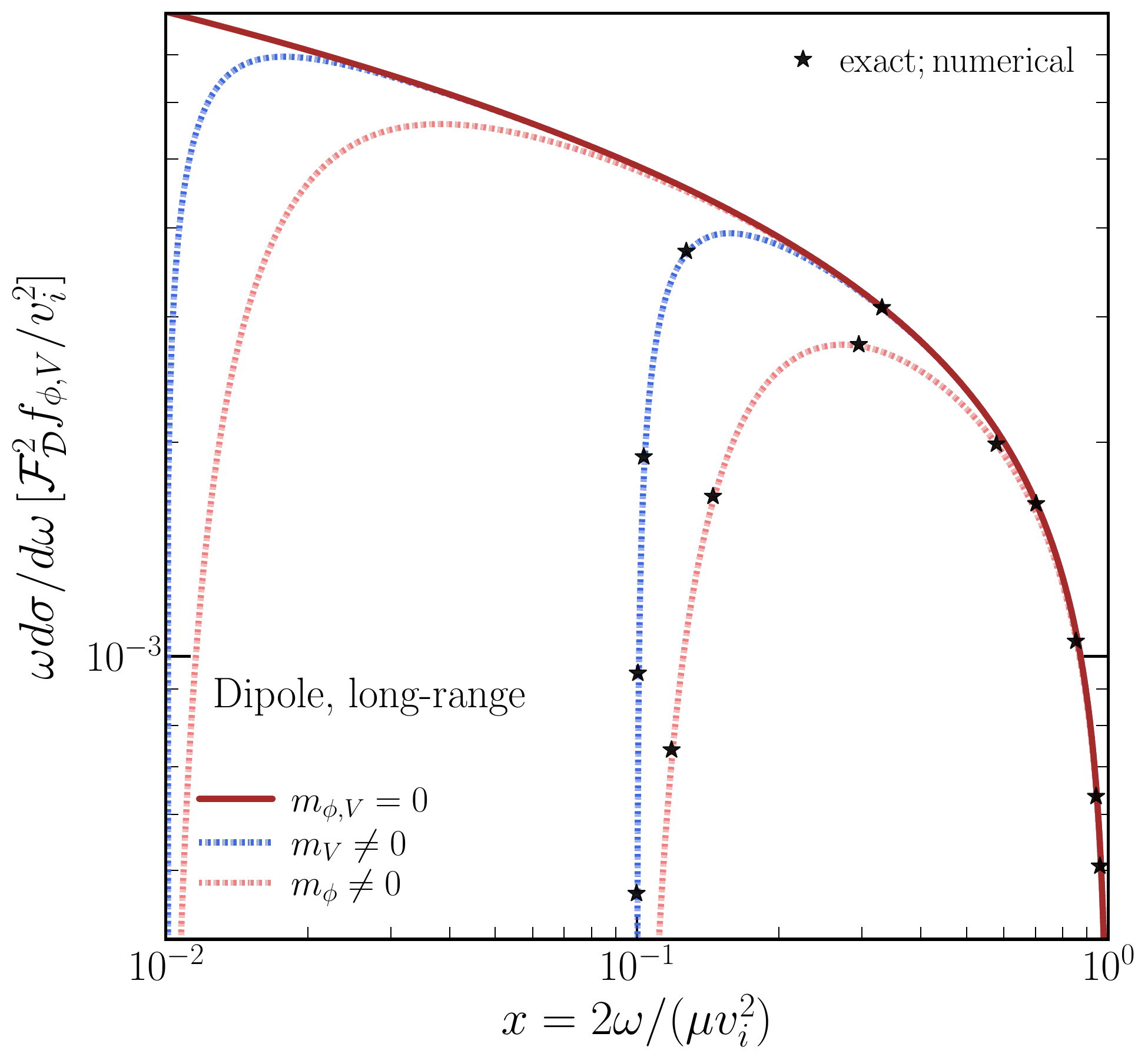}
\includegraphics[width=0.49\linewidth]{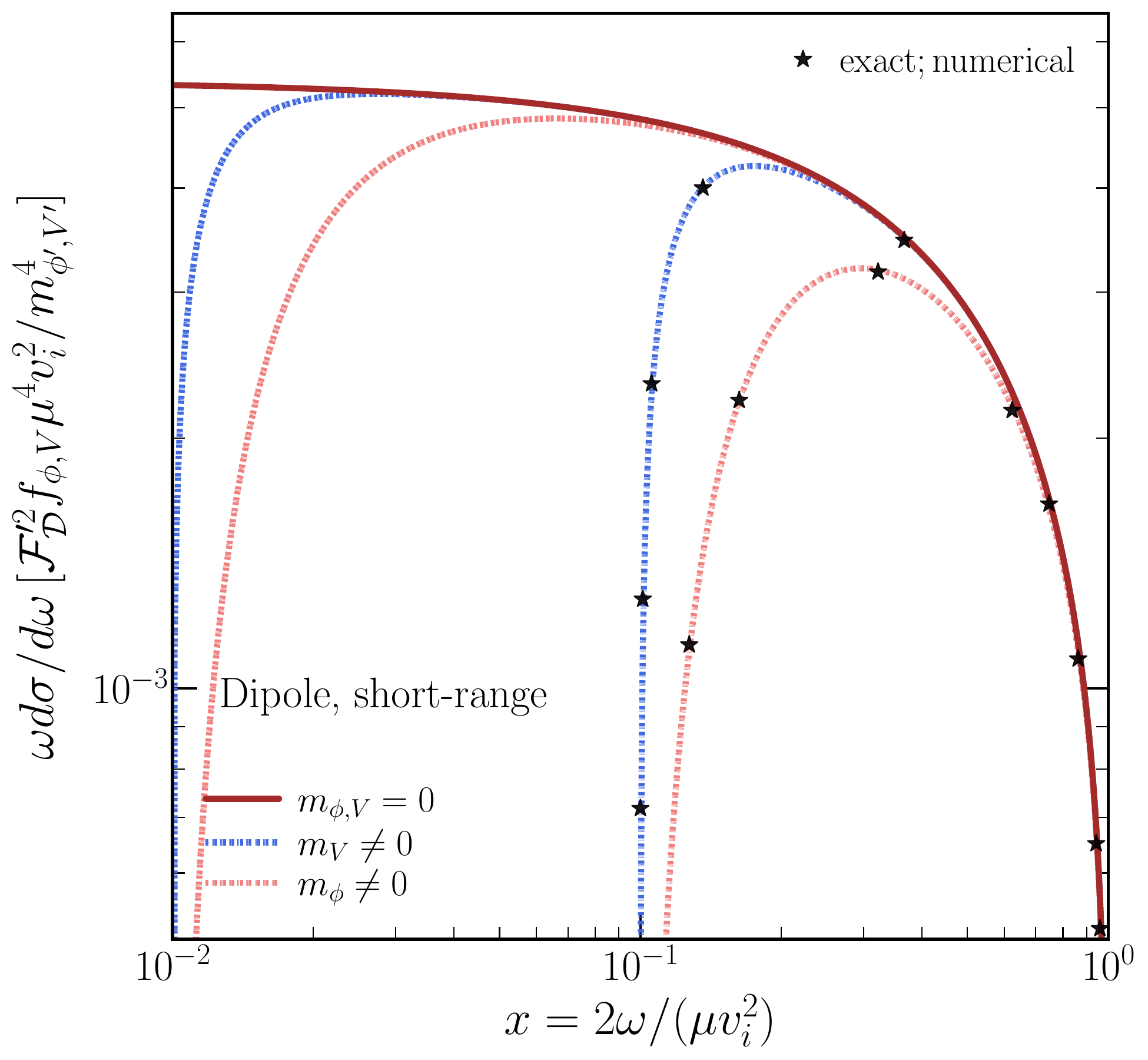}
    \caption{Energy-differential cross section for dipole emission in distinguishable particle scattering. The solid brown lines correspond to massless emission.
    The dotted colored lines show massive emission with~$m_{\phi,V}/\omega_{\text{max}} = 0.01 $ (leftmost) and~$m_{\phi,V}/\omega_{\text{max}} =0.1 $ (rightmost). The black stars indicate values of the energy-differential cross section obtained from a fully numerical integration of unexpanded amplitudes. 
    \textit{Left panel}: Dipole, long-range mediation~\eqref{dipolelongrange}. Numerical integration parameters: $m_{2}=1$ MeV, $m_{1}=1$ GeV, and $v_{i} \approx 4 \times 10^{-3}$. \textit{Right panel}: Dipole, short-range mediation~\eqref{dipoleshortrange}. Numerical integration parameters: $m_{2}=1$ MeV, $m_{1}=1$ GeV, $m_{\phi',V^{\prime}}=0.1$ GeV and $v_{i} \approx 4 \times 10^{-3}$.   }
    \label{leadingcross}
\end{figure}

In the left panel of Fig.~\ref{leadingcross}, we compare the analytic expression~\eqref{dipolelongrange} with the exact numerical integration of the unexpanded amplitude. We 
assume dissipative processes involving distinguishable particles with \( m_1 \gg m_2 \) and $\left\{g_{1},y_{1},A_{1}/m_{1} \right\}\sim \left\{ g_{2},y_{2},A_{2}/m_{2} \right\}$, ensuring that dipole emission provides the leading contribution to the energy-differential cross section. We maintain identical parameter values across both cases, i.e., we fix the same value of $v_{i},m_{j},m_{\phi,V}$ and of the couplings $g_{j},y_{j}$ and $A_{j}/m_{j}$. The comparison is performed for massive emission, and we find excellent agreement between the full numerical and analytical results.

\subsubsection*{Short-range mediated force} \label{Shortrangedipole}
We continue by considering the case of a short-range mediated force.  
The squared dipole emission amplitude, averaged over the initial spins, summed over the final ones, and integrated over the direction of the emitted particle is
\begin{equation} \label{dipoleshortrangeamplitude}
  \frac{1}{d_{1} d_{2}}  \langle |\mathcal{M_{D}} |^2 \rangle_{\hat{q}}=  \mathcal{F}_{\mathcal D}'^2 m_{{1}}^2m_{{2}}^2  \frac{ 16 f_{\phi,V} \mu^2 |\Vec{v}_{i}-\Vec{v}_{f}|^2 }{3 \omega^2 m_{\phi',V'}^4} \quad (r \ll 1,\, m_{\phi',V'} < \sqrt{s} ).
\end{equation} 
Similarly to the case of long-range mediation, we find that the kinematic part of the dipole emission amplitude is model-independent.
In the current scenario, where the mediator and the emitted particle differ, primed coupling constants $(\prime)$ are introduced that describe the mutual interaction of the colliding DM pair. Therefore, the modified dipole emission factors $\mathcal{F_D}$  for fermionic DM candidates now read
\begin{equation} \label{shortfermiondipole}
\begin{aligned}
\mathcal{F'_{D}} =
 \begin{cases}
  \displaystyle{  
   g'_{1}g'_{2} \left( \frac{g_{1}}{m_{{1}}} -  \frac{g_{{2}}}{m_{{2}}}  \right) }
      &\text{for }\ \chi_1 \chi_2\to \chi_1 \chi_2 V,\\[10pt]
  
  \displaystyle{  
   y'_{{1}}y'_{{2}} \left( \frac{y_{1}}{m_{{1}}} -  \frac{y_{2}}{m_{{2}}}  \right)   }
      &\text{for }\ \chi_1 \chi_2\to \chi_1 \chi_2 \phi,\   \tilde \chi_1 \tilde\chi_2\to \tilde\chi_1 \tilde\chi_2 \phi.\\[10pt]
 \end{cases}
\end{aligned}
\end{equation} 
For scalar DM candidates, the dipole emission factors contain both, the contribution from external leg and contact emission diagrams, Fig.~\ref{inifini} and Fig.~\ref{Contacts},

\begin{equation} \label{shortscalardipole}
\begin{aligned}
\mathcal{F^{\prime}_{D}} =
 \begin{cases}
  \displaystyle  
   
    \left(g'_{1}g'_{2} + \dfrac{m_{V'}^2 \lambda_{S}}{4 m_{1} m_{2}} \right) 
    \left( \dfrac{g_{1}}{m_{1}} -  \dfrac{g_{2}}{m_{2}} \right)
    
      & \text{for } S_1 S_2 \to S_1 S_2 V, \\[12pt]
  \displaystyle  
   
    \dfrac{1}{8} \left( \dfrac{A'_{1}}{m_{1}} \dfrac{A'_{2}}{m_{2}} - \dfrac{m_{\phi'}^2 \lambda_{S} }{m_{1} m_{2}} \right)
    \left(\frac{A_{1}}{m_{1}^2}-\frac{A_{2}}{m_{2}^2} \right) 
     & \text{for }
     \begin{array}[t]{l}
     S_1 S_2 \to S_1 S_2 \phi, \\
     \tilde{S}_1 \tilde{S}_2 \to \tilde{S}_1 \tilde{S}_2 \phi.
     \end{array}
 \end{cases}
\end{aligned}
\end{equation}

Corrections to the squared amplitude for massive particle emission are identical to \eqref{dipolescalar} and~\eqref{dipolegauge}, and the factors $C_{\phi,V}$ remain unaltered. The dipole energy-differential cross section in the short-range mediated force regime is then given by
\begin{equation} \label{dipoleshortrange}
\omega \frac{d \sigma }{d \omega} = f_{\phi,V} C_{\phi,V}    \mathcal{F}_{\mathcal D}'^2  \sqrt{1-\frac{m_{\phi,V}^2}{\omega^2}}      \frac{\mu^4  v_{i}^2 \sqrt{1-x} \left(2-x\right)}{12 \pi^3 m_{\phi',V'}^4  } \quad (r \ll 1, \, m_{\phi',V'} < \sqrt{s}) .
\end{equation}

In the right panel of Fig.~\ref{leadingcross}, we compare the analytic expression~\eqref{dipoleshortrange} with the exact numerical integration of the full, unexpanded amplitude. The comparison is performed for massive emission, assuming dissipative processes involving distinguishable particles with \( m_1 \gg m_2 \) and $\left\{g_{1},y_{1},A_{1}/m_{1} \right\}\sim \left\{ g_{2},y_{2},A_{2}/m_{2} \right\}$, such that dipole emission dominates the energy-differential cross section. Identical parameter values are used in both cases, with fixed values of \( v_i \), \( m_j \), $m_{\phi,V}, \,m_{\phi',V'}$ and the couplings $g_{j}^{(\prime)},y_{j}^{(\prime)},A^{(\prime)}_{j}/m_{j}$ and $\lambda_{S}$. The numerical results are found to match the analytical predictions, thereby validating the latter.

\subsection{Quadrupole transitions} \label{quadrupoleemission}
When the scattering DM particles have the same coupling-to-mass ratio,  the dipole emission amplitudes derived in the previous section vanish, and the process is of higher order in relative velocity. 
\subsubsection{Scattering of distinguishable particles} \label{quadrudisting}
We start by discussing quadrupole radiation for distinguishable DM particle scattering, i.e., the limit of canceling coupling-to-mass ratios of the DM pair.
\paragraph{Long-range mediation} In the case of a long-range mediated force,  the mediator is the same as the emitted particle. The squared emission amplitude, averaged over  initial spins, summed over  final spins, and averaged over the direction of the emitted particle reads
\begin{align} \label{generalquadrupole}
   \frac{1}{d_{1} d_{2}} \langle |\mathcal{M_{Q}}|^2 \rangle_{\hat{q}} & = 16 f_{\phi,V}m_{{1}}^2m_{{2}}^2\mathcal{F}_{\mathcal Q}^2 \frac{1 }{15 \omega^2 |\Vec{v}_{i}-\Vec{v}_{f}|^4} \nonumber \\
   & \times  \left\{ \alpha\left[|\Vec{v}_{f}|^4+ |\Vec{v}_{i}|^4 + |\Vec{v}_{i}|^2 |\Vec{v}_{f}|^2 \left(2- 4( \oldhat{\Vec{v}}_{i} \cdot \oldhat{\Vec{v}}_{f})^2\right) \right] + \beta \left(|\Vec{v}_{i}|^2-|\Vec{v}_{f}|^2\right)^2  \right\}.
\end{align}
Here, $\Hat{v}_{i,f} = \vec{v}_{i,f} /|\vec{v}_{i,f}|$ and $(\alpha, \beta)$ are model-dependent numerical coefficients that are provided below.
The squared amplitudes are obtained by taking the transverse polarization sum in the case of vector boson emission.  The averaging over the direction of the emitted particle involves the identities given in~\eqref{averagingquadrupole1} and~\eqref{averagingquadrupole2}.

At quadrupole order, the universality of emission amplitudes in the NR expansion across the various cases is now lost, so that the additional coefficients $\alpha$ and $\beta$ parameterizing the differences are introduced. The model dependence stems from differing NR currents for scalar and vector emission from fermions at this order in the $v_i$-expansion and a loss of symmetry under the exchange of $\vec q $ and $\vec{e}^{*}$; see the respective equations~\ref{withemissionYukaq} and~\ref{withemissionQED} in App.~\ref{scalarvsgauge}.
The quadrupole emission factors, together with those coefficients, read
\begin{equation} \label{factorQ}
\begin{aligned}
\left\{\mathcal{F_{Q}},\alpha,\beta \right\} =
 \begin{cases}
  \displaystyle{  \left\{ g_{{1}} g_{{2}} \left( \frac{g_{{1}}}{m_{{1}}^2} + \frac{g_{{2}}}{m_{{2}}^2} \right)  , \frac{3}{4} , \frac{13}{4} \right\} }
      &\text{for }\ \chi_1 \chi_2\to \chi_1 \chi_2 V,\ S_1 S_2\to S_1 S_2 V ,\\[10pt]
       \displaystyle{ 
      \left\{
     y_{{1}} y_{{2}}  \left( \frac{y_{{1}}}{m_{{1}}^2} + \frac{y_{{2}}}{m_{{2}}^2}  \right), 1,6 \right\} }
      &\text{for }\ \chi_1 \chi_2\to \chi_1 \chi_2 \phi,\   \tilde \chi_1 \tilde\chi_2\to \tilde\chi_1 \tilde\chi_2 \phi ,\\[10pt]
    \displaystyle{  \left\{ \frac{1}{8} \frac{A_{{1}}}{m_{{1}}} \frac{A_{{2} } }{m_{{2}}}\left(\frac{A_{1 }}{m_{1}^3} + \frac{A_{2}}{m_{2}^3}   \right) , 1, 11 \right\} }
     &
     \parbox[t]{0.6\textwidth}{
     \text{for $\ S_1 S_2\to S_1 S_2 \phi , \ \tilde S_1 \tilde S_2\to \tilde S_1 \tilde S_2 \phi  $} \\[8pt] \text{\qquad and $ (A_{j}/m_{j})^2 \gg \lambda_{j}$ .}}.
 \end{cases}
\end{aligned}
\end{equation} 
The energy-differential cross section associated with the quadrupole emission amplitude~\eqref{generalquadrupole} is given by
\begin{equation} \label{crosssectionquadrupole}
   \omega \frac{d\sigma}{d\omega} = f_{\phi,V} \mathcal{F}_{\mathcal Q}^2 \frac{  \mu^2   }{60 \pi^3 } \left[ (\beta-\alpha)\sqrt{1- x } + 
   \alpha(2-x) \ln{\left(\frac{ 1 + \sqrt{1-x}}{1-\sqrt{1-x}}\right)} \right] .
\end{equation}
For vector emission, this expression agrees with the Born limit of~\cite{Pradler:2020znn}.

In the case of scalar emission $\phi$ from scalar DM candidates, we now consider the regime where quartic interactions between the mediator and the colliding particles dominate, $\lambda_{j} \gg (A_{j}/m_{j})^2 $. In this limit, the leading-order of the emission amplitude is dominated by the contribution of the vertex diagrams illustrated in Fig.~\ref{vertex},  
\begin{equation} \label{limitbasse}
\begin{aligned}
\langle |\mathcal{M}_{\text{vertex}}|^2 \rangle_{\hat{q}} &= \frac{\left(A_{1} \lambda_{2} + A_{2}\lambda_{1}\right)^2}{\mu^4|\Vec{v}_{i}- \Vec{v}_{f}|^4}\qquad (\lambda_{j} \gg (A_{j}/m_{j})^2 ).
\end{aligned}
\end{equation}
It is important to emphasize that this contribution, although scaling as \( \mathcal{O}(v_i^{-4}) \), does not originate from quadrupole emission. Its overall coupling dependence is reduced by two powers compared to the latter.
The associated energy-differential cross section reads
\begin{equation} \label{crossvertexlong}
   \omega \frac{d \sigma }{d \omega } = \left(A_{1} \lambda_{2} + A_{2}\lambda_{1}\right)^2\frac{ \sqrt{1-x} }{256 \pi^3 m_{{1}}^2 m_{{2}}^2  }  \qquad ( \lambda_{j} \gg (A_{j}/m_{j})^2 ).
\end{equation}
Expressions for the full amplitudes and differential cross sections that are not in any limit of $(A_{j}/m_{j})^2$ relative to $\lambda_{j}$ are given in App.~\ref{scalarscalarquad}.

As in the case of dipole emission, we may take into account the emitted particle mass~$m_{\phi,V}$. In contrast to~\eqref{dipolescalar} and~\eqref{dipolegauge}, there is no universal mass-dependent factor that could be pulled out in front of an amplitude of vanishing emitted particle mass. However, {\it prior} to squaring the full amplitude, the $m_{\phi,V}$-dependent terms in $\phi$-emission and in the longitudinal mode of $V$-emission agree; see App.~\ref{Massiveemission} for a detailed discussion.
We further note that in the case of massive scalar emission $\phi$, in the limit $ (A_{j}/m_{j})^2 \ll \lambda_{j}$, the NR expansion of the emission amplitude does not gain any dependency on~$m_\phi$, so that any mass-dependent corrections are solely due to phase space in the cross section. 

In the left panel of Fig.~\ref{quadcrossplotlong}, we plot the energy-differential cross sections given in~\eqref{crosssectionquadrupole} and~\eqref{crossvertexlong}. The mass-dependent contribution of the emitted particle, derived in App.~\ref{Massiveemission}, is included and compared with the numerical integration of the full, unexpanded amplitude. To perform the comparison, we consider a system of distinguishable particles with equal coupling-to-mass ratios, ensuring that quadrupole emission provides the dominant contribution to the energy-differential cross section. Identical parameter values are used in both cases prior to any normalization, with fixed values of \( v_i, m_j, m_{\phi,V} \), and the coupling constant \(g_{j},y_{j},A_{j}/m_{j}\) and \(\lambda_{j}\). The numerical results are in agreement with the analytical expressions, thereby validating the latter.

\paragraph{Short-range mediation} Turning now to the case of a short-range interaction mediated by $\phi'$ or $V'^{\mu}$, the quadrupole emission amplitudes remain model-dependent.  
In the limits \((r \ll 1,\, m_{\phi',V'} < \sqrt{s} )\), the squared, averaged over the initials spins, summed over the final spins and averaged over the direction of the emitted particles amplitudes are obtained from~\eqref{generalquadrupole} with the replacement \(|\Vec{v}_{i}-\Vec{v}_{f}|^4 \rightarrow m_{\phi',V'}^4/\mu^4\) and using the following quadrupole emission factors and coefficients for fermionic DM candidates instead,
\begin{equation} \label{quadshortfermion}
\begin{aligned}
\left\{\mathcal{F'_{Q}},\alpha,\beta \right\} =
 \begin{cases}
  \displaystyle{  \left\{
  g_1' g_2'\left(
  \frac{g_{{1}}}{m_{{1}}^2} + \frac{g_{{2}}}{m_{{2}}^2} \right)  , \frac{3}{4} , \frac{1}{4}  \right\} }
      &\text{for }\ \chi_1 \chi_2\to \chi_1 \chi_2 V,\\[10pt] 
       \displaystyle{ 
      \left\{
     y'_{{1}} y'_{{2}}  \left( \frac{y_{{1}}}{m_{{1}}^2} + \frac{y_{{2}}}{m_{{2}}^2}  \right), 1,7 \right\} }
      &\text{for }\ \chi_1 \chi_2\to \chi_1 \chi_2 \phi,\   \tilde \chi_1 \tilde\chi_2\to \tilde\chi_1 \tilde\chi_2 \phi.\\[10pt]
 \end{cases}
\end{aligned}
\end{equation} 
For scalar DM candidates, the contribution from the contact and the vertex diagrams should also be considered, yielding for  $S_1 S_2 \to S_1 S_2 V$
\begin{equation} \label{quadshortscalar1}
\begin{aligned}
\left\{\mathcal{F'_{Q}},\alpha,\beta \right\} =
  \displaystyle{  \left\{
  \left(g_1' g_2' + \frac{m_{V'}^2\lambda_{S}}{4m_{1}m_{2}} \right)\left(
  \frac{g_{{1}}}{m_{{1}}^2} + \frac{g_{{2}}}{m_{{2}}^2} \right)  , \frac{3}{4} , \frac{1}{4}  \right\} },
\end{aligned}
\end{equation}
while for $S_1 S_2 \to S_1 S_2 \phi$ and $\tilde{S}_1 \tilde{S}_2 \to \tilde{S}_1 \tilde{S}_2 \phi$, in external leg-dominated emission (i.e., $\langle |\mathcal{M}_{\text{external leg}}|^2 \rangle_{\vec{\hat{q}}} \gg \langle |\mathcal{M}_{\text{vertex}}|^2 \rangle_{\vec{\hat{q}}}$; see App.~\ref{scalarscalarquad} for details), 
\begin{equation} \label{quadshortscalar2}
\begin{aligned}
\left\{\mathcal{F'_{Q}},\alpha,\beta \right\} =
    \displaystyle{ \left\{   \frac{1}{8} \left( \frac{A'_{1}}{m_{1}} \frac{A'_{2}}{m_{2}} - \frac{m_{\phi'}^2 \lambda_{S}}{m_{1}m_{2}} \right)\left(\frac{A_{1 }}{m_{1}^3} + \frac{A_{2}}{m_{2}^3}   \right) , 1, 2 \right\} }.
\end{aligned}    
\end{equation}
These emission amplitudes lead to the energy-differential cross sections
\begin{equation} \label{shortrangequad}
  \omega \frac{d \sigma }{d \omega} =   f_{\phi,V} \mathcal{F}_{\mathcal Q}'^2  \frac{\mu^6  v_{i}^4 \sqrt{1-x} }{180 \pi^3  m_{\phi',V'}^4} \left[ 8\alpha(1-x) + 3x^2(\alpha+\beta) \right]\quad (r \ll 1, \, m_{\phi',V'} < \sqrt{s})  .
\end{equation}

In the case of scalar emission from scalar DM candidates, when the emission is vertex-dominated (see App.~\ref{scalarscalarquad}), the squared emission amplitude is obtained from~\eqref{limitbasse} via the substitutions $|\vec{v}_{i} - \vec{v}_{f}|^4 \rightarrow m_{\phi',V'}^4/\mu^4$, $\lambda_j \to \lambda'_j$, and $A_j \to A'_j$. Similarly to the long-range mediation case, this squared emission amplitude does not originate from quadrupole emission, despite exhibiting the same $r$-dependence. The associated energy-differential cross section is
\begin{equation} \label{vertexscalarshort}
   \omega \frac{d \sigma }{d \omega } = \left(A'_{1} \lambda'_{2} + A'_{2}\lambda'_{1}\right)^2\frac{ \mu^4 v_{i}^4 x^2  \sqrt{1-x} }{256 \pi^3 m_{1}^2 m_{2}^2 m_{\phi'}^4  } \quad \text{(vertex dominated emission)}.
\end{equation}
We note that, in the contact limit, this energy-differential cross section depends only on the primed coupling of the DM to the short-range mediator and on an interaction involving both mediators simultaneously. In contrast, the quadrupole energy-differential cross sections~\eqref{shortrangequad} receive contributions from the primed couplings of the DM to the short-range mediator as well as from the unprimed couplings to the long-range mediator.
For massive emission, the agreement between the $m_{\phi,V}$-dependent term of the $\phi$-emission amplitude and the longitudinal component of $V$-emission prior squaring again holds; see App.~\ref{Massiveemission} for a comprehensive treatment. In the case of the energy-differential cross section~\eqref{vertexscalarshort}, the $m_{\phi}$-dependent correction arises exclusively from the phase space.

In the left panel of Fig.~\ref{quadcrossplotshort}, we show the energy-differential cross sections given in~\eqref{shortrangequad} and~\eqref{vertexscalarshort}, including the mass-dependent contribution of the emitted particle, as derived in App.~\ref{Massiveemission}. The latter are compared to a numerical integration of the full, unexpanded amplitude. To enable a direct comparison, we consider a system of distinguishable particles with equal coupling-to-mass ratios, for which quadrupole emission constitutes the leading contribution to the energy-differential cross section. Identical values of \( v_i \), \( m_j \), \( m_{\phi,V} \), \( m_{\phi',V'} \), and the coupling constants\footnote{We use the following shorthand notation in the case of emission process involving scalar DM candidate:~\ ~$C_{g^{\prime}_j,\lambda_S} = g_{1}^{\prime}g^{\prime}_{2} + \lambda_S m^2_{V'} / (4m_{1}m_{2})$, and~$C_{A_{j}^{\prime},\lambda_S}=(A_{1}^{\prime}/m_{1})(A^{\prime}_{2}/m_{2}) - \lambda_S m^2_{\phi'} /(m_1m_2) $.} \( g^{(\prime)}_j \), \( y^{(\prime)}_j \), \( A^{(\prime)}_j/m_j \), and \( \lambda_S \) are used in both cases, prior to any normalization. We find excellent agreement between the analytical and numerical results.

\subsubsection{Scattering of identical particles} 
\label{identicalparticles}

Now we discuss the most relevant case of indistinguishable particles, where dipole transitions are forbidden, rendering the leading-order emission to be of quadrupole nature. In the derivation of amplitudes, there are also $u$-channels to be considered, doubling the number of Feynman diagrams shown in Figs.~\ref{inifini}, \ref{vertex}, and~\ref{VIB}.

\paragraph{Long-range mediation} We start with the long-range mediated force. The analytical complexity of the NR results increases, so that we relegate the presentation of squared matrix elements to  App.~\ref{EqualmassAmplitudes}. 

For the emission of a vector boson $V^{\mu}$, in contrast to the scattering of distinguishable particles, the emission amplitudes of the dSIDM scenarios $ \chi \chi\to \chi \chi V$ and $ S S \to S S V$ are not identical. This difference arises from the statistical properties of the DM candidates, which determine the interference sign between the contributions from the $t$- and $u$-channel diagrams.
For vector emission from Dirac fermions, $ \chi \chi\to \chi \chi V$, the energy-differential cross section reads
\begin{align} \label{EMQED}
    \omega \frac{d\sigma}{d \omega}  =  \frac{ g^6\sqrt{1-x} }{240 \pi^3 m_{\chi}^2} (1-\kappa^2)^{3/2} \left[  K(17,-3) + \frac{1}{2}L(12,-7,-3) 
    +  \frac{\kappa^2}{4} \left(K(52,-8) + L(16,-11,-4)\right) \right] .
\end{align}
Here, we have already included the mass-dependent correction factor in terms of the variable \(\kappa\equiv m_{\phi,V}/\omega\).  For $\kappa=0$, the result agrees with \cite{Fedyushin1952,Garibyan1953}. To simplify the presentation, we  introduce the short-hand notation for the following $x$-dependent functions
\begin{align}
        K(a,b) &\equiv a + \frac{b x^2}{(2-x)^2}, \\ 
        L(a,b,c) & \equiv \frac{a(2-x)^4 +b(2-x)^2x^2 +c x^4 }{(2-x)^3\sqrt{1-x}} \log\left(\frac{1+\sqrt{1-x}}{1-\sqrt{1-x}}\right)   .
\end{align} 
Similarly, for vector emission from complex scalars, $ S S \to S S V$, we obtain
\begin{equation} \label{EMQEDscalar}
    \omega \frac{d\sigma}{d \omega}  =  \frac{ g^6 \sqrt{1-x} }{240 \pi^3 m_{S}^2} (1-\kappa^2)^{3/2} \left[K\left(26,6\right) + L(6,7,3) + \frac{\kappa^2}{2} \left( K(38,8) + L(8,11,4)\right) \right].
\end{equation}
In the limit of massless mediator emission, $\kappa=0$, this result agrees with \cite{Gould:1981an}.

We now turn to the emission of $\phi$ particles from fermions and scalars.
For scalar emission from fermions, both Dirac fermions, $\chi\chi\to \chi \chi \phi$ and Majorana fermions, $\tilde\chi \tilde\chi\to \tilde \chi \tilde \chi \phi$, share the same energy-differential cross section 
\begin{align} \label{EMchiscalar}
  \omega  \frac{d\sigma}{d \omega} = \frac{y^6\sqrt{1-x}}{240 \pi^3 m_{\chi}^2} \sqrt{1-\kappa^2} &\left[\vphantom{\frac{y^2}{4} }  K(18,-2) + L(4,-4,-1) + \kappa^2( K(-16,4) + L(-8,3,2)) \right.
 \nonumber \\  & \left.  + \frac{\kappa^4}{4}\left( K(52,-8) + L(16,-11,-4) \right) \right] .
\end{align}
Here, we are in disagreement with the results of \cite{2303.03123} where scalar emission from electron-electron and nucleon-nucleon bremsstrahlung is derived. We ascertain our analytical result by comparing it against an exact numerical evaluation of the full 2-to-3 cross section. 

The energy-differential cross sections for $\phi$ emission from complex scalars, $SS\to SS\phi$, and real scalars $\tilde S\tilde S\to \tilde S\tilde S \phi$ are identical. In the limit $ ({A}/{m_{S}})^2 \gg \lambda$, the external leg emission diagrams dominate, and the associated energy-differential cross section reads
\begin{equation} \label{EMscalarscalar}
\begin{aligned}
  \omega  \frac{d\sigma}{d \omega} = \left(\frac{A}{m_{S}}\right)^6  \frac{\sqrt{1-x}}{1920 \pi^3 m_{S}^2} \sqrt{1-\kappa^2} & \left[    K\left(\frac{11}{2},\frac{1}{2}\right) + L\left(\frac{1}{2},\frac{9}{4},\frac{1}{4}\right) - \frac{\kappa^2}{2}\left( K(12,2) + L(2,4,1)\right) \right.  \\ & \left.  + \frac{\kappa^4}{16}\left( K(38,8) + L(8,11,4)\right)  
     \right]\qquad ( \left({A}/{m_{S}}\right)^2 \gg \lambda) .
\end{aligned}
\end{equation}
In the opposite limit, $({A}/{m_{S}})^2 \ll \lambda$, emissions from vertex-diagram
 dominate, the result simplifies and is given by 
\begin{equation} \label{energydiffvertexlong}
\begin{aligned}
     \omega  \frac{d\sigma}{d \omega} = \left(\frac{A}{m_{S}}\right)^2\lambda^2 \frac{\sqrt{1-x}}{128 \pi^3 m_{S}^2} \sqrt{1-\kappa^2}\left[ K(2,0) + L(0,1,0)  \right]\qquad (\lambda \gg \left({A}/{m_{S}}\right)^2 ).
\end{aligned}
\end{equation}

\begin{figure}[t!]
    \centering
\includegraphics[width=0.49\linewidth]{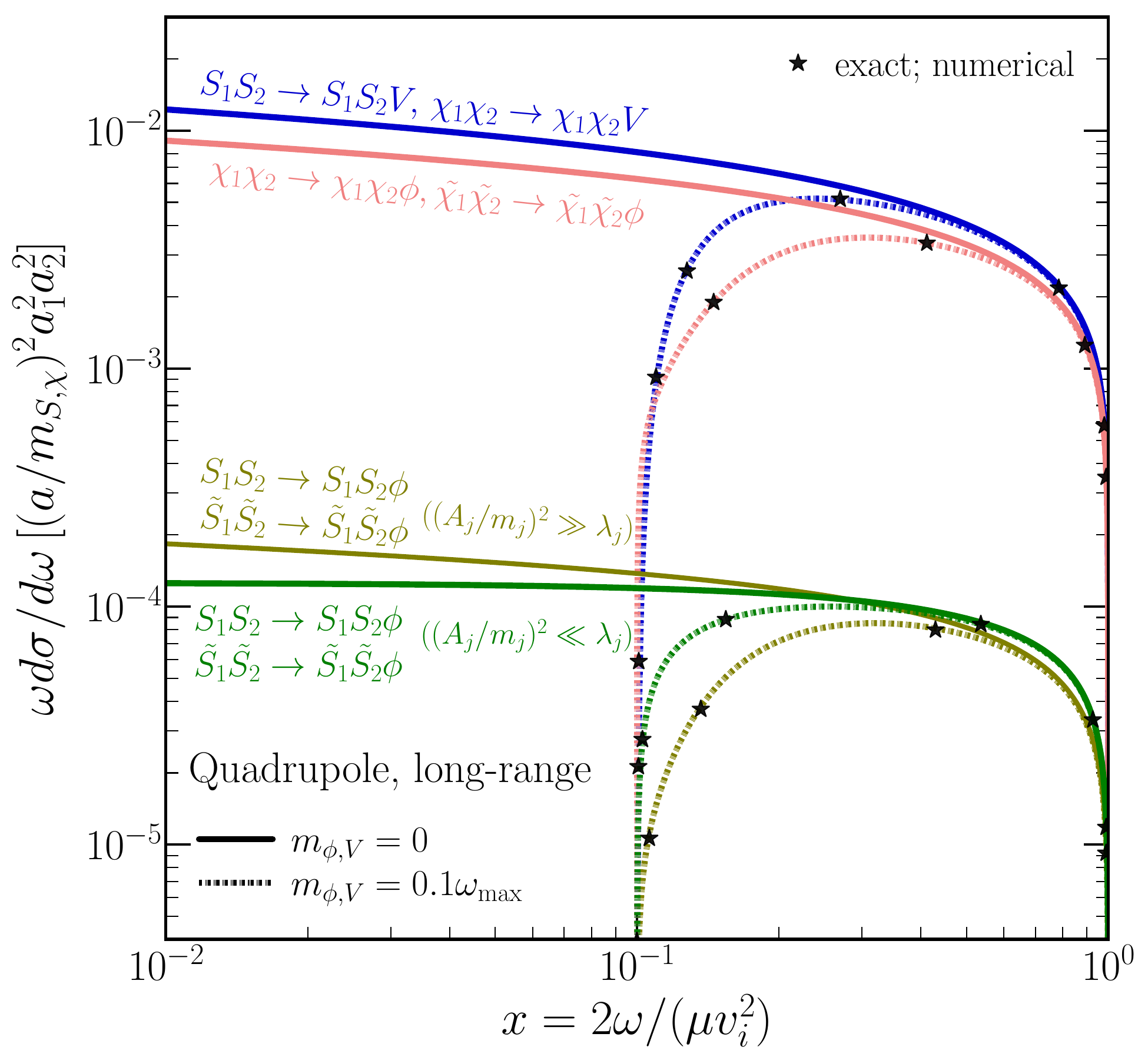}
 \includegraphics[width=0.49\linewidth]{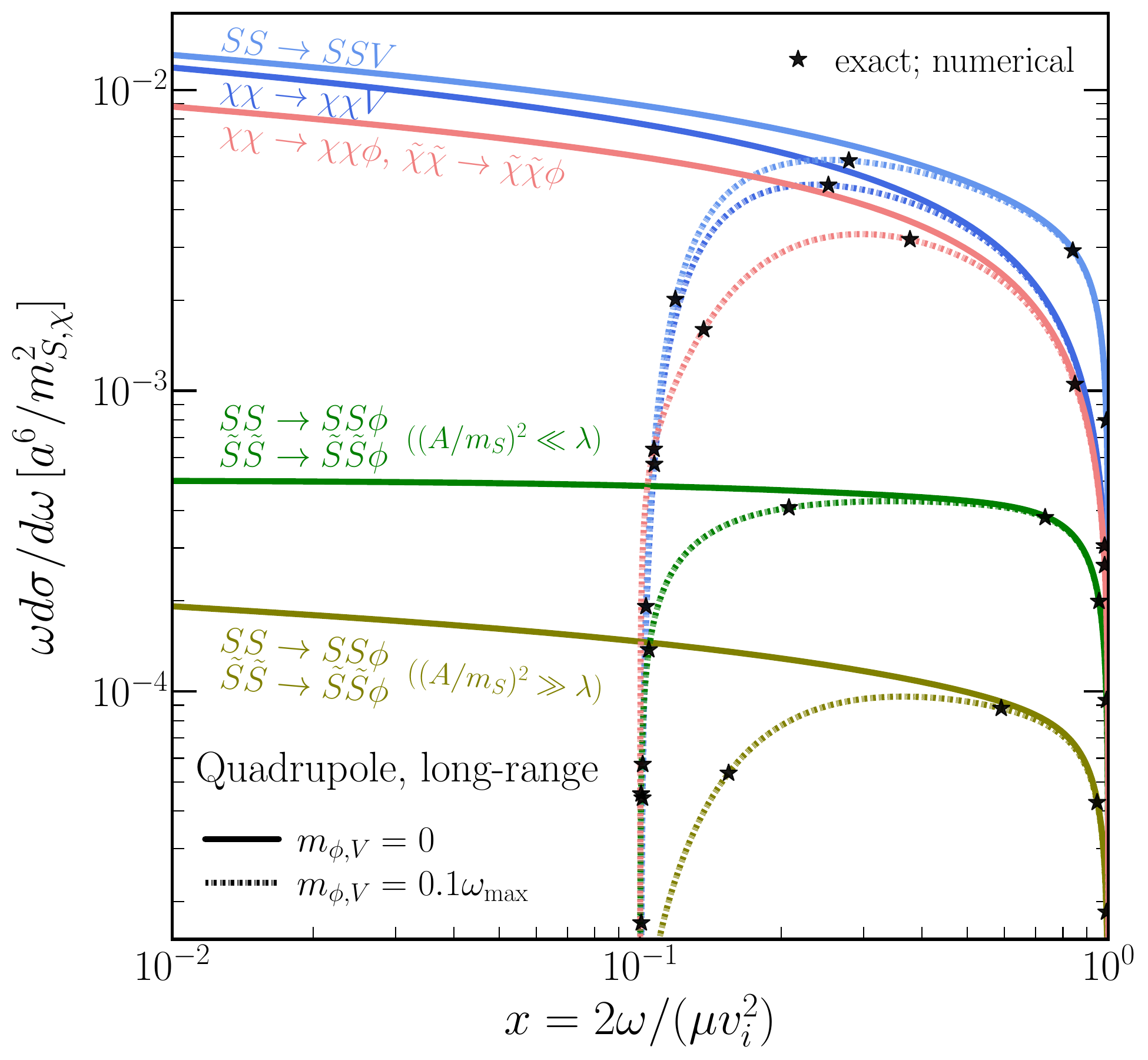}
\caption{Quadrupole energy-differential cross section for the different dSIDM scenarios with long-range mediation. The massless emission limits are shown as solid colored lines, while the dotted colored lines represent massive emission with \( m_{\phi,V}/\omega_{\text{max}} = 0.1 \). The black stars indicate values of the energy-differential cross section obtained from a fully numerical integration of unexpanded amplitudes. The set of couplings \( \{a, a_j\} \) depends on the dSIDM scenario: \textit{i}) for vector emission, \( \{g, g_j\} \); \textit{ii}) for scalar emission with $\chi (\tilde \chi)$, \( \{y, y_j\} \); \textit{iii}) for scalar emission with $S (\tilde S)$, \textit{iii.a}) in the limit  (\((A/m_{S})^2 \gg \lambda_j\)), \( \{A/m_{S}, A_j/m_j\} \) \textit{iii.b}) in the limit~(\((A/m_{S})^2\ll\lambda_j \)),~$(a/m_{\chi,S})^2a^{2}_{1}a_{2}^2 \to ((A_{2}/m_{2})\lambda_{1}/m_{1}+(A_{1}/m_{1})\lambda_{2}/m_{2})^2 $ and~$a^{6}\to (A/m_{S})^2\lambda^2 $. In the numerical analysis, these two limits were implemented by setting \( (A_{j}/m_j)^2 = 100\,\lambda_j \) and \( (A_{j}/m_j)^2 = 0.01\,\lambda_j \), respectively.
\textit{Left panel}: Distinguishable particle scattering with massless (massive) emission derived in Sec.~\ref{quadrudisting} (App.~\ref{Massiveemission}) with equal coupling-to-mass ratio. Numerical integration parameters: $m_{1}=0.1$~GeV,
$m_{2}=1$~GeV, $a_{1}=0.1\,a_{2}$ and $v_{i} \approx 4 \times 10^{-3}$. \textit{Right panel}: Identical particle scattering (Sec.~\ref{identicalparticles}). Numerical integration parameters: $m_{S,\chi} = 1$ GeV and $v_{i}\approx6 \times 10^{-3}$.
}
    \label{quadcrossplotlong}
\end{figure}

In the right panel of Fig.~\ref{quadcrossplotlong}, we plot the energy-differential cross sections derived in this section. Assuming a system of identical particles, we compare the result for massive emission with the numerical integration of the full, unexpanded amplitude. To perform a direct comparison, identical parameter values are used in both cases prior to any normalization, with fixed values of \( v_i \), \( m_{S,\chi}, m_{\phi,V} \) and of the couplings $g,y,A/m_{S}$ and $\lambda$. We observe excellent agreement between the two approaches, thereby validating our analytical results.

\paragraph{Short-range mediation} 
We now turn to the case of a short-range mediated force. In this regime, the $t$-channel and the $u$-channel contributions become identical. 

As a result, for fermionic DM candidates, the energy-differential cross section for massless emission is readily obtained by multiplying the right-hand side of~\eqref{shortrangequad} by the factor \( S_f = 1/2 \), accounting for the symmetry of identical particles in the final state, and by performing the following substitutions: \( g^{(\prime)}_j \to g^{(\prime)} \), \( y^{(\prime)}_j \to y^{(\prime)} \), and \( m_j \to m_\chi \).
In the case of massive emission, these changes must be applied in~\eqref{csshortmassiveQED} for the process \( \chi \chi \to \chi \chi V \), and in~\eqref{csshortmassiveyuka} for the processes \( \tilde{\chi} \tilde{\chi} \to \tilde{\chi} \tilde{\chi} \phi \) and \( \chi \chi \to \chi \chi \phi \).

For scalar DM candidates, the inclusion of the \( u \)-channel doubles the contribution of the external leg and vertex diagrams to the emission amplitude compared to the contact emission diagrams. %
As a result, the dissipative process $SS\to SS V$ has the following energy-differential cross section
\begin{equation}
\label{quadheavyVS}
    \omega \frac{d\sigma}{d \omega}= g^{2}\left(g'^2 + \frac{m_{V'}^2}{8m_{S}^2} \lambda_{S} \right)^2 \frac{m^2_{S}v_{i}^4(1-\kappa^2)^{3/2}}{240\pi^3m_{V'}^4}\left[ B(1,2) + \frac{\kappa^2}{24} B(21,32)   \right],
\end{equation}
where we use the following short-hand notation for the $x$-dependent function
\begin{equation}
    B(a,b)\equiv\sqrt{1-x}\left[ax^2 +b(1-x)\right].
\end{equation}

The scenarios involving the emission of a scalar boson from real or complex scalar DM candidates, i.e.,~$SS\to SS\phi$, and~$\tilde S\tilde S\to \tilde S\tilde S \phi$, share identical energy-differential cross sections across the two limiting cases previously defined in Sec.~\ref{shortrangeforce}.
For external leg-dominated emission (i.e., $\langle |\mathcal{M}_{\text{external leg}}|^2 \rangle_{\vec{\hat{q}}} \gg \langle |\mathcal{M}_{\text{vertex}}|^2 \rangle_{\vec{\hat{q}}}$; see App.~\ref{scalarscalarquad} for details), we find
\begin{equation} \label{quadheavyScalar}
    \omega \frac{d\sigma}{d \omega}= \left( \frac{A}{m_{S}}\right)^2\left[ \left(\frac{A'}{m_{S}}\right)^2  - \frac{m_{\phi'}^2}{2m_{S}^2}\lambda_{S} \right]^2 \frac{m_{S}^2v_{i}^4\sqrt{1-\kappa^2}}{92160\pi^3m_{\phi'}^4}\left[ B(9,8) - \kappa^2 B(3,16) +  \frac{\kappa^4}{4} B(21,32)  \right].
\end{equation}
For vertex-dominated emission (see App.~\ref{scalarscalarquad}) in the scenarios~$SS\to SS\phi$, and~$\tilde S\tilde S\to \tilde S\tilde S \phi$, the energy-differential cross section is
\begin{equation} \label{energydiffvertexshort}
    \omega \frac{d\sigma}{d \omega}= \left(\frac{A'}{m_{S}}\right)^2 \lambda'^{2}\frac{m_{S}^2v_{i}^4\sqrt{1-\kappa^2}}{512\pi^3m_{\phi'}^4} B(1,0).
\end{equation}

\begin{figure}[h!]
    \centering
\includegraphics[width=0.49\linewidth]{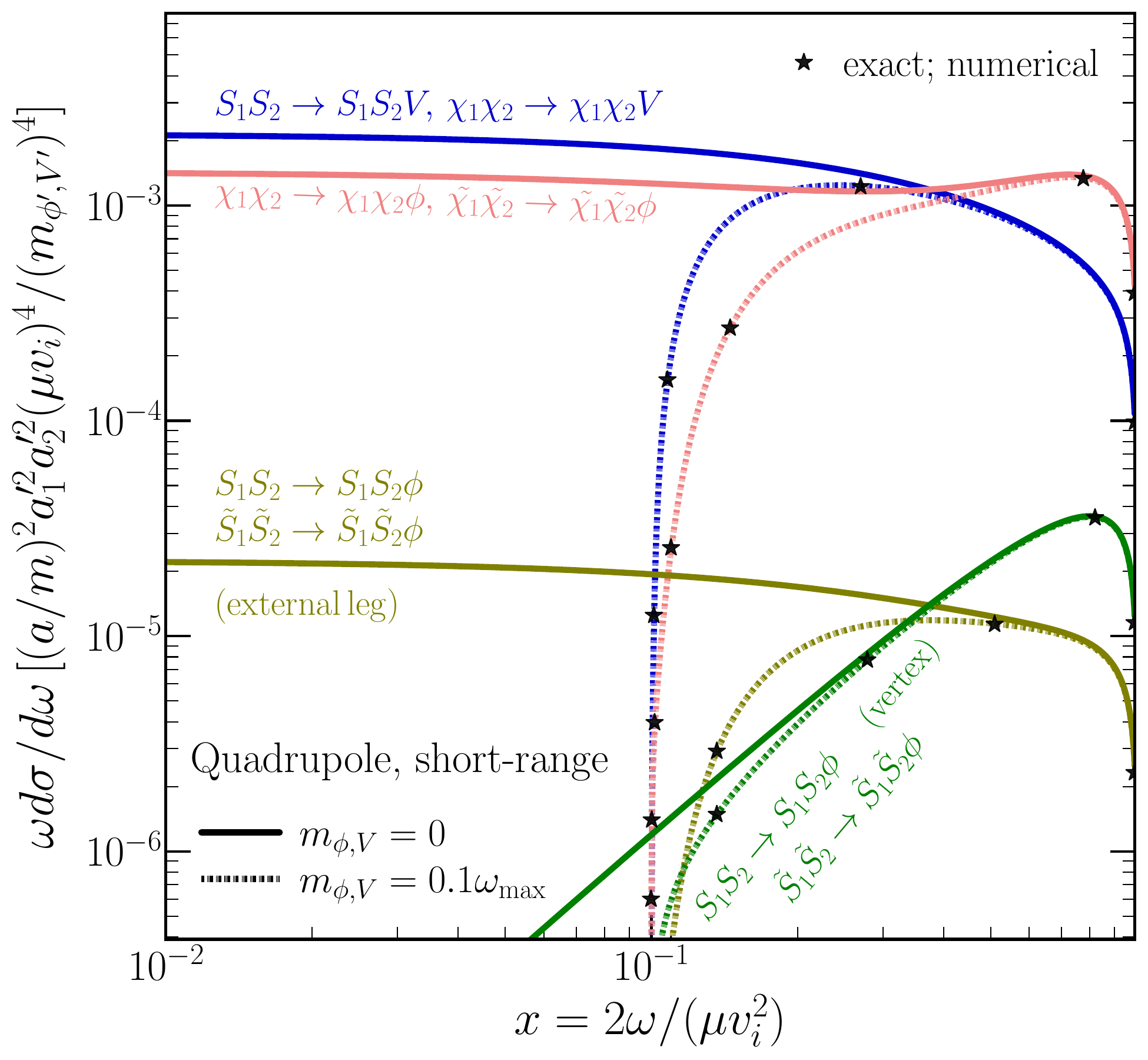}
\includegraphics[width=0.49\linewidth]{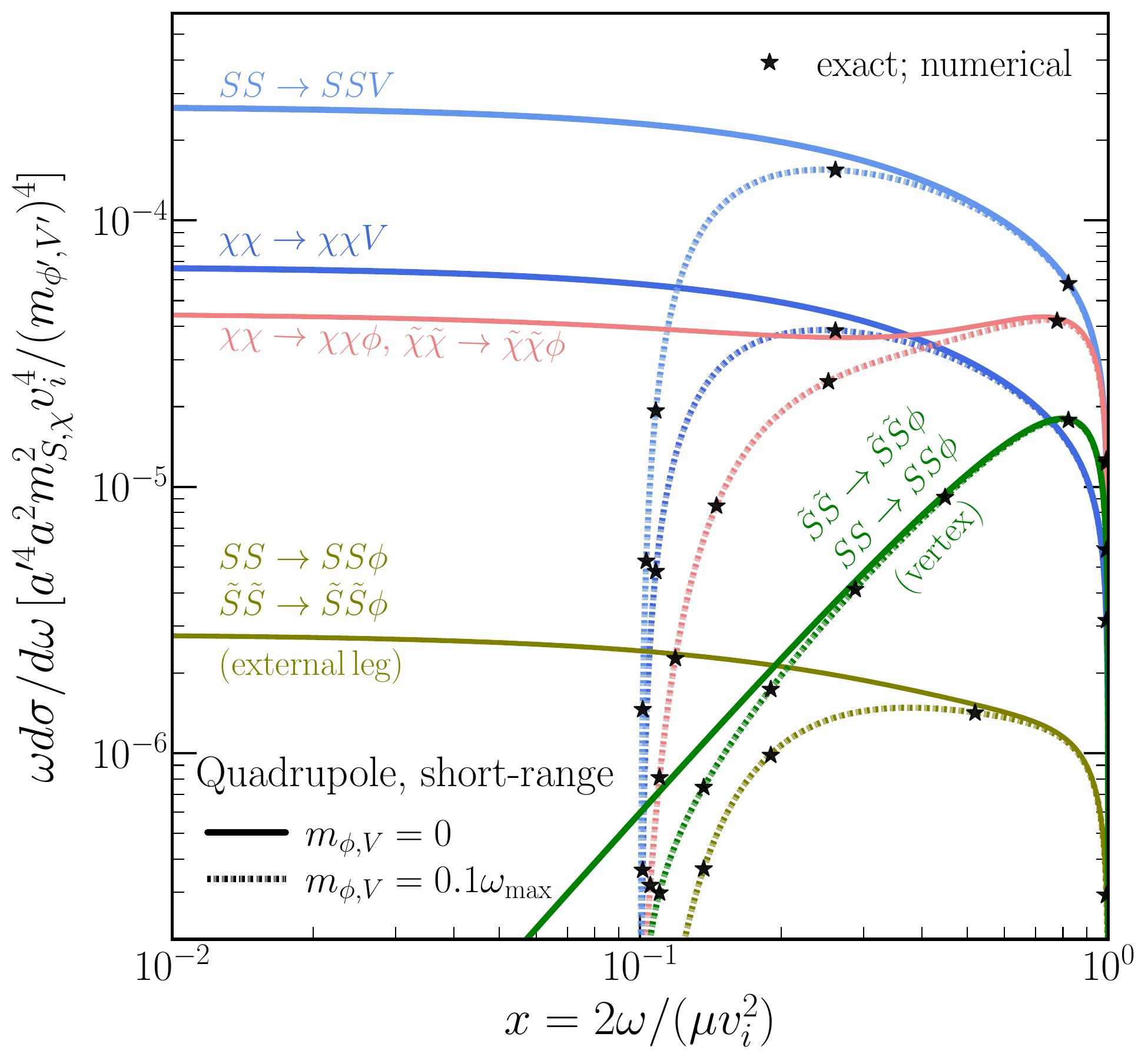}
\caption{
Quadrupole energy-differential cross section for the different dSIDM scenarios mediated by a short-range force.  The massless emission limits are shown as solid colored lines, while the dotted colored lines represent massive emission with \( m_{\phi,V}/\omega_{\text{max}} = 0.1 \). The black stars indicate values of the energy-differential cross section obtained from a fully numerical integration of unexpanded amplitudes. The relevant coupling combinations \( \{a, a', a_j' \} \) vary by scenario: \textit{i}) for \( \chi\chi \to \chi\chi V \),  \( \{g, g', g_j' \} \); 
\textit{ii}) for \( SS \to SS V \), \( \{g, C^{1/2}_{g',\lambda_S}, C^{1/2}_{g_j',\lambda_S} \} \); 
\textit{iii}) for scalar emission from \( \chi \)(\( \tilde{\chi} \)), \( \{y, y', y_j' \} \); 
\textit{iv}) for scalar emission with \( S \)(\( \tilde{S} \)) \textit{iv.a}) in the external-leg regime (see App.~\ref{scalarscalarquad}), \( \{A/m_S, C^{1/2}_{A^{\prime},\lambda_S},  C^{1/2}_{A^{\prime}_j,\lambda_S} \} \), \textit{iv.b}) in the vertex-dominated regime (see App.~\ref{scalarscalarquad}), $(a/m_{S,\chi})^2a^{\prime2}_{1}a^{\prime2}_{2}\to ((A'_{2}/m_{2})\lambda'_{1}/m_{1}+(A'_{1}/m_{1})\lambda'_{2}/m_{2})^2 $. 
\textit{Left panel}: Distinguishable particle scattering (massless emission in Sec.~\ref{quadrudisting} and massive emission in App.~\ref{Massiveemission}) with equal coupling-to-mass ratio $a/m$. Numerical integration parameters: $m_{1}=0.1$~GeV, $m_{2}=1$~GeV, $a_{1}=0.1\,a_{2}$, $v_{i} \approx 4 \times 10^{-3}$, and $m_{\phi',V'}=0.1$~GeV. \textit{Right panel}: Identical particle scattering (Sec.~\ref{identicalparticles}). Numerical integration parameters: $m_{S,\chi} = 1$~GeV, $v_{i} \approx 6 \times 10^{-3}$, and $m_{\phi',V'}=0.1$~GeV. Additionally, we assume \( \lambda_S \sim g^{\prime}_j,A^{\prime}_{j}/m_{j} \).
}
    \label{quadcrossplotshort}
\end{figure}

In the right panel of Fig.~\ref{quadcrossplotshort}, we plot the energy-differential cross sections derived in this section. We compare the result for massive emission with the numerical integration of the full, unexpanded amplitude, assuming a system of identical particles. To enable a direct comparison, identical parameter values are used in both cases prior to any normalization, with fixed values of \( v_i \), \( m_{S,\chi} \), \( m_{\phi,V} \), the heavy mediator mass \( m_{\phi',V'} \) and the couplings\footnote{We use the following shorthand notation in the case of emission process involving scalar DM candidate:~\ ~$C_{g',\lambda_S} = g^{\prime2} + \lambda_S m^2_{V'} / (8m_{S}^2)$, and~$C_{A',\lambda_S}=\left(A^{\prime}/m_{S}\right)^2 - \lambda_S m^2_{\phi'} /(2m_S^2) $.} $g^{(\prime)},y^{(\prime)},A^{(\prime)}/m_{s},\lambda^{\prime}$ and $\lambda_{S}$. We find that the numerical results are consistent with our analytical derivations.

\section{Amplitudes and cross sections for soft emission}
\label{soft}

In this section, we aim to derive a simplified description of the emission process in the limit of ``soft'' particle emission, $x \ll 1$, well below the kinematic endpoint energy. The description for emission of Abelian gauge bosons $V$ is then, of course, well-grounded in the factorization theorem of soft-photon emission, but we shall also find an efficient description for the emission of scalar particles~$\phi$. 

\subsection{Emission factors and elastic scattering amplitudes}

In the emission of infrared photons, $x \ll 1$, the $2\rightarrow3$  bremsstrahlung amplitude factorizes and can be written as a product of the elastic $2 \rightarrow 2$ scattering amplitude $\mathcal{M}_{\rm elast}$ times an emission factor~\cite{Low:1958sn,Weinberg:1965nx},
\begin{equation} \label{softfactorgauge} 
\mathcal{E}_{{V}} =    \sum_{n} \eta_{n}  g_{n} \frac{p_{n}\cdot \varepsilon^{*}(q)}{p_{n} \cdot q} . \end{equation}
The sum runs over external legs where $\eta_n=+1(-1)$ accounts for the emission from final- (initial)-state leg~$n$ with four-momentum $p_{n}$ and charge $g_n$; here we are adapting our notation in which the charge of the particle has been absorbed into the definition of the gauge coupling, $g_j = Q_{\chi_j} g$. This factorization holds regardless of spin, so that~\eqref{softfactorgauge} equally well describes the emission from spinless particles with $g_j=Q_{S_j}g$.

We may also seek to establish a factorization in the description of emission of scalar particles~$\phi$. Whereas for gauge vector emission, the universality arises from Lorentz invariance and charge conservation~\cite{Weinberg:1964ew}, the emission of scalars has received correspondingly less attention~\cite{1703.07885, 2208.05023}.
Hence, in what follows, we do not base the description on any factorization theorem per~se, but nevertheless find useful results.
For $\phi$-emission from trilinear interactions with $S$ in the regime that the external leg emission processes of Fig.~\ref{inifini} dominate, we obtain the emission factors from~\eqref{softfactorgauge},
\begin{equation} \label{softfactorscalarA}
    \mathcal{E}_{{\phi S}} =\mathcal{E}_{{\phi \tilde S}}=  \sum_{n}  \eta_{n}\frac{   A_n }{2 p_{n} \cdot q } \quad \left(A_j/m_{j}\right)^2\gg \lambda_j;\, \text{external leg emission} .
\end{equation}

For $\phi$-emission from Yukawa interactions with $\chi$, we find that the decomposition of the emission amplitude necessitates the addition of a constant, $q$-independent term
\begin{equation}
     \label{softfactorscalar}
    \mathcal{E}_{{\phi \chi}} = \mathcal{E}_{{\phi\tilde\chi}} = \sum_{n}\left(  \eta_{n}y_n \frac{   m_n }{p_{n} \cdot q } + \frac{y_{n}}{2 m_{n}} \right)  . 
\end{equation}
While this additional term might appear negligible in the limit $\omega \to 0$, 
we show in the following that it plays a crucial role at quadrupole order. 
In App.~\ref{scalarvsgauge}, we trace its origin to the $\slashed q$–dependent part of the emission current, and demonstrate explicitly how this contribution arises in the soft expansion. The form of the factorization~\eqref{softfactorscalar} for scalar emission from fermions has also been observed in~\cite{2208.05023}.

The emission factors above are then to be paired with the elastic matrix elements $\mathcal{M}_{\rm elast}$ in the limit~$\vec q\to 0$. We note that there are no interference terms between the \( v_{i} \)-expansion of \( \mathcal{M}_{\rm elast} \) and those of~\eqref{softfactorgauge},~\eqref{softfactorscalarA}, and~\eqref{softfactorscalar} that are relevant for the dipole and quadrupole soft emission amplitudes; see App.~\ref{Softderivations} for details. This allows us to safely square each of these quantities individually without any loss of information.

For distinguishable particle scattering, either $\chi_1\chi_2 \to \chi_1\chi_2, \, \tilde{\chi}_1\tilde{\chi}_2 \to \tilde{\chi}_1\tilde{\chi}_2 $ for fermionic candidates or $S_1 S_2\to S_1 S_2,\,\tilde{S}_1\tilde{S}_2 \to \tilde{S}_1\tilde{S}_2$ for scalar candidates, the long-range (initial state spin-averaged) expression of the NR elastic squared matrix element is given by the Rutherford formula,
\begin{equation} \label{elastic}
 \frac{1}{d_{1}d_{2}} |\mathcal{M}_{\rm elast}|^2  =  16 \frac{m_{1}^2m_{2}^2 }{\mu^4}a_1^2a_2^2  \frac{1}{|\Vec{v}_{i}-\Vec{v}_{f}|^4},
\end{equation} 
where $a_j$ is given by $g_j,\, y_j,\, A_j/2m_j$ in the different scenarios. In the following, we use $a \equiv a_1=a_2$ for identical particles, 
while the primed quantities $a_j'$ and $a'$ denote the corresponding 
couplings in the short-range limit. In this  limit ($r\ll1$), the squared elastic matrix  element becomes instead 
\begin{equation} \label{elasticheavy}
 \frac{1}{d_{1}d_{2}} |\mathcal{M}_{\rm elast}|^2  =  \frac{16 m_{1}^2m_{2}^2}{m_{\phi',V'}^4}   \left[ a'_1a'_2 \pm \delta_{s,0}\frac{m_{\phi',V'}^2\lambda_{S}}{4m_{1}m_{2}} \right]^2,
\end{equation} 
where the contribution from contact diagrams (Fig.~\ref{Contacts}) is included for scalar DM candidates (\( s = 0 \)), in contrast to their fermionic counterparts (\( s = 1/2 \)). The $+$ sign applies to vector emission, while the $-$ sign applies to scalar emission.
 
For the scattering of identical particles, the Coulomb expression reads
\begin{equation} 
\label{elastic2}
  \frac{1}{d_{1}d_{2}} |\mathcal{M}_{\rm elast}|^2  =  16 \frac{m^4}{\mu^4}  a^4 \left[ \frac{1}{|\Vec{v}_{i}-\Vec{v}_{f}|^4} +  \frac{1}{|\Vec{v}_{i}+\Vec{v}_{f}|^4} + 2 \frac{(-1)^{2s}}{2s+1} \frac{1}{|\Vec{v}_{i}-\Vec{v}_{f}|^2|\Vec{v}_{i}+\Vec{v}_{f}|^2}   \right],
\end{equation}  
while in the case of short-range interactions, $r\ll 1$, the squared matrix element is given by
\begin{equation}
\label{elasticheavy2}
\frac{1}{d_1 d_2} |\mathcal{M}_{\rm elast}|^2 = 
16\frac{m^4}{m_{\phi',V'}^4} \left[
\frac{2}{2s+1}a'^2
\pm \delta_{s,0}  \frac{m_{\phi',V'}^2 \lambda_{S}}{4 m^2}
\right]^2.
\end{equation}
With these ingredients at hand, we may now obtain the dipole and quadrupole expressions of the soft energy-differential cross sections. 

\subsection{Soft dipole emission}
In the case of non-equal coupling-to-mass ratios, dipole emission dominates. The corresponding soft emission factors $\mathcal{E_D}$ are obtained by  
expanding the emission factors in~\eqref{softfactorgauge},~\eqref{softfactorscalarA}, and~\eqref{softfactorscalar} to leading order in the NR \( v_i- \) and $r$-expansions. 
The detailed steps of the latter derivations are presented in App.~\ref{Softderivations}. In the following, we only state the obtained final expressions.

For vector emission, applying the $v_{i}$-expansion to~\eqref{softfactorgauge} and isolating the leading order, yields, after squaring, applying~\eqref{averageqrelationsvector}
for the polarization sum, and the angular averaging over the emission directions,
\begin{equation} \label{eikonaldipole}
\begin{aligned}
  & \langle |\mathcal{E_D}_{{,V}}|^2 \rangle_{\hat{q}} = \frac{2}{3 \omega^2} \left(\frac{g_{1}}{m_{{1}}}- \frac{g_{{2}}}{m_{{2}}} \right)^2 \mu^2 |\Vec{v}_{i} - \Vec{v}_{f}|^2 .
    \end{aligned} 
\end{equation}
This result is in agreement with the textbook derivation presented in \cite{Berestetskii:1982qgu}.
For scalar emission, applying the same procedure to~\eqref{softfactorscalarA}, and~\eqref{softfactorscalar} results in 
\begin{align} 
   \langle |\mathcal{E_D}_{{,\phi\chi}}|^2 \rangle_{\hat{q}}  &  = \langle |\mathcal{E_D}_{{,\phi\Tilde{\chi}}}|^2  \rangle_{\hat{q}} =\frac{1}{2} \left.   \langle |\mathcal{E_D}_{{,V}}|^2 \rangle_{\hat{q}}   \right|_{g_{j} \to y_{j}  } , 
   \\
 \langle\mathcal{|E_D}_{{,\phi S}}|^2\rangle_{\hat{q}}  &= \langle\mathcal{|E_D}_{{,\phi \Tilde{S}}}|^2\rangle_{\hat{q}} =  \frac{1}{8}  \left.    \langle| \mathcal{E_D}_{{,V}}|^2 \rangle_{\hat{q}}  \right|_{g_{j} \to A_j/m_j } \label{eikonaldipolescalar} .
\end{align}
We note that applying the NR expansion in the contact limit $r\ll1$ reproduces the same results as~\eqref{eikonaldipole}-\eqref{eikonaldipolescalar}.

Multiplying by the appropriate squared elastic matrix element, i.e.,~\eqref{elastic} for long-range and~\eqref{elasticheavy} for short-range mediation, one obtains, in each case, a single formula for the soft dipole emission matrix element, $\left. \langle |\mathcal{M_D}|^2 \rangle_{\hat{q}} \right|_{\rm soft}$,
where the prefactors assemble nicely to the dipole emission factors \( \mathcal{F_D} \) defined in~\eqref{prefactordipole} and \( \mathcal{F_D'} \) defined in~\eqref{shortfermiondipole}-\eqref{shortscalardipole}, respectively. In fact, the squared amplitudes in the soft limit become identical to the full result of~\eqref{generaldipole} (resp.~\eqref{dipoleshortrangeamplitude}) for long- (short-) range mediation
\begin{equation} \label{dipolesoftgeneral}
 \left. \langle |\mathcal{M_D}|^2 \rangle_{\hat{q}} \right|_{\rm soft} = \langle |\mathcal{M_D}|^2 \rangle_{\hat{q}}.
\end{equation}
This equality arises from the NR power counting scheme for which $|\Vec{q}|/\mu \sim \mathcal{O}(v_{i}^2)$; cf.~\eqref{energyconservation}. In other words, any corrections due to the emission of quanta are of sub-leading order in the velocity (and contact limit) expansion when it concerns dipole radiation. As we shall see below, this simplification does not hold at quadrupole order where the leading dipole velocity terms cancel. 

\begin{figure}[t]
    \centering    
    \includegraphics[width=0.49\linewidth]{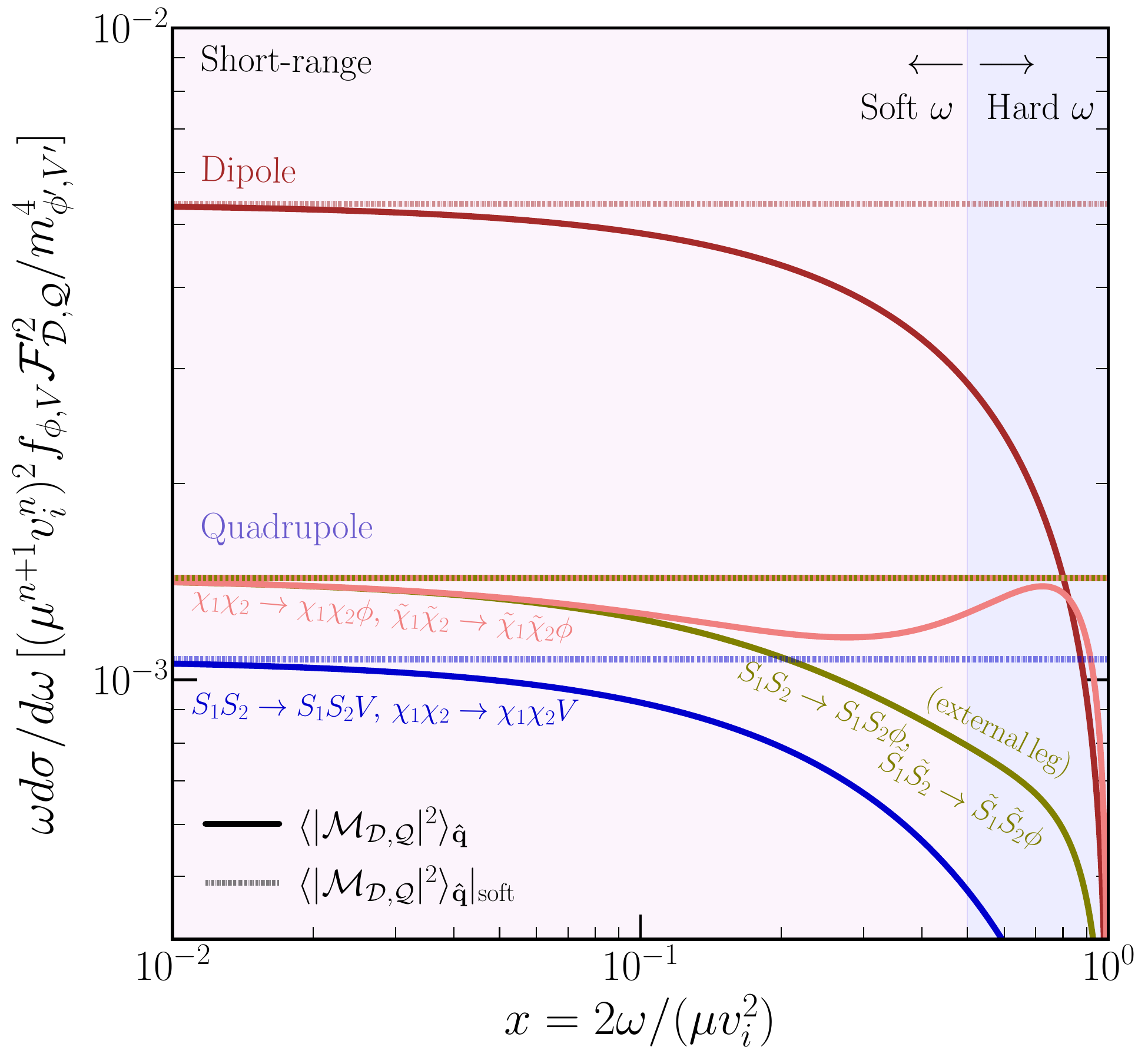}
    \includegraphics[width=0.49\linewidth]{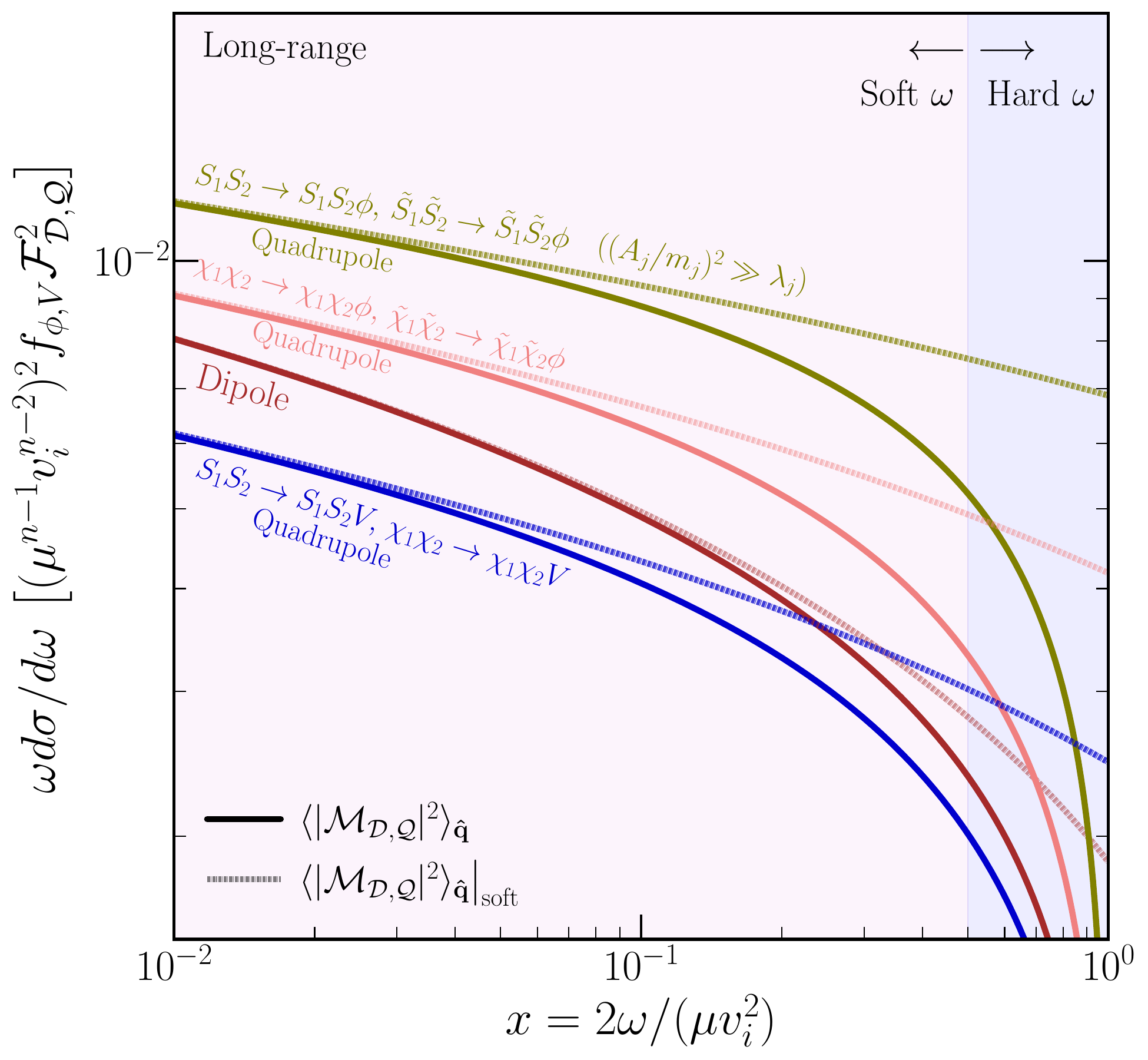}
    \caption{Dipole \((n=1)\) and quadrupole \((n=2)\) energy-differential cross sections for distinguishable particle scattering. Solid lines are the massless limits of our analytical energy-differential cross section expressions; dotted lines show the associated soft emission limits. A separation scale between ``hard'' and ``soft'' emission at half the available CM energy \(\Lambda \sim \mu v_i^2/4\) is introduced; soft modes (\(\omega \lesssim \Lambda\)) are shaded light purple, hard modes (\(\omega \gtrsim \Lambda\)) light blue.
    {\it Left:} short-range mediation. The brown line shows the dipole result~\eqref{dipoleshortrange}, while the brown dotted line shows its soft limit~\eqref{softdipole:short}. Quadrupole lines follow~\eqref{shortrangequad} (full) and~\eqref{softquadheay} (soft). 
    {\it Right:} long-range mediation. The brown line shows the dipole result~\eqref{dipolelongrange}, and the corresponding dotted line shows its soft limit~\eqref{softdipole:long}. For quadrupole emission, solid colored lines show the full results~\eqref{crosssectionquadrupole}, and the dotted colored lines the soft limit~\eqref{softquadlight}. }
    \label{softfigdipole}
\end{figure}

We are now in a position to state the energy-differential cross sections in the soft emission limit using~\eqref{dipolesoftgeneral}. Performing the phase space integrations in~\eqref{crosssection} in the limit $x\ll 1$, i.e., in the limit of $\Vec{v}_{i}\rightarrow \Vec{v}_{f}$ yields
\begin{subnumcases}{\left. \omega \frac{d \sigma}{d \omega} \right|_{\rm soft} = \label{softdipole}}
  \displaystyle 
  \frac{\mathcal{F}_{\mathcal D}'^2 f_{\phi,V}\mu^4 v_{i}^2}{6\pi^3 m_{\phi',V'}^4}
  & for short-range mediation, \label{softdipole:short} \\[5pt]
  \displaystyle 
  \frac{\mathcal{F}_{\mathcal D}^2 f_{\phi,V}}{24 \pi^3 v_{i}^2}
  \ln\left(\frac{2\mu v_{i}^2}{\omega}\right)
  & for long-range mediation. \label{softdipole:long}
\end{subnumcases}
Those expressions agree with a leading-order in \( x \) expansion of the full dipole emission formula~\eqref{dipolelongrange} (resp.~\eqref{dipoleshortrange}) for long- (short-) range mediation. The differences between the full and soft results~\eqref{softdipole} are from phase space alone. 
In Fig.~\ref{softfigdipole}, we compare the soft and the full expressions, both for short-range (left panel) and long-range (right panel) mediation. 

\subsection{Soft quadrupole emission: amplitudes}  \label{softquadrupole}
The factorization that we alluded to in the previous section, of course, continues to hold at higher order in a relative velocity expansion in~$v_i$ and hence for quadrupole emission. However, a naive application, i.e.,~multiplying the eikonal factors \eqref{softfactorgauge}, \eqref{softfactorscalarA}, and~\eqref{softfactorscalar} with the $|\vec q|=0$ elastic scattering squared matrix elements~\eqref{elastic} and~\eqref{elastic2} and expanding to appropriate order in~$v_i$ turns out to be impractical for long-range mediation: the requirement on the softness for gauge vector emission is so stringent, $x\lll 1$, that although the factorization formally holds, it already yields inaccurate results for $x \ll 1$. 

For long-range forces, we therefore introduce corrections detailed below, in order to retrieve an accurate $x\ll1$ behavior of the soft energy-differential cross section. It results in quadrupole soft factors that differ between long-range and short-range mediation. Moreover, for both types of mediation, long- and short-range, the full kinematics\footnote{A similar observation seems to be made in~\cite{2401.16066}, where it was noted that the original Low theorem \cite{Low:1958sn} is formulated in terms of the physical momenta of the emission process, without relying on the $|\vec{q}| \to 0$ limit of exchanged momenta.} of the emission process must be retained in the derivation of the soft factors~\eqref{softfactorgauge},~\eqref{softfactorscalarA} and~\eqref{softfactorscalar}, as finite \(\vec{q}\)-corrections become essential at quadrupole order.

Even if such procedures appear somewhat cumbersome, deriving emission rates this way is still significantly easier than performing the full Feynman-diagrammatic computations and ensuring the correct low-energy expansion.

\paragraph{Short-range mediation} 
For short-range mediation, the quadrupole soft factors are obtained by performing the $r$-expansion of the eikonal factors~\eqref{softfactorgauge},~\eqref{softfactorscalarA} and~\eqref{softfactorscalar}, using the physical momenta of the emission process and isolating the next-to-leading order contribution. 

We first consider vector emission. The quadrupolar part of the emission factor after squaring, summing over the polarization of $V$ and averaging over its direction, denoted by a subscript~\(\mathcal{Q}\), reads
\begin{equation} \label{softVquadrupoleheavy}
  \langle \mathcal{|E_Q}_{,V}|^2 \rangle_{\hat{q}} =   \left( \frac{g_{{1}}}{m_{{1}}^2} +   \frac{g_{{2}}}{m_{{2}}^2}\right)^2 
  \frac{ \mu^4}{30 \omega^2} \left( 3T_{1} + T_{2} \right). 
\end{equation}
To simplify the presentation, we introduce the short-hand notation for the following functions, 
\begin{align} 
T_{1} & \equiv |\Vec{v}_{f}|^4+ |\Vec{v}_{i}|^4 + |\Vec{v}_{i}|^2 |\Vec{v}_{f}|^2 \left(2- 4(\oldhat{\Vec{v}}_{i} \cdot \oldhat{\Vec{v}}_{f})^2\right), \label{T1}\\
T_{2} & \equiv \left(|\Vec{v}_{i}|^2-|\Vec{v}_{f}|^2\right)^2. \label{T2}
\end{align}
With this result established, we now turn to the expression of the soft quadrupole emission amplitudes. These are obtained by multiplying~\eqref{softVquadrupoleheavy} with the elastic matrix element squared~\eqref{elasticheavy} and~\eqref{elasticheavy2}, for distinguishable and identical particle scattering, respectively. In the former case, the result matches the full expression presented in Sec.~\ref{quadrudisting},

\begin{equation} 
\label{eq:equalityQuad}
 \left. \langle |\mathcal{M_Q}_{,V}|^2 \rangle_{\hat{q}} \right|_{\rm soft} = \langle |\mathcal{M_Q}_{,V}|^2 \rangle_{\hat{q}}.
\end{equation}

For identical particle scattering, the equality~\eqref{eq:equalityQuad} still holds for both fermionic and scalar DM candidates. Indeed, the only difference compared to the squared matrix element in the distinguishable scattering case arises in the process $SS \to SS V$, where the contributions from the external-leg diagrams acquire a factor of 2 relative to the contact diagram—an enhancement already accounted for in the prefactor of~\eqref{elasticheavy2}.

The same procedure is to be repeated in the case of $\phi$-radiation. Starting with the scenarios $S_{1}S_{2}\to S_{1}S_{2}\phi$ and $\tilde{S}_{1}\tilde{S}_{2}\to \tilde{S}_{1}\tilde{S}_{2}\phi$, we expand the eikonal factors~\eqref{softfactorscalarA}  in~$r \ll 1$ using the momenta of the emission process, and isolate the next-to-leading order piece. The corresponding quadrupole emission soft factor, squared and averaged over the direction of $\phi$ is given by, 
\begin{equation}
 \label{softscalarAheavy}
\langle \mathcal{|E}_{\mathcal{Q},\phi S} |^2\rangle_{\vec{\hat{q}}} =
\left(\frac{A_{1 }}{m_{1}^3} + \frac{A_{2}}{m_{2}^3}   \right)^2 
  \frac{  \mu^4}{60 \omega^2} \left( T_{1} + 2T_{2} \right) + \text{additional terms}. %
\end{equation}
Importantly, we note this expression contains the quadrupole soft factor, together with additional terms of the same order in~$r$. The complete expression is given in~\eqref{scalarscalarfactorheavy}. One may proceed as before and obtain the expression of the soft emission amplitudes by multiplying the factor~\eqref{softscalarAheavy} with the elastic matrix element squared~\eqref{elasticheavy} in the case of distinguishable particle scattering. This product not only reproduces the quadrupole emission amplitude~\eqref{quadscalarheavyApp}, but also matches the full result~\eqref{externallegscalarheavy} in the limit $\langle|\mathcal{M}_{\text{vertex}}|^2\rangle \ll \langle|\mathcal{M}_{\text{external leg}}|^2\rangle$,
\begin{equation} 
\label{eq:equalityQuadScalar}
\left. \langle |\mathcal{M}_{\mathcal{Q},\phi S}|^2 \rangle_{\hat{q}} \right|_{\rm soft} = \langle |\mathcal{M}_{\mathcal{Q},\phi S}|^2 \rangle_{\hat{q}}.
\end{equation}
We emphasize that the ``additional terms'' in~\eqref{softscalarAheavy} are essential for establishing this correspondence in the case of distinguishable scattering. They originate from the full $\vec{q}$-dependent kinematics and vanish for equal mass-to-coupling ratios. Similarly, for identical particle scattering, the total soft emission amplitude squared is obtained by multiplying~\eqref{softscalarAheavy} with the squared elastic amplitude~\eqref{elasticheavy2}. As in the process \(SS \to SSV\), the only difference from the distinguishable case lies in the prefactor, which is already accounted for in~\eqref{elasticheavy2}, and thus the equality~\eqref{eq:equalityQuadScalar} remains valid.

For the processes \(\chi_1 \chi_2 \to \chi_1 \chi_2 \phi\) and \(\tilde{\chi}_1 \tilde{\chi}_2 \to \tilde{\chi}_1 \tilde{\chi}_2 \phi\), we carry out the same steps, this time using the soft factor~\eqref{softfactorscalar}, which yields
\begin{equation} \label{quadrupoleheavyfermionscalar}
  \langle \mathcal{|E_Q}_{,\phi \chi}|^2 \rangle_{\hat{q}} =   \left( \frac{y_{{1}}}{m_{{1}}^2} +   \frac{y_{{2}}}{m_{{2}}^2}\right)^2 
  \frac{ \mu^4}{15 \omega^2} \left( T_{1} + 7T_{2} \right). 
\end{equation}
To derive the soft quadrupole emission amplitude, we proceed by taking the quadrupole soft factor~\eqref{quadrupoleheavyfermionscalar} and multiplying it by the elastic amplitudes squared~\eqref{elasticheavy} (distinguishable particle scattering) and~\eqref{elasticheavy2} (identical particle scattering). In both cases, the resulting soft expression reproduces the full quadrupole emission amplitudes, namely,
\begin{equation} \label{equalityQuadScalarfermion} \left. \langle |\mathcal{M}_{\mathcal{Q},\phi \chi}|^2 \rangle_{\hat{q}} \right|_{\rm soft} = \langle |\mathcal{M}_{\mathcal{Q},\phi \chi}|^2 \rangle_{\hat{q}}. \end{equation}
Here, we note that had we only kept  the leading soft terms in~\eqref{softfactorscalar}
we would have obtained identical soft emission factors for $S(\tilde{S})$ and $\chi(\tilde{\chi})$ under the substitution $A_j/m_j \to 2  y_j$. In this case, instead of the correct quadrupole soft factor~\eqref{quadrupoleheavyfermionscalar} we would have obtained~\eqref{softscalarAheavy}. The  equality~\eqref{equalityQuadScalarfermion} would not be valid, and the full NR quadrupole emission amplitude would not be recovered. However, for the special case of short-range mediation, the soft limit remains unaffected: both $\chi_{1}\chi_{2}\to\chi_{1}\chi_{2}\phi$ and $S_{1}S_{2}\to S_{1}S_{2} \phi$ share the same analytical part of their quadrupole amplitudes that is enhanced is the limit $x \ll 1$; see App.~\ref{scalarvsgauge} for more details.

Before presenting the soft energy-differential cross sections associated with the squared amplitudes derived in this section, we examine their counterparts in the case of long-range mediated interactions.

\paragraph{Long-range mediation and limitations of the soft emission approach}

For long-range mediation, although the factorization of amplitudes formally connects to the exact result, for quadrupole emission it does so only for extremely small values, $x \lll 1$, making it effectively useless in practice and producing wrong results when used outside the narrow range of validity. This observation was first made in \cite{Pradler:2020znn} for the QED quadrupole bremsstrahlung process, where, to mitigate this practical limitation, the eikonal factor~\eqref{softfactorgauge} is modified by introducing a correction term 
\begin{equation} \label{correctionsoft}
  k_{\text{elastic}}^2/k_n^2 , 
\end{equation}
where $k_n$ represents the actual transferred momentum from $n$-th external leg during the emission, while $k_{\text{elastic}}$ corresponds to the elastic transferred momentum in the limit $|\vec{q}| \to 0$. Such factors are to be applied prior to a velocity expansion.
This way, $\vec q$-dependent corrections to the elastic process are being reinstated through the eikonal factors while retaining the simple analytic structure of the elastic amplitudes. Such prescriptions only take effect at quadrupole order, see App.~\ref{Softderivations} for more details. We emphasize that the correction factor~\eqref{correctionsoft} is not needed in the contact limit, since the momentum transfer in the elastic scattering is negligible compared to the mediator mass. 

With those considerations in place, we can now improve~\eqref{softfactorgauge},~\eqref{softfactorscalarA} and~\eqref{softfactorscalar} by the prescription explained above. We first consider the case of $V$-emission.  Multiplying~\eqref{softfactorgauge} by the correction factor~\eqref{correctionsoft}, applying the $v_i$-expansion using the physical momenta of the emission process, and isolating the next-to-leading order contribution yields the quadrupole term, squared, summed over the polarization of $V$ and averaged over its direction,
\begin{equation} \label{softVquadrupole}
  \langle \mathcal{|E_Q}_{,V}|^2 \rangle_{\hat{q}} =   \left( \frac{g_{{1}}}{m_{{1}}^2} +   \frac{g_{{2}}}{m_{{2}}^2}\right)^2 
  \frac{ \mu^4}{30 \omega^2} \left( 3T_{1} + 13T_{2} \right), 
\end{equation}
which is in agreement with \cite{Pradler:2020znn}. 
For distinguishable particle scattering, multiplying the soft factor~\eqref{softVquadrupole} by the Rutherford squared amplitude~\eqref{elastic} leads, as in the short-range mediation case, to the equality~\eqref{eq:equalityQuad} between the soft emission amplitude and the full result~\eqref{generalquadrupole}.
In the case of identical particle scattering, the soft emission amplitude is readily obtained by multiplying~\eqref{softVquadrupole} with the elastic matrix element squared~\eqref{elastic2}. Although this product correctly captures the \(x \ll 1\) behavior of the full emission amplitudes~\eqref{EMQEDampli}, it cannot reproduce the full expressions, specifically the interference terms between the \(t\)- and \(u\)-channel.

Turning to \(\phi\)-radiation, the same steps as above are applied. One combines the soft factors~\eqref{softfactorscalarA} and~\eqref{softfactorscalar} with the correction term~\eqref{correctionsoft}, keeps the full physical momenta of the emission process, and applies the~$v_i$-expansion. For the processes \(S_1 S_2 \to S_1 S_2 \phi\) and \(\tilde{S}_1 \tilde{S}_2 \to \tilde{S}_1 \tilde{S}_2 \phi\), the corresponding quadrupole soft factor, squared and averaged over the direction of $\phi$, takes the form
\begin{equation} \label{softscalarlong}
\langle |\mathcal{E}_{\mathcal{Q},\phi S} |^2\rangle_{\vec{\hat{q}}} =
\left(\frac{A_{1 }}{m_{1}^3} + \frac{A_{2}}{m_{2}^3}   \right)^2  
  \frac{  \mu^4}{60 \omega^2} \left(T_{1} +  11T_{2} \right) + \text{additional terms}.
\end{equation}
Similarly to the short-range mediation case,~\eqref{softscalarlong} is composed of a quadrupole component and some additional terms that are of the same order in~$v_{i}$, which cancel in the case of equal coupling to mass ratio; the full expression is found in~\eqref{scalarscalarfactor}.  The quadrupole soft  factor of  $\chi_1 \chi_2 \to \chi_1 \chi_2 \phi$ and $\tilde{\chi}_1 \tilde{\chi}_2 \to \tilde{\chi}_1 \tilde{\chi}_2 \phi$, squared and averaged over the direction of $\phi$, is given by
\begin{equation} \label{lightsoftfermion}
  \langle |\mathcal{E}_{\mathcal{Q},\phi\chi} |^2\rangle_{\vec{\hat{q}}} =   \left( \frac{y_{{1}}}{m_{{1}}^2} +   \frac{y_{{2}}}{m_{{2}}^2}\right)^2 
  \frac{ \mu^4}{15 \omega^2} \left( T_{1} + 6T_{2} \right). 
\end{equation}
In contrast to short-range mediation, it is crucial to carry along the $\mathcal{O}(q^{0})$-term in~\eqref{softfactorscalar} in order to obtain the correct $x\ll1$ behavior of the full NR amplitude; see App.~\ref{scalarvsgauge} for a detailed discussion.

With these factors at hand, we now obtain the soft quadrupole emission amplitudes. In the case of distinguishable particle scattering, each amplitude is obtained by multiplying the quadrupole soft factors~\eqref{softscalarlong} and~\eqref{lightsoftfermion} with the squared Rutherford matrix element~\eqref{elastic}. These two products reproduce, respectively, the quadrupole emission amplitude~\eqref{quadscalarlongapp} and the full result~\eqref{externallegscalarlong} in the limit $\lambda_j \ll (A_j/m_j)^2$, and the full quadrupole amplitude~\eqref{generalquadrupole}, thereby yielding the equalities~\eqref{eq:equalityQuadScalar} and~\eqref{equalityQuadScalarfermion}. For identical particle scattering, the situation parallels the other scenarios. Multiplying the quadrupole soft factor~\eqref{softscalarlong} with the squared Coulomb amplitude~\eqref{elastic2} reproduces the full quadrupole amplitude~\eqref{EMscalar} (for $\kappa=0$), while using the soft factor~\eqref{lightsoftfermion} instead reproduces~\eqref{EMyuka}, up to the interferences between the $t$- and $u$-channel contributions, which are not fully captured. Nevertheless, the results coincide with the full expressions in the $x \ll 1$ limit.

We conclude the derivations of the soft amplitudes with a few remarks, applicable to both short- and long-range mediation. For vector emission, the equality~\eqref{eq:equalityQuad} arises from gauge invariance: the gauge constraint cancels the $\slashed{q}$-contribution in the emission current at quadrupole order, reducing the full dissipative amplitude to the factorized product of the elastic amplitude and the soft factor. For scalar emission from scalar DM candidates, verifying~\eqref{eq:equalityQuadScalar} is straightforward: the general amplitude of the process reduces in the NR limit to its soft expression, as the emission current does not present any momentum dependencies for scalar candidates. Finally, for scalar emission from fermionic DM candidates, the equality~\eqref{equalityQuadScalarfermion} arises once a subleading $\mathcal{O}(q^{0})$-contribution is included in the soft factor. The need for this subleading term directly follows from the absence of any underlying symmetry constraining scalar emission; for further details, see App.~\ref{scalarvsgauge}.

\subsection{Soft quadrupole emission: energy-differential cross sections} \label{softenergydifferentialcross-sectionquad}

We are now in a position to obtain the energy-differential cross sections for quadrupole emission. For identical particle scattering, these are the leading order in $v_i$ and $r$ results. 
As mentioned several times, for non-identical particles, they are sub-leading to the dipole order, unless the coupling-to-mass ratios in the dipole factors~$\mathcal{F_D}$ in~\eqref{prefactordipole} for long-range mediation or $\mathcal{F_D'}$ in~\eqref{shortfermiondipole}-\eqref{shortscalardipole} for short-range mediation, happen to cancel.

\paragraph{Distinguishable particles}
In the case of distinguishable particle scattering, we perform the phase-space integrations in~\eqref{crosssection} in the $x \ll 1$ limit, using the soft quadrupole emission amplitudes defined in Sec.~\ref{softquadrupole}. At leading order in~$x$, this yields the following soft energy-differential cross sections for short-range mediation,
\begin{equation} \label{softquadheay}
\left. \omega \frac{d \sigma}{d \omega} \right|_{\rm soft} =
  \mathcal{F}_{\mathcal Q}'^2\frac{2 \alpha f_{\phi,V}  \mu^6 v_{i}^4}{45\pi^3 m_{\phi'}^4} \quad \text{(external legs emission)},
\end{equation}
where the parameters $\alpha$ and the pre-factors $\mathcal{F}'_{\mathcal{Q}}$ are defined in~\eqref{quadshortfermion}-\eqref{quadshortscalar2}. 
While, in the case of long-range mediation, the corresponding soft energy-differential cross sections read,
\begin{equation}  \label{softquadlight}
   \left. \omega \frac{d \sigma}{ d \omega} \right|_{\rm soft}= \mathcal{F}_{\mathcal Q}^2 \frac{ f_{\phi,V} \mu^2 }{60 \pi^3} \left[ (\beta - \alpha) +  2 \alpha\ln{\left(\frac{2\mu v_{i}^2}{\omega}\right)} \right]\quad  (A_j/m_{j})^2\gg \lambda_j ,  
\end{equation}
where the parameters $\alpha,\beta$ and the pre-factors $\mathcal{F}_{\mathcal{Q}}$ are given in~\eqref{factorQ}. 

We note that both~\eqref{softquadheay} and~\eqref{softquadlight} agree at leading order in $x$ with the full quadrupole energy-differential cross sections, respectively~\eqref{shortrangequad} and~\eqref{crosssectionquadrupole}, for all the scenarios considered in this work.
In Fig.~\ref{softfigdipole}, as for the dipole emission case, we compare the soft and the full expressions, both for short-range (left panel) and long-range (right panel) mediation.

\paragraph{Identical particles}
When treating scattering of identical particles, the soft quadrupole emission amplitude squared is given by the product of the quadrupole soft factor defined in Sec.~\ref{softquadrupole} and the squared elastic amplitude~\eqref{elastic2} (long-range mediation) and~\eqref{elasticheavy2} (short-range mediation). 

For long-range mediation, substituting these products into~\eqref{crosssection} and integrating in the $x \ll 1$ limit yields, for vector emission,
\begin{subequations}\label{SoftEDcsVector} %
\begin{align}
\omega\frac{d\sigma}{d\omega}\Big|_{\text{soft}}
  &= \frac{g^{6}}{240\pi^{3} m_{\chi}^{2}}
     \left[17+12\ln\!\left(\frac{2\mu v_{i}^{2}}{\omega}\right)\right]
     \quad \text{for} \quad \chi\chi \to \chi\chi V,
     \label{SoftEDcsVector:chi}\\[4pt]
\omega\frac{d\sigma}{d\omega}\Big|_{\text{soft}}
  &= \frac{g^{6}}{240\pi^{3} m_{S}^{2}}
     \left[26+12\ln\!\left(\frac{2\mu v_{i}^{2}}{\omega}\right)\right]
     \quad\text{for }\quad SS \to SSV.
     \label{SoftEDcsVector:S}
\end{align}
\end{subequations}
These soft energy-differential cross sections agree with their full quadrupole counterparts,~\eqref{EMQED} and~\eqref{EMQEDscalar}, for $\kappa=0$, once the latter are expanded to first order in $x$. 
\begin{figure}[t!]
    \centering
\includegraphics[width=1\linewidth]{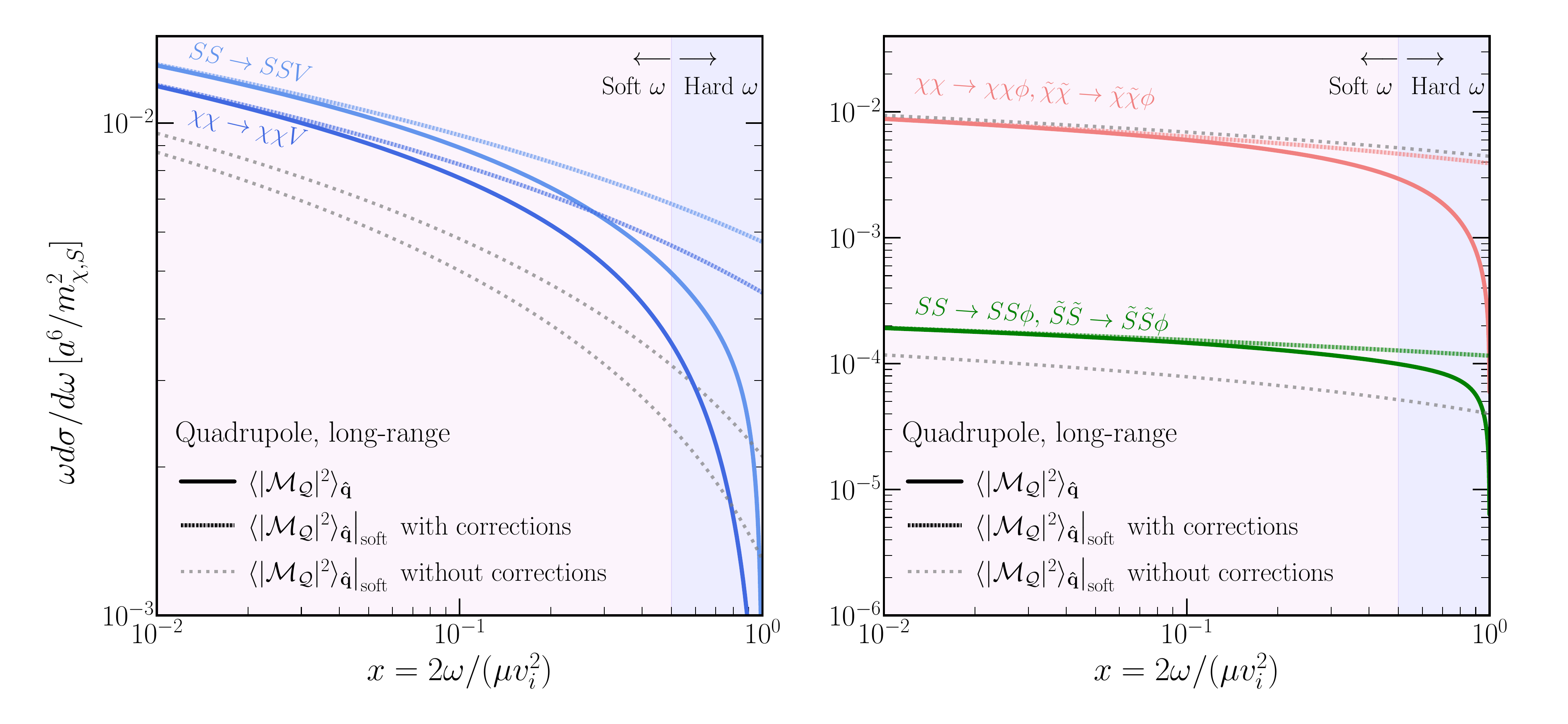}
    \caption{ Energy-differential cross section for identical particle scattering mediated by a long-range interaction. Here, $a$ corresponds to the couplings $g,y$ or $A/m$ depending on the dSIDM scenario considered. A separation scale between ``hard'' and ``soft'' emission at half the available CM energy \(\Lambda \sim \mu v_i^2/4\) is introduced; soft modes (\(\omega \lesssim \Lambda\)) are shaded light purple, hard modes (\(\omega \gtrsim \Lambda\)) light blue. The solid colored lines represent the massless emission limit ($\kappa=0$) of the energy-differential cross sections from Sec.~\ref{identicalparticles}, while the colored dotted lines correspond to~\eqref{SoftEDcsVector} (vector emission) and to~\eqref{SoftEDcsScalar}-\eqref{softfermionscalar} (scalar emission), obtained using the soft quadrupole emission amplitude squared corrected by the factor~\eqref{correctionsoft}. The light grey dotted lines show the uncorrected soft approximation.    
    \textit{Left}: vector emission. \textit{Right}: scalar emission.}
    \label{softfig}
\end{figure}

Applying the same methodology to scalar emission, we obtain the following result for scalar DM candidates, in the limit $(A/m)^2 \gg \lambda$ and at leading order in $x$,
\begin{equation} \label{SoftEDcsScalar}
    \left. \omega \frac{d \sigma}{d \omega} \right|_{\text{soft}}= \left(\frac{A}{m_{S}}\right)^6 \frac{1}{3840 \pi^3 m_{S}^2}\left[11+2\ln{\left(\frac{2 \mu v_{i}^2}{\omega}\right)} \right] \quad \text{for} \quad SS \to SS \phi, \, \Tilde{ S} \Tilde{ S} \to \Tilde{S} \Tilde{S} \phi,
\end{equation}
which agrees with the full quadrupole counterpart~\eqref{EMscalarscalar}, for $\kappa=0$, when expanded to leading order in~$x$. Similarly, for fermionic DM candidates, we obtain the soft energy-differential cross section 
\begin{equation}\label{softfermionscalar}
  \left.  \omega \frac{d \sigma}{d \omega}\right|_{\text{soft}}= \frac{y^6}{120 \pi^3 m_{\chi}^2}  \left[9 +4 \ln{\left( \frac{2 \mu v_{i}^2}{\omega}\right)} \right] \quad \text{for} \quad \chi \chi \to \chi \chi \phi,\, \Tilde{\chi} \Tilde{\chi} \to \Tilde{\chi} \Tilde{\chi} \phi,
\end{equation}
which reproduces the correct $x \ll 1$ behavior of~\eqref{EMchiscalar} for $\kappa =0$. In Fig.~\ref{softfig}, we compare the soft energy-differential cross sections obtained with and without the correction factor~\eqref{correctionsoft} for the case of identical particle scattering.

Finally, in the case of short-range mediation, the resulting soft energy-differential cross sections are in perfect agreement with the full results ($\kappa=0$) presented in Sec.~\ref{identicalparticles}, once the latter are expanded at first order in $x$. Since the resulting expressions can be directly inferred from the full results, we refrain from presenting them explicitly.

\section{Energy loss rate} \label{energylossrate}

A central quantity to be derived from the above results is the thermally averaged energy loss rate of a NR Maxwellian gas of DM particles. On general grounds, the total emitted energy into $\phi$ or $V$ particles from binary DM collisions per time and unit volume is given by 
\begin{equation} \label{ELR}
\Dot{\epsilon} \equiv \int \prod_{k}d\Pi_{k}\  (2\pi)^4 \delta^{(4)} \left(p_{1}+p_{2}- p_{3}-p_{4}-q \right)
S_{i}S_{f}\ \omega |\mathcal{M}|^2 
f_{1}f_{2} \left(1 \pm f_{3} \right) \left(1 \pm f_{4}\right) \left(1 + f_{q} \right) ,
\end{equation}
where $\mathcal{|M|}^2$ is the squared matrix element summed over initial and final spins; $S_{i(f)}$ are the symmetry factors for identical particles in the initial (final) state as before. The product of integrations runs over all particles involved and $d\Pi_{k}= d^{3}\vec p_{k}/[(2 \pi)^3 2 E_{k}]$ is the Lorentz-invariant phase space element with $E_k = (\vec p_k^2 + m_k^2)^{1/2}$ as usual. We again reserve $\omega \equiv E_q =  (\vec q^2 + m_{\phi,V}^2)^{1/2}$ for the emitted particle energy. The phase space occupation numbers are given by~$f_k$. In the above equation we neglect the inverse process of absorption, which, by the principle of detailed balancing, can be obtained from the emission rate.

Here we shall treat only the most relevant case of a non-degenerate Maxwellian DM gas of, potentially, two non-identical particle species $1$ and $2$ of common temperature $T=T_{1,2}$ and respective number densities $n_1$ and $n_2$, or of a single particle species of temperature $T$ and number density~$n$. We may then neglect the final state Bose enhancement or Pauli blocking factors $(1\pm f_{3,4})$ and, for all practical purposes, neglect the effect of stimulated mediator emission $(1+f_q)\approx 1$. The DM occupation number is then given by 
\begin{equation} \label{occupationnb}
f_j=\frac{n_{j}}{d_j} \left(\frac{2\pi}{m_{j}T} \right)^{3/2}e^{-\frac{|\vec{p}_{j}|^2}{2m_{j}T}}.   
\end{equation}
Using a standard reduction of the phase space integrals\footnote{This reduction holds under the assumption that the CM frame and the medium frame coincide in the NR framework \cite{Pradler:2021ppc}. It introduces an error of order $\mathcal{O}(\vec{v}_{\text{CM}})$, with $\vec{v}_{\text{CM}}$ the velocity of the CM relative to the medium frame.}, the energy loss rate~\eqref{ELR} can then be brought to an integral over the CM energy-differential cross sections derived above,
\begin{equation} \label{Dot(e)}
 \dot{ \epsilon }= \frac{\sqrt{2}\mu^{3/2}n_{1}n_{2}S_{i}}{\sqrt{ \pi} T^{3/2}} \int_{v_{\text{min}}}^{\infty} dv_{i}\, v_{i}^3 e^{-\frac{\mu v_{i}^2}{2 T}} \int_{m_{\phi,V}}^{\mu v_{i}^2/2} d\omega\, \omega \frac{d \sigma}{d \omega},
\end{equation}
with $v_{\text{min}}=\sqrt{2m_{\phi,V}/\mu}$. In particular, in a gas composed of two non-identical particles, the total energy loss rate decomposes as
\begin{equation}
   \dot{\epsilon}_{\text{total}}= \dot{\epsilon}_{11} + \dot{\epsilon}_{22} + \dot{\epsilon}_{12}, 
\end{equation}
where the subscript ${jk}$ identifies the species of the two incoming particles in the scattering.
The first two terms correspond to scattering among particles of the same species (identical-particle scattering), while the last term arises from scattering between the two different species (distinguishable particles scattering). The expressions for the latter case are provided in Sec.~\ref{edotdisting}, whereas those in the former case are given in Sec.~\ref{edotidentic}. A summary of all the expressions is provided in Tab.~\ref{tab:energylossratefinal}.
The energy loss rate can be evaluated analytically for massless emission using~\eqref{Dot(e)}. For the emission of a massive particle, however, a numerical evaluation is required. In this case, for dipole emission and the scattering of identical particles, the energy-differential cross section, defined in Sec.~\ref{Leading order} and Sec.~\ref{identicalparticles}, must be inserted in~\eqref{Dot(e)}, while for quadrupole emission from the scattering of distinguishable particles, the energy-differential cross section provided in App.~\ref{Massiveemission} must be used. 

Moreover, as can be inferred from~\eqref{Dot(e)}, the emission of a massive boson becomes exponentially suppressed once $m_{\phi,V} \gtrsim T$. In addition to this suppression, it follows from the previous sections that the mass of the emitted particle appears in the energy-differential cross section in a non-trivial, model-dependent way, altering the detailed shape of the overall suppression. Therefore, to quantify these effects, we define the ratio
\begin{equation} \label{R}
    R = \frac{\dot{\epsilon}}{\dot{\epsilon}_0},
\end{equation}
where the subscript $0$ indicates the emission of a massless particle, $m_{\phi,V} = 0$, but for otherwise identical model-parameters.

\subsection{Scattering of distinguishable particles} \label{edotdisting}
We start by computing the energy loss rate corresponding to distinguishable particle scattering. Here, dipole radiation always dominates, unless there is an accidental cancellation in the coupling-to-mass ratios between the two particle species. We provide the general expressions for the emission rate in both long-range and short-range mediation, derived from the energy-differential cross sections of Secs.~\ref{Leading order} and~\ref{quadrupoleemission}, which correspond to massless emission. We then discuss the differences between the various dissipative models for a given type of emission (dipole or quadrupole), including the impacts of emitted particle mass on the emission rate.

\paragraph{Long-range mediation}
In the case of interactions mediated by a long-range force, by substituting the energy-differential cross sections~\eqref{dipolelongrange} (for $m_{\phi,V}=0$) and~\eqref{crosssectionquadrupole} in~\eqref{Dot(e)}, we obtain for the fermionic dSIDM options, $\chi_1\chi_2\to \chi_1\chi_2 \phi/V$, $\tilde \chi_1\tilde \chi_2\to \tilde \chi_1\tilde\chi_2 \phi$, as well as for the scalar dSIDM option $S_1 S_2\to S_1S_2\phi/V$, and  $\tilde S_1 \tilde S_2\to \tilde S_1\tilde  S_2\phi$ under the premise $(A_{j}/m_{j})^2 \gg \lambda_j $, the emission rate
\begin{equation} \label{NRJlossrate}
\centering
\dot{\epsilon}=n_{1}n_{2} \frac{f_{\phi,V}}{6\pi^{7/2} \sqrt{2}} \left[\mathcal{F}_{\mathcal D}^2 (\mu T)^{1/2}+ \frac{24}{45}\mathcal{F}_{\mathcal Q}^2  (\mu T)^{3/2} \left( 4 \alpha + \beta \right)  \right] ,
\end{equation}
where the dipole and quadrupole emission factors $\mathcal{F_D}$ and $\mathcal{F_Q}$ are found in~\eqref{prefactordipole} and~\eqref{factorQ}, respectively. For quadrupole emission, the model-dependent coefficients $\alpha$ and $\beta$ are found in Sec.~\ref{quadrupoleemission}.  When the scalar quartic coupling dominates in the processes \( S_1 S_2 \to S_1 S_2 \phi \) and \( \tilde{S}_1 \tilde{S}_2 \to \tilde{S}_1 \tilde{S}_2 \phi \), i.e., for \(\lambda_j \gg (A_{j}/m_{j})^2  \), the emission proceeds via the vertex diagram (Fig.~\ref{vertex}) and the associated energy loss rate exhibits the same temperature dependence as for quadrupole radiation and reads
\begin{equation} \label{longvecterNRJ}
\dot{\epsilon}_{\textrm{vertex}}=n_{1}n_{2} \left(A_{{1}}\lambda_{2} + A_{{2}}\lambda_{{1}}\right)^2\frac{  T^{3/2}}{48  \pi^{7/2} \sqrt{2 \mu} m_{1}^2m_{2}^2}.
\end{equation}

\paragraph{Short-range mediated force}
We now evaluate the energy loss rate arising from short-range mediated forces, using the energy-differential cross sections given in~\eqref{dipoleshortrange} (for $m_{\phi,V}=0$) for dipole and~\eqref{shortrangequad} for quadrupole emission. 
For the various dSIDM scenarios considered, including the processes $S_1 S_2 \to S_1 S_2 \phi$ and $\tilde S_1 \tilde S_2 \to \tilde S_1 \tilde S_2 \phi$ in the limit $\langle|\mathcal{M}_{\text{vertex}}|^2\rangle \ll \langle|\mathcal{M}_{\text{external leg}}|^2\rangle$ (see App.~\ref{scalarscalarquad}), we obtain
\begin{equation}  \label{NRJ2}
\dot{\epsilon} = 
n_{{1}}n_{{2}} \frac{32 \sqrt{2}f_{\phi,V}}{15 m_{\phi',V'}^4 \pi^{7/2} } \left[ \mathcal{F}_{\mathcal D}'^2(\mu T)^{5/2} +  \frac{8}{35} \mathcal{F}_{\mathcal Q}'^2  (\mu T)^{7/2}  \left( 8 \alpha + \beta \right) \right],
\end{equation}
where the dipole and quadrupole emission factors $\mathcal{F'_D}$ and $\mathcal{F'_Q}$ are defined in~\eqref{shortfermiondipole}-\eqref{shortscalardipole} and in~\eqref{quadshortfermion}-\eqref{quadshortscalar2} respectively. Furthermore, for the processes~$S_1 S_2\to S_1S_2\phi$, and  $\tilde S_1 \tilde S_2\to \tilde S_1\tilde S_2\phi$, we may again obtain the opposite case of vertex-dominated emission, i.e., $\langle|\mathcal{M}_{\text{vertex}}|^2\rangle \gg \langle|\mathcal{M}_{\text{external leg}}|^2\rangle$ (see App.~\ref{scalarscalarquad}),
\begin{equation}\label{shortvecterNRJ}
\dot{\epsilon}_{\textrm{vertex}}=n_{1}n_{2}\frac{4 \sqrt{2}}{35}\left(A'_{{1}}\lambda_{2}' + A'_{{2}}\lambda_{{1}}'\right)^2  \frac{ \mu^{3/2}  T^{7/2}}{\pi^{7/2}  m_{{1}}^2m_{{2}}^2 m_{\phi',V'}^4 }.
\end{equation}

The parametric scaling of energy loss rates of short-range over long-range mediation, up to coupling constant prefactors, goes as 
\begin{equation}
\frac{\dot{\epsilon}|_{\text{short-range}}}{\dot{\epsilon}|_{\text{long-range}}} \sim \frac{(\mu T)^2 }{m_{\phi',V'}^4} .
\end{equation}
On the account that $\mu T \sim (\mu \langle v_i\rangle)^2 $, we see that energy loss due to short-range mediation is suppressed, because of the premise there, that the typical momentum transfer $\mu \langle v_i\rangle$ in the scattering is smaller than the mediator mass~$m_{\phi',V'}$.

\subsubsection{Dipole emission}

The lowest powers in temperature $T^{1/2}$ in~\eqref{NRJlossrate} (long-range mediation) and $T^{5/2}$ in~\eqref{NRJ2} (short-range mediation) are associated with dipole emission. 

\paragraph{Massless emission} For both types of mediation, the energy loss rate for massless $V$-emission is enhanced by a factor of~2 compared to massless $\phi$-emission. This enhancement follows directly from the squared dipole emission amplitudes~\eqref{generaldipole} and~\eqref{dipoleshortrangeamplitude}. In the case of long-range mediation, for massless vector emission, in the limit $m_{1} \ll m_{2}$, our result aligns with the standard textbook treatments \cite{Rybicki:2004hfl,2011piim.book.....D}, assuming a free–free Gaunt factor $g_{\text{ff}}\!\approx\!1$, as was also adopted in related dark-sector applications \cite{Foot:2014uba,1705.10341}. For scalar---both massive and massless---emission, from fermionic DM candidates, our dipole energy loss rate agrees with the results of~\cite{2303.00778,2303.03123}.

\paragraph{Massive emission} If the emitted particle is massive, the energy loss rate~\eqref{Dot(e)} must be evaluated numerically using the dipole energy-differential cross sections~\eqref{dipolelongrange} (long-range mediation) and~\eqref{dipoleshortrange}  (short-range mediation). As discussed in Sec.~\ref{Leading order}, we emphasize that in the case of dipole emission, the form of the $m_{\phi,V}$-dependence appearing in the unpolarized emission amplitudes is the same for both types of mediation and depends only on the nature of the emitted particle, not on the spin of the DM candidates. 
\begin{figure}[t!] 
    \centering
\includegraphics[width=1\linewidth]{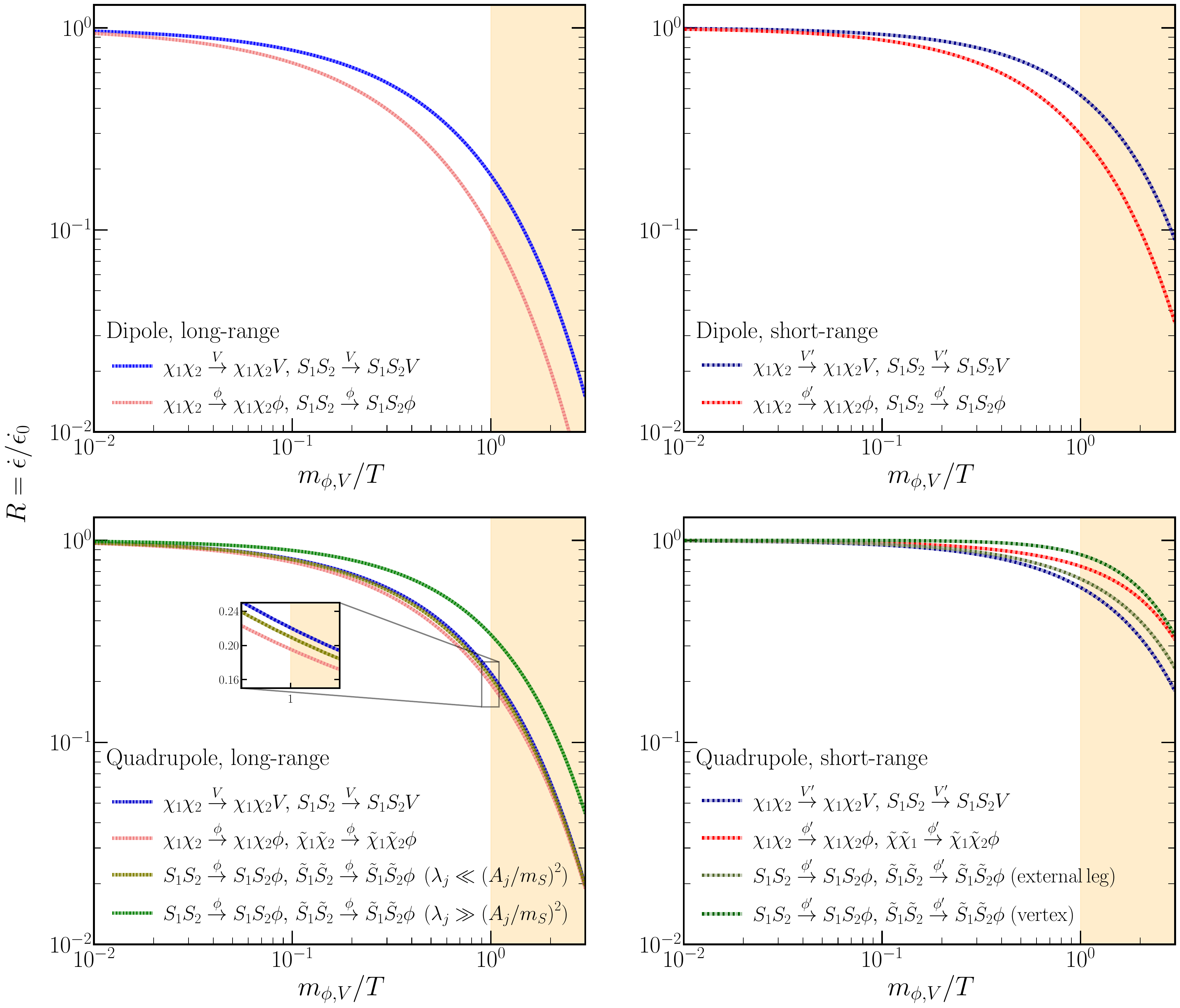}
    \caption{ Effect of the mass of the emitted particle on the energy loss rate, shown relative to the massless case~\eqref{R}, for distinguishable particle scattering, in the case of long-range ({\it left}) and short-range ({\it right}) mediated interactions. {\it Upper panels:} Dipole emission. {\it Lower panels:} Quadrupole emission and, for $S_{1}S_{2}\to S_{1}S_{2} \phi$ and $\tilde S_{1} \tilde S_{2}\to \tilde S_{1} \tilde S_{2} \phi$, vertex emission~\eqref{longvecterNRJ} and~\eqref{shortvecterNRJ}. 
    The orange region corresponds to the kinematically suppressed regime \( m_{\phi, V} \geq \mu \langle v_{i}^2 \rangle \).}
\label{energylossratequadanddipolesl}
\end{figure}

In the upper panels of Fig.~\ref{energylossratequadanddipolesl}, we plot the ratio $R$ defined in~\eqref{R} which compares the energy loss rates between massless and massive emissions.  For both types of mediation, the difference in the effects of the emitted particle mass between $\phi$- and $V$-emission is mild up to $m_{\phi,V}\sim T$.
When $m_{\phi,V}=3T$, the mass of the emitted particle already exceeds the typical kinetic energy of the plasma, and the energy loss rate is exponentially suppressed. In this regime, the energy loss rate from massive $\phi$-emission is more suppressed than in the case of massive $V$-emission. Additionally, we note that the impact of $m_{\phi,V}$ on the energy loss rate of short-range mediation is weaker by an order of magnitude compared to the long-range mediation scenario.

\subsubsection{Quadrupole emission}
Higher powers in $T$ of the energy loss rates, $T^{3/2}$ for long-range mediation in~\eqref{NRJlossrate} and $T^{7/2}$ for short-range mediation in~\eqref{NRJ2}, arise from quadrupole emission. Their suppression relative to dipole emission scales with the typical kinetic energy in the system, \(T \sim \mu \langle v_i^2 \rangle\). At this order, the energy loss rates become model-dependent for both mediations, reflecting the structure of the corresponding energy-differential cross sections~\eqref{crosssectionquadrupole} and~\eqref{shortrangequad}. This model dependence can be qualitatively evaluated by comparing the energy loss rates across different dSIDM scenarios.

 \paragraph{Massless emission}
Starting with long-range mediated processes, $\chi_1 \chi_2 \to \chi_1 \chi_2 V$, $\chi_1 \chi_2 \to \chi_1 \chi_2 \phi$, $\tilde{\chi}_1 \tilde{\chi}_2 \to \tilde{\chi}_1 \tilde{\chi}_2 \phi$, and $S_1 S_2 \to S_1 S_2 V$ all give rise to quadrupole radiation. In the first three cases, the dominant contribution arises from external leg emission, while in the last process, there is an additional vertex emission contribution. For the processes $S_1 S_2\to S_1S_2\phi$ and $\tilde S_1 \tilde S_2\to \tilde S_1\tilde  S_2\phi$, two limits must be considered. In the limit $(A_{j}/m_{j})^2\gg \lambda_{j}$, the latter processes are dominated by external leg emission, yielding quadrupole emission as well. Consequently, by assuming a common reduced mass across all the DM candidates and massless emission, the ratios of the energy loss rates between the different dSIDM scenarios reduce to
\begin{equation} \label{quadrupolecompalongrange}
\begin{aligned}
\left.\frac{ \dot{\epsilon}_{\mathcal{Q},l}  } {\dot{\epsilon}_{\mathcal{Q},k} }  \frac{ \mathcal{F}_{\mathcal{Q},k}^2 } { \mathcal{F}_{\mathcal{Q},l}^2 }\right|_{\text{long-range}}=
 \begin{cases}   
   \displaystyle{ \frac{2}{3}\quad \text{for}\quad  l =\phi \,\chi_{1,2}(\Tilde{\chi}_{1,2}),\,\, k= \phi \, S_{1,2}(\Tilde{S}_{1,2})  }, \\[10pt]
      \displaystyle{  \frac{5}{6}\quad \text{for}\quad  l= V \, \chi_{1,2}({S}_{1,2}),\,\, k= \phi \, S_{1,2}(\Tilde{S}_{1,2})  },\\[10pt]
      \displaystyle{ \frac{4}{5}  \quad \text{for}\quad l =\phi \,\chi_{1,2}(\Tilde{\chi}_{1,2}),\,\, k = V \, \chi_{1,2}({S}_{1,2})   }.
 \end{cases}
\end{aligned}
\end{equation}
Here, the subscript $\mathcal{Q}$ indicates the quadrupole nature of the emission. The indices $l,k$ label the scenarios, with the shorthand notation listing the emitted particle first, followed by the DM candidates; for the latter, one may, individually for $l$ and/or $k$, choose either the unbracketed or bracketed DM candidate for the equation to hold.
When the processes $S_1 S_2 \to S_1 S_2 \phi$ and $\tilde{S}_1 \tilde{S}_2 \to \tilde{S}_1 \tilde{S}_2 \phi$ are dominated by vertex emission, i.e., in the regime $(A_{j}/m_{j})^2\ll \lambda_{j}$, the energy loss rate ratio becomes
\begin{equation} \label{quadrupolecompalongrange2}
\frac{\dot{\epsilon}_{\mathcal{Q},k}}{\dot{\epsilon}_{\text{vertex}}}
= C_{k} \,  \frac{ \mu^2 m_1^2 m_2^2 \mathcal{F}_{\mathcal Q}^2 }{ \left(A_1 \lambda_2 + A_2 \lambda_1\right)^2  } \quad \quad (A_{j}/m_{j})^2\ll \lambda_{j} ,
\end{equation}
with $C_k = 128/3$ for $k = \phi \,\chi_{1,2}(\Tilde{\chi}_{1,2}) $, and $C_k = 160/3$ for~$k = V \, \chi_{1,2}({S}_{1,2}) $.

Turning now to the case of short-range mediation, the scenarios $\chi_1 \chi_2\to \chi_1\chi_2 V$, $\chi_1 \chi_2\to \chi_1\chi_2 \phi$, and $\tilde{\chi}_1 \tilde{\chi}_2\to \tilde{\chi}_1\tilde{\chi}_2 \phi$
are dominated by external leg emission, while $S_1 S_2\to S_1S_2 V$ gains additional contributions from the vertex and contact emission diagrams, Fig.~\ref{inifini}-\ref{Contacts}. These dissipative channels lead to quadrupole emission. For the processes $S_1 S_2 \to S_1 S_2 \phi$ and $\tilde S_1 \tilde S_2 \to \tilde S_1 \tilde S_2 \phi$, two limiting cases should again be considered. In the limit $\langle|\mathcal{M}_{\text{vertex}}|^2\rangle \ll \langle|\mathcal{M}_{\text{external leg}}|^2\rangle$ (see App.~\ref{scalarscalarquad})—i.e., when external leg emission, and hence quadrupole emission, dominates—and assuming a common reduced mass across all DM candidates, the differences between the various energy loss rates reduce to (with the notation as in~\eqref{quadrupolecompalongrange})
\begin{equation} 
\begin{aligned}
 \left. \frac{ \dot{\epsilon}_{\mathcal{Q},l}  } {\dot{\epsilon}_{\mathcal{Q},k} }  \frac{ {\mathcal{F}_{\mathcal{Q},k}^{\prime 2}} } { {\mathcal{F}_{\mathcal{Q},l}^{\prime 2}}  }\right|_{\text{short-range}} =
 \begin{cases}   
   \displaystyle{ \frac{3}{2}\quad \text{for}\quad  l =\phi \, \chi_{1,2}(\Tilde{\chi}_{1,2}),\,\, k= \phi \, S_{1,2}(\Tilde{S}_{1,2})  }, \\[10pt]
      \displaystyle{ \frac{5}{4}\quad \text{for}\quad  l= V\, \chi_{1,2}({S}_{1,2}),\,\, k= \phi \, S_{1,2}(\Tilde{S}_{1,2}) },\\[10pt]
      \displaystyle{ \frac{6}{5}  \quad \text{for}\quad l =\phi \,\chi_{1,2}(\Tilde{\chi}_{1,2}),\,\, k = V\, \chi_{1,2}({S}_{1,2})  }.
 \end{cases} 
\end{aligned}
\label{quadrupolecompashortrange}
\end{equation}
For $S_1 S_2\to S_1S_2\phi$, and  $\tilde S_1 \tilde S_2\to \tilde S_1\tilde  S_2\phi$, in the limit $\langle|\mathcal{M}_{\text{vertex}}|^2\rangle \gg \langle|\mathcal{M}_{\text{external leg}}|^2\rangle$ (see App.~\ref{scalarscalarquad}), the following subsitution should be done in the ratio~\eqref{quadrupolecompalongrange2}: $\mathcal{F_{Q}}\to \mathcal{F^{\prime}_{Q}}\ $, $A_{1,2}\to A^{\prime}_{1,2}$,$\lambda_{1,2}\to\lambda^{\prime}_{1,2}$, and, $C_k = 64$ for $k = \phi \, \chi_{1,2}(\Tilde{\chi}_{1,2}) $, and $C_k =160/3$ for~$k = V \, \chi_{1,2}({S}_{1,2}) $.

We note that while vertex-type emissions are suppressed by prefactors smaller by one to two orders of magnitude compared to quadrupole-like processes, their weaker dependence on the coupling constants may render them relevant in certain regions of parameter space.

\paragraph{Massive emission}

For massive emission, we evaluate numerically the energy loss rate using the energy-differential cross sections defined in App.~\ref{Massiveemission}.  We plot the corresponding ratio $R$, as defined in~\eqref{R}, in the lower panels of Fig.~\ref{energylossratequadanddipolesl}.

For both types of mediation, the processes $S_1 S_2 \to S_1 S_2 \phi$ and $\tilde S_1 \tilde S_2 \to \tilde S_1 \tilde S_2 \phi$ in the limit $(A_{j}/m_{j})^2 \ll \lambda_{j} $ (long-range mediation) or $\langle|\mathcal{M}_{\text{vertex}}|^2\rangle \gg \langle|\mathcal{M}_{\text{external leg}}|^2\rangle$ (short-range mediation), have $m_{\phi,V}$-contributions arising only through the phase-space factor in~\eqref{ELR}, leading to a weaker suppression of the energy loss rate compared to the other dSIDM scenarios. In contrast, applying the limit $(A_{j}/m_{j})^2 \gg \lambda_{j} $ (long-range mediation) and  $\langle|\mathcal{M}_{\text{vertex}}|^2\rangle \ll \langle|\mathcal{M}_{\text{external leg}}|^2\rangle$ (short-range mediation) to $S_1 S_2 \to S_1 S_2 \phi$ and $\tilde S_1 \tilde S_2 \to \tilde S_1 \tilde S_2 \phi$, and for all of the other dissipative scenarios explored, the mass of the emitted particle also enters the quadrupole emission amplitudes. Nevertheless, for long-range mediation, the resulting impact on the energy loss rate remains broadly comparable across the different dissipative processes. For short-range mediation, however, $m_{\phi,V}$-dependencies lead to a comparatively milder suppression of the energy loss rate, with small differences emerging between the various scenarios, among which 
$V$-emission is the most suppressed. We postpone a detailed discussion to the section on identical particle scattering, since for short-range mediation the difference from distinguishable scattering amounts only to an overall prefactor in the energy loss rates.

\begin{figure}
    \centering
    \includegraphics[width=0.49\linewidth]{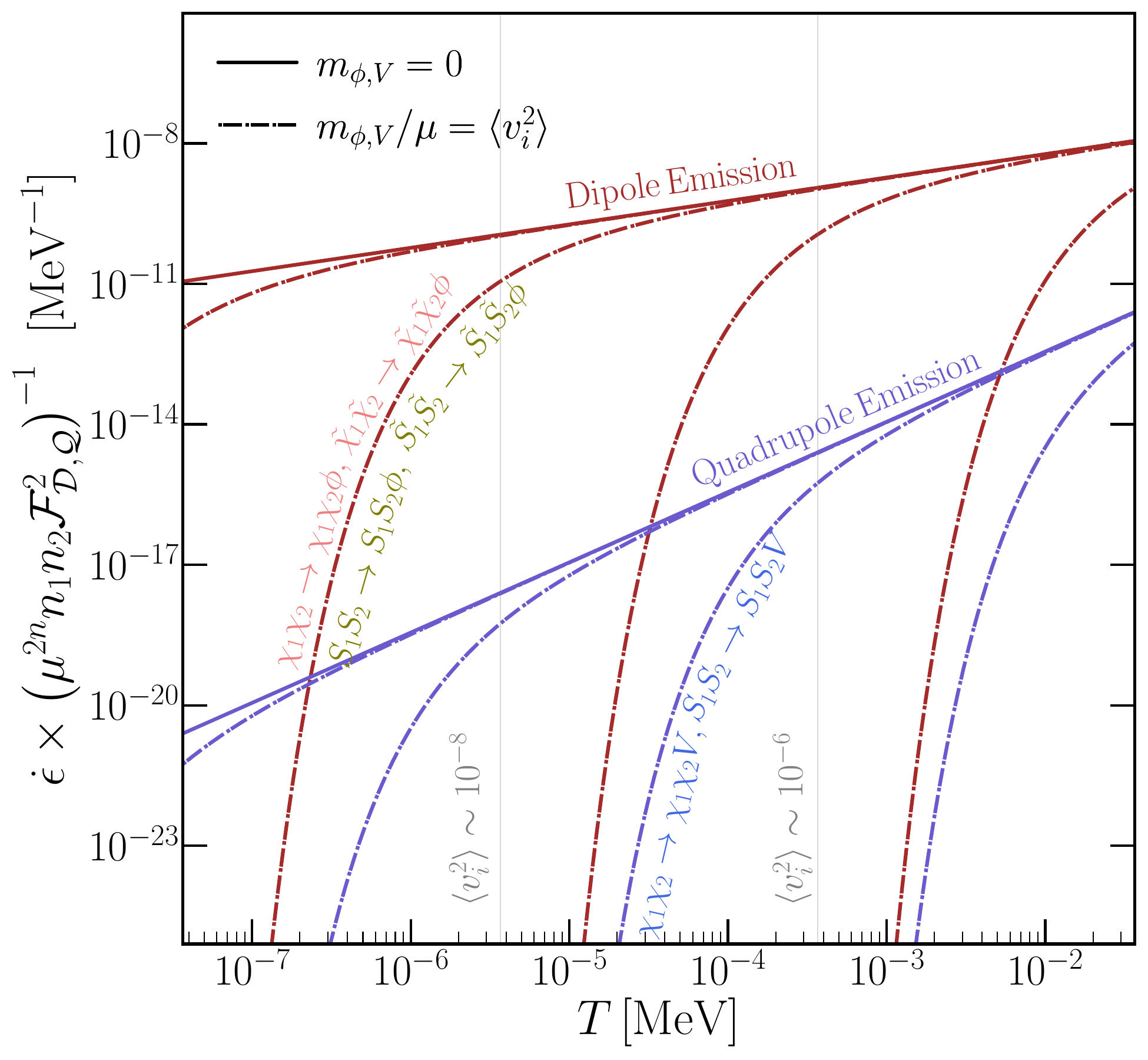}
\includegraphics[width=0.49\linewidth]{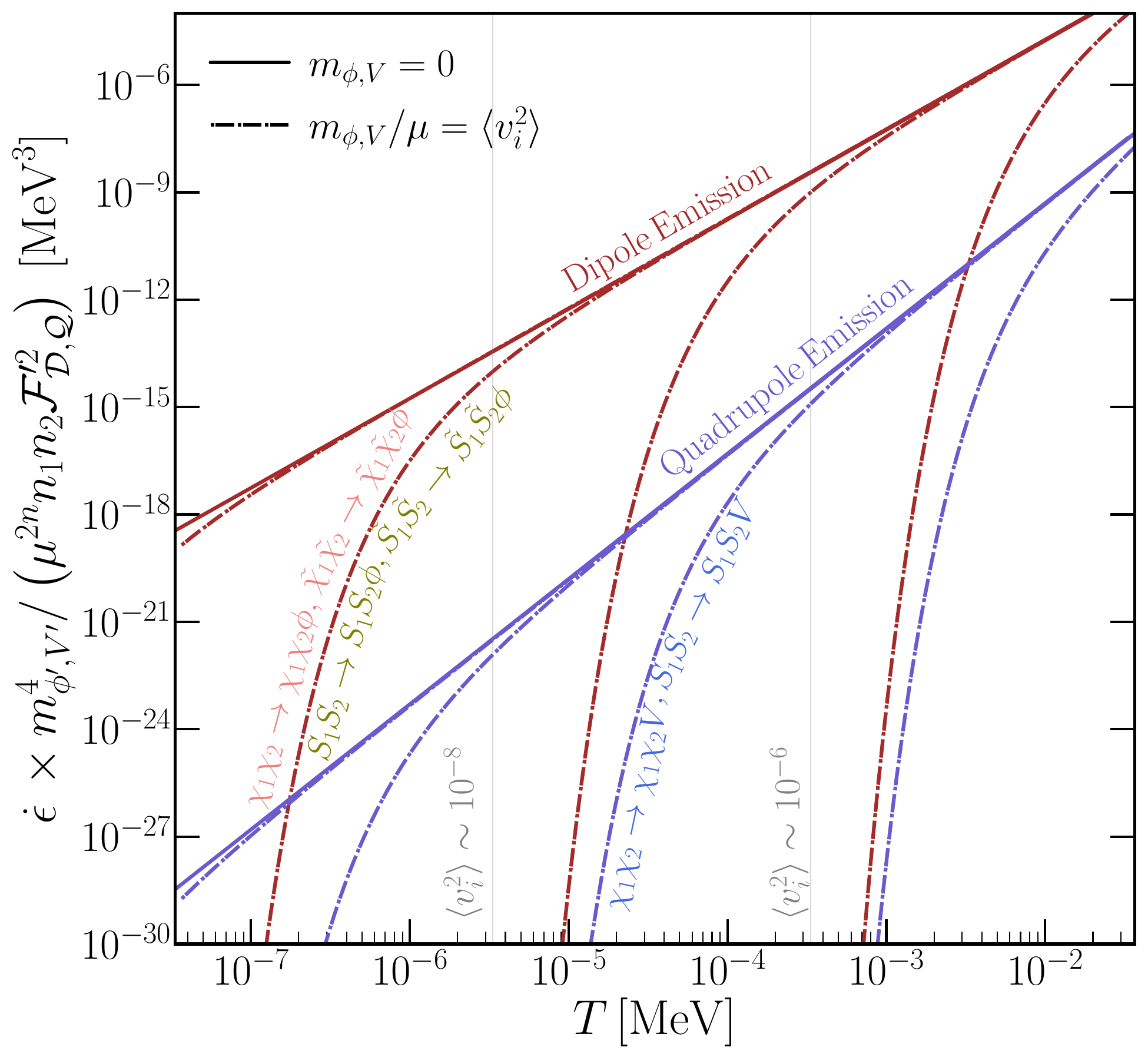}
    \caption{ Energy loss rate from long-range~\eqref{NRJlossrate} (\textit{left panel}) and short-range~\eqref{NRJ2} (\textit{right panel}) mediated dissipative processes. In both cases, the dipole channel is from $\phi$-emission, while quadrupole emission is taken from $V$-emission.
Rates are normalized by the dipole $n=1$ (quadrupole $n=2$) prefactor $\mathcal{F}^{(\prime)}_{\mathcal{D}}$ ($\mathcal{F}^{(\prime)}_{\mathcal{Q}}$), the square (fourth power) of the reduced mass $\mu$ and the number density $n_{1,2}$. For short-range mediation, an additional scaling by $m_{\phi',V'}^4$ is applied. The reduced mass is set to $\mu = 1.1~\mathrm{GeV}$. 
Massless emission is shown with solid lines; massive emission is shown with dotted lines. In the latter case, from left to right: $m_{\phi,V}/\mu=\{10^{-10},10^{-8},10^{-6},10^{-4}\}$.
The vertical grey lines represent $\mu \langle v_i^2 \rangle = 3  T$. }
    \label{finaldistinguishable}
\end{figure}

Having established the relevant distinctions between each dSIDM scenario, we now present in Fig.~\ref{finaldistinguishable} the energy loss rate from long-range-mediated processes~\eqref{NRJlossrate} (left panel) and short-range-mediated processes~\eqref{NRJ2} (right panel), plotted according to:
\begin{itemize}
    \item For dipole emission, we plot the rate corresponding to $\phi$-emission for both types of mediation, as it undergoes the strongest suppression when the emitted particle is massive, despite the small difference with $V$-emission. 
    \item For quadrupole emission, we have shown that model-dependent differences remain mild in both the long-range and short-range cases for massless emission; see~\eqref{quadrupolecompalongrange} and~\eqref{quadrupolecompashortrange}, respectively. For massive emission, the same conclusion holds for long-range mediation. In contrast, for short-range mediation, small differences start to arise for $m_{\phi,V}\sim T$, with the largest suppression observed for $V$-emission. We therefore use the latter as a representative example.
\end{itemize}

Finally, we emphasize that adding a small but finite mass to the emitted particle provides a natural way to modulate the strength of the dissipative effects across different halo sizes. Once typical CM energy---associated with the typical mean speed of DM particles---lies below the emitted particle mass scale, the dissipative effects are strongly suppressed.

\subsection{Scattering of identical particles: quadrupole dominance} \label{edotidentic}
Next, we derive the energy loss rate for dSIDM processes involving identical particle scattering, which results in quadrupole emission. We use to evaluate~\eqref{Dot(e)} the energy-differential cross sections derived in Sec.~\ref{identicalparticles}, which take into account the $t$-channel, the $u$-channel, and their interference terms.

\subsubsection{Massless emission}
If the particle emitted is massless, the emission rate can be computed analytically using the energy-differential cross sections from Sec.~\ref{identicalparticles} for $\kappa=0$.
\paragraph{Long-range mediation}
We start by considering processes of identical particle scattering mediated by a long-range force. In the case of vector emission, the processes $\chi\chi \to \chi\chi V$ and $SS \to SS V$ exhibit distinct energy-differential cross sections, leading to differences in their respective energy loss rates. In contrast, the processes $\chi\chi \to \chi\chi\phi$ and $\tilde{\chi}\tilde{\chi} \to \tilde{\chi}\tilde{\chi}\phi$ and $SS \to SS\phi$ and $\tilde{S}\tilde{S} \to \tilde{S}\tilde{S}\phi$ yield, respectively, the same energy loss rate. 

For the scenarios involving fermionic DM candidates, we obtain the following energy loss rates,

\begin{subnumcases}{\dot{\epsilon}= \frac{\left(44-3\pi^2\right) n^2 T^{3/2}}{m_{\chi}^{5/2}\pi^{7/2}} \times }
   g^6\frac{5}{144} & \textrm{for} \quad $\chi \chi \to \chi \chi V $, \label{fermionV}
   \\
   y^6\frac{1}{36} & \textrm{for} \quad $\chi \chi \to \chi \chi \phi, (\chi \leftrightarrow \tilde \chi ). $  \label{fermionphi}
\end{subnumcases}
Our results for $\chi \chi \to \chi \chi V $ is in agreement with~\cite{1812.07000}, which follows the classical treatment of~\cite{Haug1975}. 

For the processes involving scalar DM candidates, the energy loss rates read 
\begin{subnumcases}{\dot{\epsilon}= \frac{\left(3\pi^2-20\right) n^2 T^{3/2}}{m_{S}^{5/2}\pi^{7/2}} \times }
g^6\frac{5}{72} & \textrm{for} \, $S S \to S S V $, \label{scalarV}
   \\
\left( \frac{A}{m_{S}}\right)^6 \frac{1}{768} & \textrm{for} \, $ S S \to S S \phi, \,  (S\leftrightarrow \tilde S) $:  $\left( \frac{A}{m_{S}}\right)^2 \gg \lambda$,   \label{scalarphi1} \\
  \lambda^2 \left( \frac{A}{m_{S}}\right)^2  \frac{1}{192} & \textrm{for} \, $ S S \to S S \phi, \, (S\leftrightarrow \tilde S) $:  $\left( \frac{A}{m_{S}}\right)^2 \ll \lambda$. \label{scalarphi2}
\end{subnumcases}
We note that the leading contribution to the energy loss rate for $SS\to SS \phi$ and $\tilde S \tilde S\to \tilde S \tilde S \phi$ depends on the relative magnitude of the couplings $A$ and $\lambda$. In the limit where $(A/m_S)^2 \gg \lambda$, external leg emission dominates while in the opposite limit, $(A/m_S)^2 \ll \lambda$, the dominant contribution comes from the vertex diagrams.

\paragraph{Short-range mediation} We now turn to the case of short-range mediation. For fermionic DM candidates, the energy loss rates for the processes 
\(\chi\chi \to \chi\chi V\), \(\chi\chi \to \chi\chi \phi\), and 
\(\tilde{\chi}\tilde{\chi} \to \tilde{\chi}\tilde{\chi} \phi\) are 
\begin{subnumcases}{\dot{\epsilon}= \frac{n^{2}\,T^{7/2}}{\pi^{7/2}\,m_{\chi}^{1/2}} \times} 
    \displaystyle
g^{2}g'^{4} \frac{16\,}{21m_{V'}^{4}}
    & for $\chi\chi \to \chi\chi V$, \label{fermionVshort} \\
    \displaystyle
   y^{2}y'^{4} \frac{32}{35m_{\phi'}^{4}}
    & for $\chi\chi \to \chi\chi\phi,\,
    \left(\tilde{\chi} \leftrightarrow \chi \right) $.
      \label{fermionphishort}
\end{subnumcases}

For scalar DM candidates, the emission amplitudes receive contributions from the contact dissipative channels, Fig.~\ref{Contacts}, in addition to those from the external leg and vertex emission diagrams. While the latter contributions double in the emission amplitude for identical particle scattering, this enhancement does not apply to the contact diagram contributions. For the scenario \( S S \to S S V \), and for the processes 
\( S S \to S S \phi \) and \( \tilde{S} \tilde{S} \to \tilde{S} \tilde{S} \phi \) in the external leg-dominated emission regime ($\langle|\mathcal{M}_{\text{vertex}}|^2\rangle \ll \langle|\mathcal{M}_{\text{external leg}}|^2\rangle$; see App.~\ref{scalarscalarquad} for details), the energy loss rates are given by
\begin{subnumcases}{\dot{\epsilon}= \frac{n^{2}\,T^{7/2}}{\pi^{7/2}\,m_{S}^{1/2}}\times}
    \displaystyle
g^2\left(g^{\prime2}  + \frac{m_{V'}^2\lambda_{S}}{8m_{S}^2} \right)^2 \frac{64\,}{21m_{V'}^{4}}
    & for $ S S \to S S V$, \label{scalarVshort} \\
    \displaystyle
   \left( \frac{A}{m_{S}} \right)^2  \left[ \left(\frac{A'}{m_{S}}\right)^2 - \frac{m_{\phi'}^2 \lambda_{S}}{2m_{S}^2} \right]^2 \frac{4}{105m_{\phi'}^{4}}
    & for $S S \to S S \phi,\;
           (\tilde{S} \leftrightarrow S)$.
      \label{scalarphishort1}
\end{subnumcases}

In the case where vertex emission diagrams dominate the processes \( SS \to SS \phi \) and \( \tilde{S} \tilde{S} \to \tilde{S} \tilde{S} \phi \), i.e., when~$\langle|\mathcal{M}_{\text{vertex}}|^2\rangle \gg \langle|\mathcal{M}_{\text{external leg}}|^2\rangle$ (see App.~\ref{scalarscalarquad}), the energy loss rate is
\begin{equation}\label{scalarphishort2}
\dot{\epsilon}_{\text{vertex}}=\left(\frac{A'}{m_{S}}\right)^2\lambda'^2  \frac{ 8 n^2 T^{7/2}}{35\pi^{7/2} m_{S}^{1/2}  m_{\phi'}^4 }.
\end{equation}

\begin{table}[t]
    \centering
    \small
    \begin{tabular}{r|cccccc}
    \toprule
    dSIDM $(k,l)$ & $(\phi\chi, V\chi)$ & $(\phi\chi, \phi S)$ & $(V\chi, VS)$ & $(V\chi, \phi S)$ & $(VS, \phi S)$ & $(\phi\chi, VS)$ \\
    \midrule
    \textbf{$\dot{\epsilon}_k a_l^6 / \dot{\epsilon}_l a_k^6$} 
    & $0.8$ 
    & $ 31.9$ 
    & $ 0.7$ 
    & $ 39.9$ 
    & $ 53.3$ 
    & $ 0.6$ \\
    
    $\dot{\epsilon}_k \left( A/m_\chi \right)^2 \lambda^2 / \dot{\epsilon}_{\rm{vertex}} a_k^6$
    & — 
    & $ 7.9$
    & — 
    & $ 9.9$ 
    &  $ 13.3$
    & — \\
    \bottomrule
    \end{tabular}
    \caption{Ratios of the energy loss rates between the different dissipative scenarios, involving identical particle scattering mediated by a long-range interaction. Identical fermionic and scalar DM masses are assumed. The indices $k,l$ label the scenarios, with the shorthand notation listing the emitted particle first, followed by the DM candidates. Accordingly, the coupling $a_{l,k}$ corresponds to $g$, $y$, or $A/m$. {\it First row}: the limit $(A/m_{S})^2 \gg \lambda$ is applied to $SS\to S S \phi$. {\it Second row}: the limit $(A/m_{S})^2 \ll \lambda$ is applied to $SS\to S S \phi$.} 
    \label{ratiosEMlong}
\end{table}

\

We compare in Tab.~\ref{ratiosEMlong} (long-range mediation), and in Tab.~\ref{compaquadshort} (short-range mediation), the energy loss rates obtained in this section for the various dSIDM scenarios, assuming equal DM candidate masses. %
For $\phi$-emission, we only show processes involving $\chi$ and $S$, since $\tilde\chi$ and $\tilde S$ yield identical energy loss rates to their respective counterparts.

\subsubsection{Massive emission} 
In the case of a massive emitted particle, the energy loss rate~\eqref{Dot(e)} has to be computed numerically using the energy-differential cross section from Sec.~\ref{identicalparticles} for $\kappa \neq 0$. 

\begin{figure}
    \centering
    \includegraphics[width=0.49\linewidth]{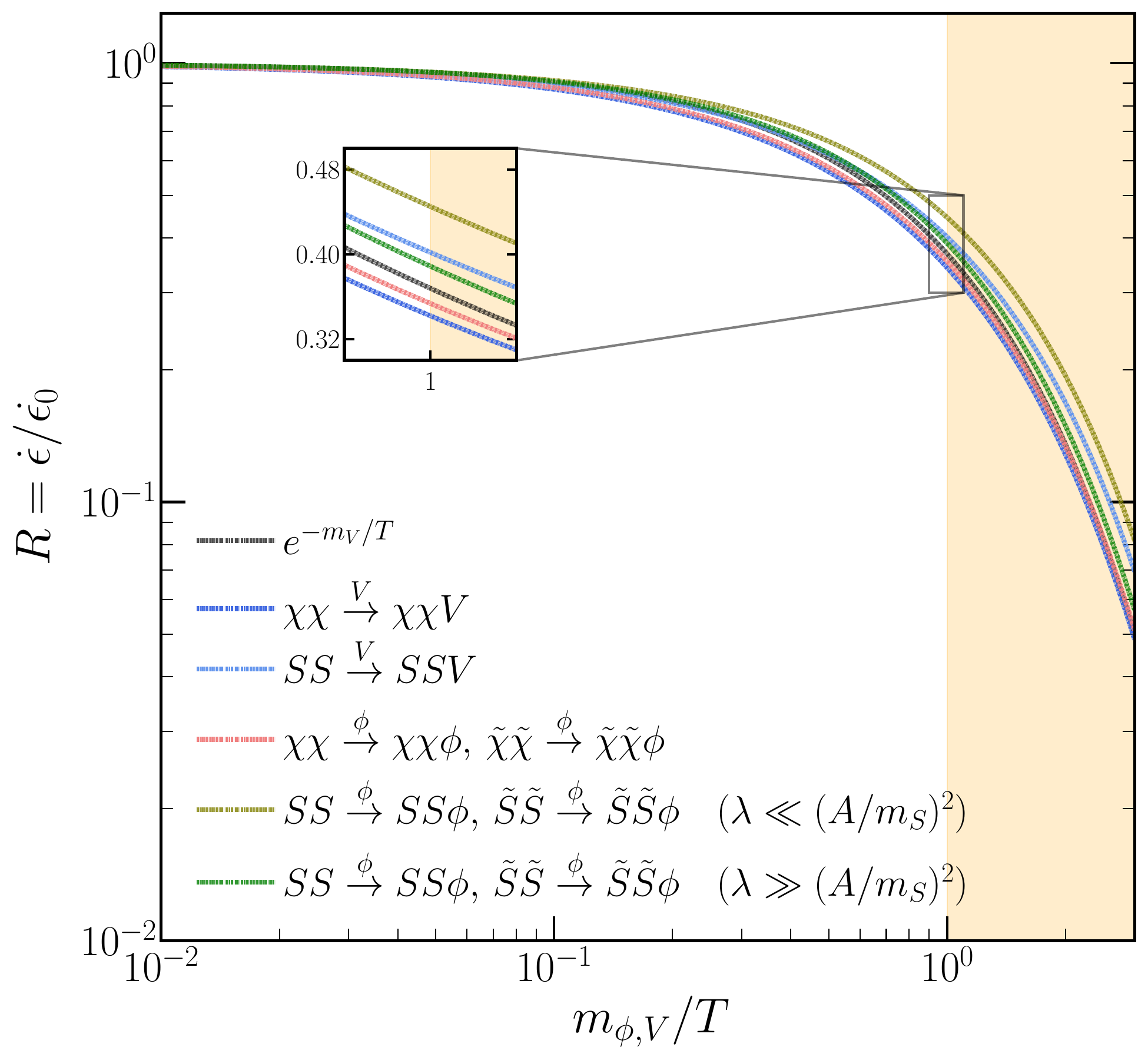}
    \includegraphics[width=0.49\linewidth]{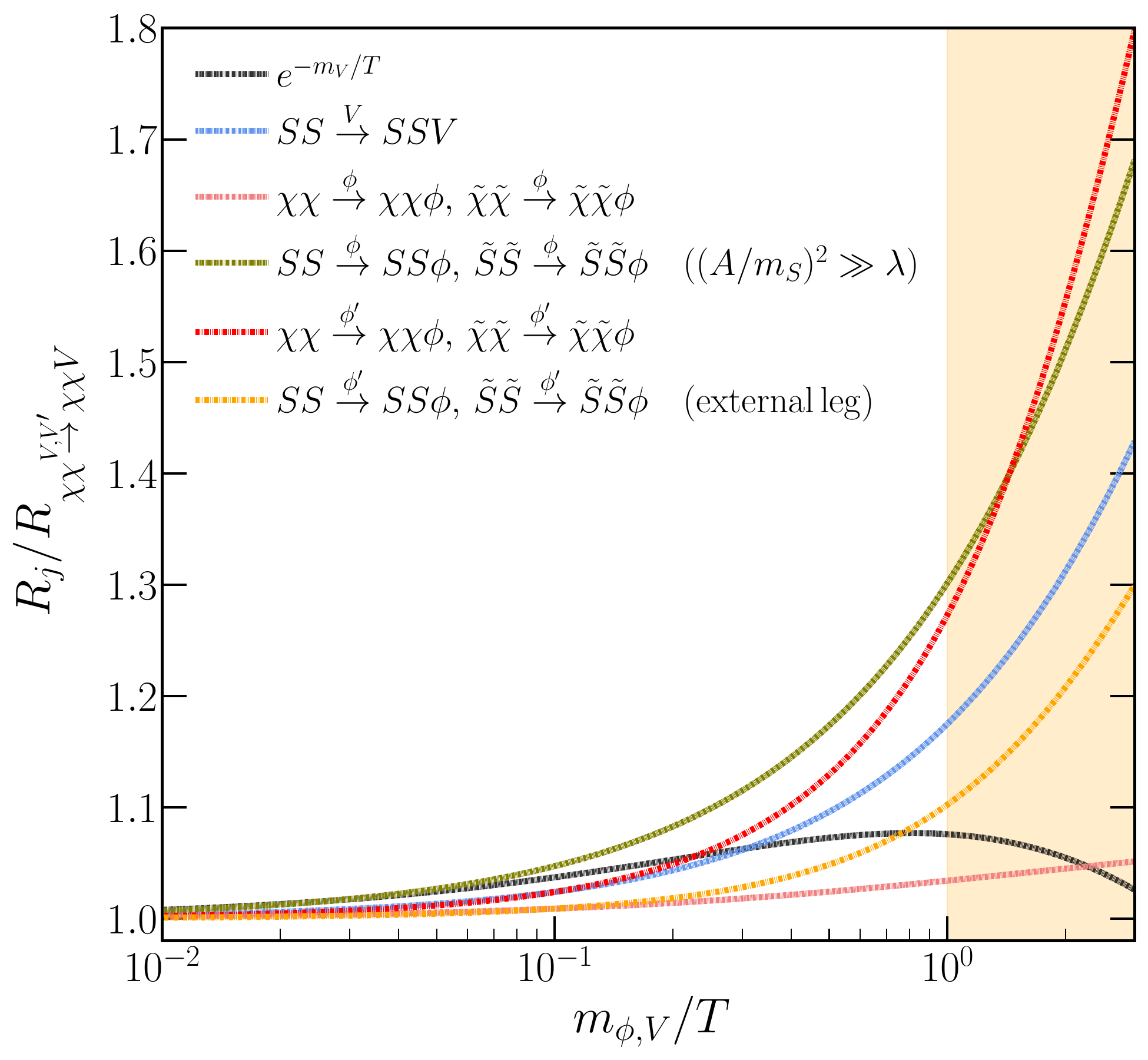}
    \caption{Effect of the mass of the emitted particle on the energy loss rate, for identical particle scattering, compared to a simple exponential suppression estimate~\cite{1812.07000} (black). {\it Left panel}: ratio $R$~\eqref{R} of the massive energy loss rate compared to the massless case, for long-range mediation. {\it Right panel}: Comparison of the ratio \( R_j \) in the different \( j \)-dSIDM scenarios to the benchmark result from the process \( \chi \chi \to \chi \chi V \), under both long-range and short-range mediation. The limits \( \left(A/m_S \right)^2 \gg \lambda  \) and \( \langle|\mathcal{M}_{\text{external leg}}|^2\rangle \gg \langle|\mathcal{M}_{\text{vertex}}|^2\rangle \) (see App.~\ref{scalarscalarquad}) are applied to the processes \( SS \to SS\phi \) and \( \tilde S \tilde S \to \tilde S \tilde S \phi \). The orange region corresponds to \( m_{\phi, V} \geq \mu \langle v_{i}^2 \rangle \). }
    \label{edotcompaidentical}
\end{figure}

We start with the case of dissipative processes with a light mediator. In the left panel of Fig.~\ref{edotcompaidentical}, we plot the ratio \( R \), as defined in~\eqref{R}, for the various dissipative scenarios. Similarly to the distinguishable particle scattering case, the impact of the emitted particle's mass is broadly comparable across the different dissipative processes. Notably, our analytical results closely follow the simple exponential suppression assumed in~\cite{1812.07000}. For scattering mediated by a short-range force, the suppressions of the rates are the same as in the case of distinguishable particle scattering (see lower left panel of Fig.~\ref{energylossratequadanddipolesl}). As a result, in contrast to long-range mediation, the processes $\chi\chi \to \chi \chi V $ and $SS\to SS V$ exhibit the same $m_V$-dependence.

Moreover, for both types of mediation, we compare in the right panel of Fig.~\ref{edotcompaidentical} the ratio \( R_{j} \) for the various dissipative scenarios $j$ with that of \( \chi \chi \to \chi \chi V \), denoted \( R_{\chi V} \), a benchmark scenario extensively studied in the literature. We focus exclusively on processes dominated by quadrupole emission. That is, we apply the limit \( (A/m_S)^2 \gg \lambda \) for long-range mediation, and \( \langle|\mathcal{M}_{\text{external leg}}|^2\rangle \gg \langle|\mathcal{M}_{\text{vertex}}|^2\rangle \) for short-range mediation (see App.~\ref{scalarscalarquad}), to the processes \( SS \to SS\phi \) and \( \tilde{S} \tilde{S} \to \tilde{S} \tilde{S} \phi \). In the case of long-range mediation, for fermionic DM candidates, the mass of the emitted particle induces comparable effects across the processes $\chi\chi \to \chi\chi V$, $\chi\chi \to \chi\chi \phi$, and $\tilde{\chi} \tilde{\chi} \to \tilde{\chi} \tilde{\chi} \phi$, with deviations of only a few percent. In contrast, for scalar candidates, the suppression associated with the emission of a massive vector or scalar boson is reduced compared to the benchmark process \( \chi \chi \to \chi \chi V \). For short-range mediation, massive $\phi$-emission is comparatively less suppressed than $V$-emission. We note that these effects are most pronounced in the regime where the rates are already exponentially suppressed.

\begin{table}[t] 
    \centering \small
    \begin{tabular}{r|cccccc}
    \toprule
   dSIDM  \( (k,l) \) & \( \left(\phi\chi, V\chi \right) \) & \( (\phi\chi, \phi S) \) & \( (V\chi   , VS  ) \) & \( ( V\chi, \phi S) \) & \( (  V S , S \phi) \) & \( (\phi\chi, VS ) \)   \\ 
    \midrule
    \textbf{\( \dot{\epsilon}_{k} a_{l}^{2}a_{l}^{\prime 4} /\dot{\epsilon}_{l} a_{k}^{2}a_{k}^{\prime 4}   \)} 
    & \(  6/5  \) & \( 24  \) & \( 1/4 \) & \(  20 \) & 80 & 3/10 \\ 
    $\dot{\epsilon}_k \left(\frac{A'}{m_{S}}\right)^2\lambda'^2 / \dot{\epsilon}_{\rm{vertex}} a_{k}^{2}a_{k}^{\prime 4}$
    & — 
    & $ 4$
    & — 
    & $10/3$ 
    &  $40/3$
    & — \\
    \bottomrule
    \end{tabular}
    \caption{ 
  Ratios of the energy loss rates between the different dissipative scenarios, involving identical particle scattering mediated by a short-range interaction. Comparable DM candidate masses are assumed. The indices $k,l$ label the scenarios, with the shorthand notation listing the emitted particle first, followed by the DM candidates. The parameter $a^2 a^{\prime 4}$ denotes the following configurations of coupling constants: $g^2 g^{\prime 4}$ for $\chi\chi \to \chi\chi V$, $y^2 y^{\prime 4}$ for $\chi\chi \to \chi\chi \phi$, $(A/m_{S})^2((A'/m_{S})^2- {m_{\phi'}^2\lambda_{S}}/2{m_{S}^2})^2$ for $SS \to SS \phi$, and $g^2(g'^2+ {m_{V'}^2\lambda_{S}}/{8m_{S}^2})^2  $ for $SS \to SS V$. {\it First row}: external-leg-dominated emission for $SS \to SS \phi$. {\it Second row}: vertex-dominated emission for $SS \to SS \phi$.}
    \label{compaquadshort}
\end{table}

\subsection{Soft emission}

To conclude the discussion of the energy loss rate, we estimate, for long-range and short-range force mediated scattering, the contribution to the energy loss rates from soft emission using the soft energy-differential cross sections derived in Sec.~\ref{soft}. We use the fiducial cut-off $\Lambda= \mu v_{i}^2/4$, such that particles emitted with an energy $\omega \lesssim \Lambda $ are considered soft. 
\subsubsection{Dipole soft emission}
For dipole emission, the energy-differential cross sections~\eqref{softdipole} describe the $x\ll1$ emission spectrum of the various dSIDM scenarios. For long-range mediation, substituting the corresponding expression into~\eqref{Dot(e)} and integrating over $\omega$ up to the cut-off $\Lambda$ yields the soft energy loss rate
\begin{equation}
    \left. \dot{\epsilon}_{\mathcal{D}}\right|_{\text{soft}}= n_{1} n_{2}f_{\phi,V}\mathcal{F}_{\mathcal D}^2 \frac{ \left(1+\log8\right) \sqrt{\mu T}}{24\sqrt{2}\pi^{7/2}},
\end{equation}
where we added the subscript $\mathcal{D}$ to label the dipole nature of the emission. Of course, up to numerical prefactors, the rate exhibits the same parametric dependencies (temperature, mass, coupling constants) as the full result.
Taking the ratio of this soft energy loss rate with the one obtained by integrating over the {\it full} energy range of dipole emission, i.e., the leading-order term in~\eqref{NRJlossrate}, denoted here as \( \dot{\epsilon}_{\mathcal{D}} \), yields 
\begin{equation} \label{pourcentage1}
\dot{\epsilon}_{\mathcal{D}}\vert_{\text{soft}} \approx 0.76 ~ \dot{\epsilon}_{\mathcal{D}}.
\end{equation}

Applying the same procedure in the case of short-range force yields the corresponding soft energy loss rate,
\begin{equation} \label{pourcentage2}
\dot{\epsilon}_{\mathcal{D}}|_{\text{soft}}= n_{1} n_{2} f_{\phi,V}\mathcal{F}_{\mathcal D}'^{2} \frac{   2\sqrt{2} (\mu T)^{5/2}}{\pi^{7/2}m_{\phi',V'}^4}.   
\end{equation}
The ratio between the soft energy loss rate and the leading-order dipole result given by the first term in~\eqref{NRJ2} is found to be
\begin{equation}
\dot{\epsilon}_{\mathcal{D}}\vert_{\text{soft}} \approx 0.93 ~ \dot{\epsilon}_{\mathcal{D}}.
\end{equation}

The ratios in~\eqref{pourcentage1} and~\eqref{pourcentage2} imply that roughly \(76\%\) (long-range) and \(93\%\) (short-range) of the total dipole energy loss rate arises from bosons whose individual energies remain at or below half of the system’s total kinetic energy.

\subsubsection{Quadrupole soft emission} 

We now turn to quadrupole emission, starting with long-range mediation. For distinguishable particle scattering, evaluating~\eqref{Dot(e)} up to the cutoff $\Lambda$ with the soft quadrupole-differential cross section~\eqref{softquadlight}, we find
\begin{equation}
\dot{\epsilon}_{\mathcal{Q}}|_{\text{soft}}= n_{1} n_{2}f_{\phi,V}\mathcal{F}_{\mathcal Q}^2 \frac{   \left[ \alpha(1+ \log64)  + \beta  \right](\mu T)^{3/2} }{15 \sqrt{2}\pi^{7/2}}.   
\end{equation}
We added the subscript \( \mathcal{Q} \) to label the quadrupole nature of the emission. The factor \( \mathcal{F}_{\mathcal{Q}} \), along with the coefficients \( \alpha \) and \( \beta \), are defined in~\eqref{factorQ}. We compare this soft energy loss rate to the full result obtained by integrating over the entire energy spectrum of quadrupole emission—i.e., the second term of~\eqref{NRJlossrate}, which we denote \( \dot{\epsilon}_{\mathcal{Q}} \), in the first row of Tab.~\ref{softtable}. For all the scenarios considered, between \(80\%\text{ to }85\%\) of the total energy loss rate is carried by bosons with energies $\omega \lesssim \mu v_{i}^2/4$.

For soft dissipative processes involving identical particles, the corresponding soft energy loss rates are trivially obtained by inserting the energy-differential cross sections~\eqref{SoftEDcsVector}--\eqref{softfermionscalar} 
into \eqref{Dot(e)} and integrating up to the cut-off \(\Lambda\).
In the last row of Tab.~\ref{softtable}, we compare these soft energy loss rates with those presented in~\eqref{fermionV}-\eqref{scalarphi1}. Relative to distinguishable-particle scattering, the soft-emission fraction is slightly larger for processes involving fermionic DM candidates and smaller for scalar DM candidates, the overall variation amounting to only a few percent. This is coming from the fact that in the $x\ll 1$ limit,  some of the interference terms between the $t$- and $u$-channel are suppressed.

We now proceed to the case of short-range mediating force. We evaluate~\eqref{Dot(e)} up to the cut-off $\Lambda$ using the soft quadrupole energy-differential cross section~\eqref{softquadheay}, and obtain 
\begin{equation} \label{softquad}
\dot{\epsilon}_{\mathcal{Q}}|_{\text{soft}}= n_{1} n_{2}f_{\phi,V}\mathcal{F}_{\mathcal{Q}}^{\prime2} \frac{ 64  \sqrt{2}   \alpha(\mu T)^{7/2}}{15 \pi^{7/2}m_{\phi',V'}^4}, 
\end{equation}
where $\alpha$ and $\mathcal{F}_{\mathcal{Q}}^{\prime2}$ are defined in~\eqref{quadshortfermion}-\eqref{quadshortscalar2}. The ratios\footnote{These ratios remain unchanged for identical particle scattering as both the soft and the full energy loss rates~\eqref{fermionVshort}--\eqref{scalarphishort1} acquire the same additional prefactors, which cancels in their quotient.} of this soft expression compared to the full expression of the energy loss rate (second term in~\eqref{NRJ2}) are derived in the second row of Tab.~\ref{softtable}. As can be inferred from the behaviour of the soft energy-differential cross section in Fig.~\ref{softfigdipole}, the soft expression \eqref{softquad} being constant, overestimates the contribution of gauge bosons carrying energy close to $ \Lambda$, yielding a ratio larger than unity. For $\chi_{1}\chi_{2}\to\chi_{1}\chi_{2}\phi$ and $\tilde \chi_{1} \tilde \chi_{2}\to \tilde \chi_{1} \tilde \chi_{2}\phi$, the same approximation suppresses the non-trivial energy dependence of the full energy-differential cross section that becomes important near $\Lambda$, leading to a lower ratio compared to the other scenarios.

\begin{table}[t] 
    \centering \small
    \begin{tabular}{r|c|c|c|c|c}
    \toprule
    \( {\left. \dot{\epsilon}_{\mathcal{Q}}\right|_{\text{soft}}}/{\dot{\epsilon}_{\mathcal{Q}}}   \) & \multicolumn{2}{c|}{\textbf{Gauge emission $V$  }}
    & \multicolumn{2}{c}{\textbf{Scalar emission $\phi$ }} \\
     \midrule
    dSIDM
      & $\chi_{1}\chi_{2}\!\to\!\chi_{1}\chi_{2}V$
      & $S_{1}S_{2}\!\to\!S_{1}S_{2}V$
      & \makecell{$\chi_{1}\chi_{2}\!\to\!\chi_{1}\chi_{2}\phi$,\\
                   $\tilde\chi_{1}\tilde\chi_{2}\!\to\!\tilde\chi_{1}\tilde\chi_{2}\phi$}
      & \makecell{$S_{1}S_{2}\!\to\!S_{1}S_{2}\phi$,\\
                   $\tilde S_{1}\tilde S_{2}\!\to\!\tilde S_{1}\tilde S_{2}\phi$} \\[2pt]
    \midrule
   \midrule
  
    \multicolumn{5}{l}{\textit{Distinguishable particle scattering}} \\
     \midrule
     \multicolumn{5}{l}{Mediation:} \\
     Long-range  
    & \(   0.85   \) & \(  0.85  \) & \( 0.83 \) &  $  0.80$ \\ 
     Short-range  
    & \(  1.05   \) & \(  1.05  \) & \(  0.58  \) &  $  0.87  $ \\ 
\midrule
\multicolumn{5}{l}{\textit{Identical particle scattering}} \\
 \midrule
     \multicolumn{5}{l}{Mediation:} \\    
 Long-range    
    & \(  0.89  \) & \(  0.78   \) & \(    0.88 \) &  $  0.71 $ \\

    \bottomrule
    \end{tabular}
    \caption{Comparison of the soft energy loss rates obtained by integrating the expression of the soft energy-differential cross sections from Sec.~\ref{soft} up to $\Lambda= \mu v_{i}^2/4$ with the full expressions presented in Sec.~\ref{edotdisting} (distinguishable particles) and Sec.~\ref{edotidentic} (identical particles), integrated to their kinematic endpoints.}
    \label{softtable}
\end{table}

\section{Conclusions}

In this work, we investigate a comprehensive set of minimal dissipative scenarios, mediated either by long- or short-range forces. Two distinct DM species are included in each scenario, allowing the study of both distinguishable and identical scatterers. The discussion is carried out in the perturbative Born limit, where dissipative processes can be systematically described by tree-level diagrams. Within this framework, the diversity of DM candidates considered gives rise to a rich set of dissipative channels, illustrated in Figs.~\ref{Diagrams}.

Assuming non-relativistic DM candidates, we introduce a systematic power-counting scheme that organises the contributions of these different dissipative channels to the total emission amplitudes by their dependence on either their initial relative velocity $v_{i}$ (long-range mediation), or the small parameter $r= {\mu v_{i}}/{m_{\phi',V'}}$ (short-range mediation). Following these prescriptions, we obtain compact analytic expressions for the emission amplitudes to leading and next-to-leading order in those parameters.  

For distinguishable particle scattering, the emission is governed by the leading order in $v_{i}$ or $r$ corresponding to dipole radiation.
At this order, the external leg dissipative channels are velocity enhanced, and their contributions to the emission amplitudes exhibit a model-independent analytic structure in the case of massless emission---\eqref{generaldipole} for long-range mediation and~\eqref{dipoleshortrangeamplitude} for short-range mediation---times a universal dipole prefactor. 

If the two DM species are identical, or share the same coupling-to-mass ratio, the dipole pieces cancel and the next-to-leading-order in $v_{i}$ or $r$ of the emission amplitudes is the dominant contribution. The latter are composed of quadrupole emission and other types of radiation processes coming from the vertex emission diagram (Fig.~\ref{vertex}) for $S_{1}S_{2}\to S_{1} S_{2}\phi$ and $\tilde S_{1} \tilde S_{2}\to \tilde S_{1} \tilde S_{2}\phi$. For the former, corrections that reflect the intrinsic properties of the DM candidates and of the emitted quanta arise, making the emission amplitude scenario-dependent (see Sec.~\ref{quadrupoleemission}) apart from a common quadrupole pre-factor. However, the kinematic structures coincide pairwise: those of $S_{1}S_{2}\to S_{1}S_{2} \phi$, and $\tilde S_{1} \tilde S_{2}\to \tilde S_{1} \tilde S_{2} \phi$ are identical, as are the structures for $\chi_{1}\chi_{2}\to \chi_{1}\chi_{2} \phi$ and $\tilde \chi_{1} \tilde \chi_{2}\to \tilde \chi_{1} \tilde \chi_{2} \phi$ and likewise for $S_{1}S_{2}\to S_{1}S_{2} V$, and $ \chi_{1}\chi_{2}\to \chi_{1}\chi_{2} V$. 
Moreover, at either order in the expansion in small parameters, we find that the combined effects of the Ward identity, obeyed by the longitudinal mode in $V$-emission, together with the NR reduction of the emission amplitudes play a crucial role when the emitted particle carries a light mass. They give rise to kinematic mass-dependent corrections to emission amplitudes that are common to $\phi$-emission and the longitudinal mode of $V$-emission.  Once the emission amplitudes are squared and polarization-summed, the apparent universality is lost. 

We then explore in detail the soft regime, where the emitted particle energy is much smaller than the available center-of-mass energy. Specifically, for each dissipative scenario, we investigate whether the correct low-energy emission amplitude can be reconstructed from the elastic scattering matrix element multiplied by an emission factor. We find that the factorised dipole-level amplitudes exactly match the full results. In contrast, quadrupole emission remains sensitive to the full dissipative kinematics and requires additional corrections within the factorization framework to recover the accurate low-energy behavior.
Building on these amplitude-level results, we summarize the energy-differential cross sections in Tab.~\ref{tab:energydifferentialcross-sectionfinal}. Equipped with the latter, we compute in a final section the energy loss rates assuming that the DM candidates form an NR Maxwell–Boltzmann gas. An overview of the corresponding expressions is provided in Tab.~\ref{tab:energylossratefinal}. 

Parametrically, for long-range mediation, dipole emission constitutes the most efficient dissipative process and scales with the temperature as $T^{1/2}$, whereas quadrupole and other next‑to‑leading‑order contributions scale as $T^{3/2}$. In contrast, when the typical momentum transfer in the scattering stays below the mediator mass $m_{\phi',V'}$, entering the regime of short-range mediation, the corresponding energy loss rates are suppressed by a factor \({(\mu T)^2 }/{m_{\phi',V'}^4}\) relative to their long-range counterparts for each type of emission (dipole or quadrupole).

Moreover, beyond these universal scalings, the numerical prefactors of the energy loss rate differ from one process to another. At dipole order, $\phi$-emission is suppressed by a factor 2 compared to $V$-emission. At quadrupole order, for both type of mediations, the processes $S_{1}S_{2}\to S_{1}S_{2} \phi$, $\tilde S_{1} \tilde S_{2}\to \tilde S_{1} \tilde S_{2} \phi$ carry the smallest of these coefficients, and therefore furnish the weakest cooling rate when the various candidates DM–emitted particles couplings are of comparable magnitude (see by e.g., Tabs.~\ref{ratiosEMlong}--\ref{compaquadshort}). 

Finally, although the emitted particle’s mass enters the energy loss rates in a model-specific way, the resulting differences in their suppression between scenarios remain mild. These differences become significant only in the regime $m_{\phi,V} \gg T$, where phase-space constraints already induce an exponential suppression of the rate.

\paragraph{Acknowledgments.} Funded by the European Union (ERC, NLO-DM, 101044443). This work was also supported by the Research Network Quantum Aspects of Spacetime (TURIS). We acknowledge the financial support by the Vienna Doctoral School in Physics (VDSP).

\begin{table}[htp]
  \centering\small
  \setlength{\tabcolsep}{4pt}
  \begin{tabular}{l|c|c|c|c}
    \toprule
   $\omega \left({d\sigma}/{d\omega}\right)$ & \multicolumn{2}{c|}{\textbf{Gauge emission $V$ $\left(f_{\phi,V}=2\right)$ }}
    & \multicolumn{2}{c}{\textbf{Scalar emission $\phi$ $\left(f_{\phi,V}=1\right)$}} \\ 
    \midrule
    dSIDM
      & $\chi_{1}\chi_{2}\!\to\!\chi_{1}\chi_{2}V$
      & $S_{1}S_{2}\!\to\!S_{1}S_{2}V$
      & \makecell{$\chi_{1}\chi_{2}\!\to\!\chi_{1}\chi_{2}\phi$,\\
                   $\tilde\chi_{1}\tilde\chi_{2}\!\to\!\tilde\chi_{1}\tilde\chi_{2}\phi$}
      & \makecell{$S_{1}S_{2}\!\to\!S_{1}S_{2}\phi$,\\
                   $\tilde S_{1}\tilde S_{2}\!\to\!\tilde S_{1}\tilde S_{2}\phi$} \\[2pt]
    \midrule
    \midrule
    \midrule
    \multicolumn{5}{l}{\textbf{Dipole radiation}}\\
    \midrule
    \midrule
    \multicolumn{5}{l}{Full spectrum (massive emission)} \\
    \midrule
    \multicolumn{5}{l}{Mediation} \\
Long-range & \multicolumn{4}{c}{\eqref{dipolelongrange}} \\
     Short-range
      & \multicolumn{4}{c}{\eqref{dipoleshortrange}}  \\
      \midrule
      \multicolumn{5}{l}{Soft spectrum}\\
        \midrule
        \multicolumn{5}{l}{Mediation}\\
     Long-range
      & \multicolumn{4}{c}{\eqref{softdipole:long}}  \\
     Short-range
      & \multicolumn{4}{c}{\eqref{softdipole:short}}  \\
    \midrule \midrule \midrule
    \multicolumn{5}{l}{\textbf{Quadrupole radiation}}\\
    \midrule
    \midrule
    \multicolumn{5}{l}{\textit{Distinguishable particle scattering $\{\alpha,\beta\}$ }}\\
    \midrule
    \multicolumn{5}{l}{Full spectrum (massive emission) }\\
    \midrule
    \multicolumn{5}{l}{Mediation}\\
Long-range
    & \multicolumn{2}{c|}{\eqref{QuadmassivelongV}} 
    & \eqref{quadmassivefermionphi} 
    & \eqref{quadmassivescalarphi} \\
Short-range
    & \multicolumn{2}{c|}{\eqref{csshortmassiveQED}}
    & \eqref{csshortmassiveyuka} 
    & \eqref{csshortmassiveyukascalar} \\

      \midrule
      \multicolumn{5}{l}{Soft spectrum}\\
      \midrule
      \multicolumn{5}{l}{Mediation}\\
     Long-range:~\eqref{softquadlight}
      &  \multicolumn{2}{c|}{$\{3/4,13/4\}$ } 
      & $\{1,6\}$ & $\{1,11\}$ \\
     Short-range:~\eqref{softquadheay} 
      &  \multicolumn{2}{c|}{$\{3/4,1/4\}$} & ${\{1,7\}}$ &  $\{1,2\}$ \\
    \midrule
    \midrule
    \multicolumn{5}{l}{\textit{Identical particle scattering}}\\
    \midrule
    \multicolumn{5}{l}{\text{Full spectrum (massive emission) } }\\
    \midrule
    \multicolumn{5}{l}{Mediation}\\
    Long-range
      & \eqref{EMQED} & \eqref{EMQEDscalar}
      & \eqref{EMchiscalar} & \eqref{EMscalarscalar} \\
   Short-range
      & Sec.~\ref{identicalparticles} & \eqref{quadheavyVS}
      & Sec.~\ref{identicalparticles} & \eqref{quadheavyScalar} \\
      \midrule
    \multicolumn{5}{l}{\text{Soft spectrum}}\\
    \midrule
    Long-range
      & \eqref{SoftEDcsVector:chi} & \eqref{SoftEDcsVector:S}
      & \eqref{softfermionscalar} & \eqref{SoftEDcsScalar} \\
      Short-range
      & \multicolumn{4}{c}{Sec.~\ref{softenergydifferentialcross-sectionquad}} \\
    \midrule \midrule \midrule
    \multicolumn{5}{l}{\textbf{Additional radiation processes}} \\
    \midrule
    \midrule
          \multicolumn{5}{l}{Mediation}\\
      Long-range
      & -- & Contact:~App.~\ref{appendixContact}
      & VIB:~\eqref{VIBenergydiff},~\eqref{eq:VIBxs:fermion} & \makecell{ Vertex:~\eqref{crossvertexlong},~\eqref{energydiffvertexlong}\\ Contact:~App.~\ref{appendixContact} \\ VIB:~\eqref{VIBenergydiff},~\eqref{eq:VIBxs:scalar} \\[2pt]}  \\
    Short-range
      & -- & --
      & VIB:~\eqref{VIBheabydcs} & \makecell{ Vertex:~\eqref{vertexscalarshort},~\eqref{energydiffvertexshort}\\ VIB:~\eqref{VIBheabydcs} } \\
    
    \bottomrule
  \end{tabular}
  
  \caption{Summary of the Born energy-differential cross sections derived in this work. ``Full spectrum'': expressions that include both hard and soft emissions. We present the massive-emission case (the massless limit follows by setting $\kappa=0$ or $C_{\phi,V}=0$). "Soft spectrum": results obtained by multiplying the elastic cross section by a soft factor (valid for $x\ll1$).
"Additional radiation processes": contributions from the dissipative channels Figs.~\ref{vertex}-\ref{VIB}.
}
  \label{tab:energydifferentialcross-sectionfinal}
\end{table}

\begin{table}[htp]
  \centering\small
  \setlength{\tabcolsep}{1.5pt}
  \begin{tabular}{l|c|c|c|c}
    \toprule
   $\dot{\epsilon}$ & \multicolumn{2}{c|}{\textbf{Gauge emission $V$ $\left(f_{\phi,V}=2\right)$ }}
    & \multicolumn{2}{c}{\textbf{Scalar emission $\phi$ $\left(f_{\phi,V}=1\right)$}} \\ 
    \midrule
    dSIDM
      & $\chi_{1}\chi_{2}\!\to\!\chi_{1}\chi_{2}V$
      & $S_{1}S_{2}\!\to\!S_{1}S_{2}V$
      & \makecell{$\chi_{1}\chi_{2}\!\to\!\chi_{1}\chi_{2}\phi,$\\
                 $\tilde{\chi}_{1}\tilde{\chi}_{2}\!\to\!\tilde{\chi}_{1}\tilde{\chi}_{2}\phi$}
      & \makecell{$S_{1}S_{2}\!\to\!S_{1}S_{2}\phi$,\\ $\tilde{S}_{1}\tilde{S}_{2}\!\to\!\tilde{S}_{1}\tilde{S}_{2}\phi$}
       \\[2pt]
    \midrule
    \midrule
    \multicolumn{5}{l}{\textbf{Dipole radiation}}\\
    \midrule
\multicolumn{5}{l}{Mediation}\\
Long-range
    & \multicolumn{4}{c}{LO \eqref{NRJlossrate}} \\
Short-range
    & \multicolumn{4}{c}{LO \eqref{NRJ2}} \\
    \midrule \midrule
    \multicolumn{5}{l}{\textbf{Quadrupole radiation}}\\[2pt]
    \midrule
    \multicolumn{5}{l}{\textit{Distinguishable particle scattering $\{\alpha,\beta\}$ }}\\[2pt]
\multicolumn{5}{l}{Mediation}\\
Long-range: NLO \eqref{NRJlossrate}
    & \multicolumn{2}{c|}{$\{3/4,\,{13}/4\}$} 
    & $\{1,\,6\}$ 
    & $\{1,\,11\}$ \\
Short-range: NLO \eqref{NRJ2}
    & \multicolumn{2}{c|}{$\{3/4,\,1/4\}$} 
    & $\{1,\,7\}$ 
    & $\{1,\,2\}$ \\[2pt]
    \midrule
    \multicolumn{5}{l}{\textit{Identical particle scattering}}\\
    \multicolumn{5}{l}{Mediation}\\
    Long-range
      & \eqref{fermionV} & \eqref{scalarV}
      & \eqref{fermionphi} & \eqref{scalarphi1} \\
    Short-range
      & \eqref{fermionVshort} & \eqref{scalarVshort}
      & \eqref{fermionphishort} & \eqref{scalarphishort1} \\
      \midrule \midrule
      \multicolumn{5}{l}{\textbf{ Additional radiation processes}}\\[2pt]
      \midrule
      \multicolumn{5}{l}{Mediation}\\
      Long-range
      & -- & Contact: App.~\ref{appendixContact}
      & VIB:~\eqref{VIBedotlongdistin} & \makecell{ Vertex:~\eqref{longvecterNRJ},~\eqref{scalarphi2}\\ Contact: App.~\ref{appendixContact} \\ VIB:~\eqref{VIBedotlongdistin} \\[2pt]}  \\
    Short-range
      & -- & --
      & VIB:~\eqref{eTri} & \makecell{ Vertex:~\eqref{shortvecterNRJ}, \eqref{scalarphishort2}\\ VIB: \eqref{eTri} } \\
    \bottomrule
  \end{tabular}
  \caption{Summary of the energy loss rates derived in this work, valid in the Born (perturbative) regime. “LO” denotes the first non-vanishing term in the temperature expansion of the energy loss rate, while “NLO” denotes the next-to-leading contribution.
The rows labelled “Additional radiation processes’’ collect the contributions from the dissipative channels shown in Figs.~\ref{vertex}-\ref{VIB}. 
For these processes, results for distinguishable and identical particle scattering are reported on the same line.}
  \label{tab:energylossratefinal}
\end{table}

\newpage

\newpage
\appendix

\section{Matrix elements}

\subsection{Center of Mass frame} \label{Powercounting}
Our derivations are performed in the center-of-mass (CM) frame of the initial DM particle pair. In this frame, the momenta of the five particles are defined as:
\begin{equation} \label{CMpj}
\begin{aligned}
    \vec{p}_{1} &= \vec{P}\frac{\mu}{m_{2}} + \vec{p}_{i}, \quad 
    \vec{p}_{2} = \vec{P}\frac{\mu}{m_{1}} - \vec{p}_{i}, \\
    \vec{p}_{3} &= (\vec{P}-\vec{q})\frac{\mu}{m_{2}} + \vec{p}_{f}, \quad
    \vec{p}_{4} = (\vec{P}-\vec{q})\frac{\mu}{m_{1}} - \vec{p}_{f}, \\   
\end{aligned}
\end{equation}
where \( \vec{P} \) is the overall CM momentum and the relative initial (final) momenta \( \vec{p}_i \), \(( \vec{p}_f) \) of the DM pair are defined in Sec.~\ref{dipolevsquadrupole}.

In the non-relativistic limit, the initial relative three-momentum is of order  $\mathcal{\vec{p}}_{i}/\mu = \mathcal{O}(v_{i})$.  Conservation of energy in the CM then implies, that the energy of the emitted particle is one order higher in velocity:
\begin{equation} \label{energyconservation}
    \omega = E_{1}+E_{2}-E_{3}-E_{4} \overset{\rm CM}{\simeq} \frac{\mu}{2}\left(|\Vec{v}_{i}|^2 - |\Vec{v}_{f}|^2\right) + \mathcal{O}(v_{i}^{4}),
\end{equation}
where the label $\text{CM}$ indicates that the $v_{i}$-expansion is carried out in the CM frame. As a result,~\eqref{energyconservation} allows us to infer the velocity scalings $\omega/\mu, |\Vec{q}|/\mu = \mathcal{O}(v_{i}^2)$.

\paragraph{Fermionic candidates} \label{4spinors}
For fermionic DM candidates, the \( v_i \)-expansion of the emission amplitudes includes the expansion of spinor wave functions $u(p,s)$ and $v(p,s)$.%
\footnote{We note that, when considering the expansion of the spinors alone, the condition \(\vec{p}_{i}/m_{\phi',V'} \sim \mathcal{O}(r)\) simply reduces to \(\vec{p}_{i} \sim \mathcal{O}(v_{i})\), since the spinors themselves do not depend on \(m_{\phi',V'}\); hence, expanding in \(r\) is equivalent to expanding in \(v_{i}\) for these terms.
}
For example, in the chiral basis,
the four-component particle spinor carrying momentum \( p \), mass \( m \), and spin \( s \) is given by
\begin{equation} \label{spinorchiral}
u(p,s)=\begin{pmatrix}
          \sqrt{p\cdot \sigma} \xi^{s} \\
            \sqrt{p\cdot \overline{\sigma}} \xi^{s}
\end{pmatrix}
\simeq
\begin{pmatrix}
          \sqrt{m}\left(\mathds{1}- \frac{\vec{p}\cdot\boldsymbol{\sigma}}{2 m} + \mathcal{O}(v_{i}^2) \right )\xi^{s} \\
           \sqrt{m}\left(\mathds{1}+ \frac{\vec{p} \cdot \boldsymbol{\sigma}}{2 m} + \mathcal{O}(v_{i}^2) \right )\xi^{s}
\end{pmatrix}.
\end{equation}
Here, \( \xi^s \) are fixed-axis two-component spinors, normalized as \( \xi^{\dagger s'} \xi^s = \delta^{ss'} \) and satisfying the completeness relation \( \sum_s \xi^s \xi^{\dagger s} = \mathds{1} \). We adopt the notation \( \overline{\sigma}^\mu = (\mathds{1}, -\boldsymbol{\sigma}) \) and \( \sigma^\mu = (\mathds{1}, \boldsymbol{\sigma}) \) for the Pauli matrices.

\subsection{Angular average}
The angular average over the direction of the emitted particle \( \hat{q} = \vec q/|\vec q | \) is defined as
\begin{equation}
    \langle A \rangle_{\hat{q}} = \frac{1}{4\pi} \int d\Omega_{\hat{q}} \, A,
\end{equation}
where $d\Omega_{\hat{q}}$ is the solid-angle element centered around~$\hat{q}$. Denoting by $\hat n_j$ any unit vector, we record the following useful identities used in this work:
\begin{equation}
\begin{aligned}
&\langle \hat{q} \cdot \hat{n} \rangle_{\hat{q}} = 0, \, 
\langle (\hat{q} \cdot \hat{n})^2 \rangle_{\hat{q}} = \frac{1}{3}, \, \langle (\hat{n}_1 \cdot \hat{q}) (\hat{n}_2 \cdot \hat{q}) \rangle_{\hat{q}} = \frac{1}{3} \hat{n}_{1}\cdot\hat{n}_{2},\,
\langle (\hat{q} \cdot \hat{n})^4 \rangle_{\hat{q}} = \frac{1}{5}, \\[5pt]
&\langle (\hat{n}_1 \cdot \hat{q})^2 (\hat{n}_2 \cdot \hat{q})^2 \rangle_{\hat{q}} = \frac{1}{15} \left[ 2(\hat{n}_1 \cdot \hat{n}_2)^2 + 1 \right], \, \langle (\hat{n}_{1} \cdot \hat{q})^3 (\hat{n}_{2} \cdot \hat{q}) \rangle_{\hat{q}} = \frac{1}{5} (\hat{n}_{1} \cdot \hat{n}_{2}).
\end{aligned}
\end{equation}

\subsection{Feynman amplitudes} \label{feynmanamplitudes}
In this appendix, we explicitly present the form of the amplitude of each dissipative channel illustrated in Fig.~\ref{Diagrams}, except for the VIB diagrams treated in App.~\ref{VinternalB}.
Unless otherwise specified, it is implicitly assumed that a sum is taken over the amplitudes of all the contributing diagrams in order to obtain the emission amplitude.

To compute the Feynman amplitudes and determine the relative signs of interfering diagrams for fermionic DM, we follow the prescription of~\cite{Denner:1992vza}, which applies to both Dirac and Majorana candidates. 
In the case of vector emission, the amplitudes from the external leg emission diagrams, Fig.~\ref{inifini}, are given by (suppressing spin and polarization indices)
\begin{equation} \label{ppt}
\begin{aligned}
i\mathcal{M}_{\chi_{1},\text{final} } &=  \overline{u}(p_3)\Gamma_{1}^{\mu} F(p_{3}+q)\Gamma^{\nu}_{1} u(p_1) \Delta_{\nu\rho}(p_{2}-p_{4}) \overline{u}(p_4)\Gamma^{\rho}_{2} u(p_2) \varepsilon^{*}_{\mu}(q), \\ 
i\mathcal{M}_{\chi_{2},\text{final} } &=  \overline{u}(p_3)\Gamma^{\nu}_{1} u(p_1) \Delta_{\nu\rho}(p_{2}-p_{4}-q) \overline{u}(p_4) \Gamma^{\mu}_{2} F(p_{4}+q) \Gamma^{\rho}_{2} u(p_2) \varepsilon^{*}_{\mu}(q), \\
i\mathcal{M}_{\chi_{1},\text{initial} } &=  \overline{u}(p_3)\Gamma^{\nu}_{1} F\left(p_{1}-q\right) \Gamma^{\mu}_{1} u(p_1) \Delta_{\nu \rho}(p_{2}-p_{4}) \overline{u}(p_4)\Gamma^{\rho}_{2} u(p_2) \varepsilon^{*}_{\mu}(q),  \\
i\mathcal{M}_{\chi_{2},\text{initial} } &=  \overline{u}(p_3)\Gamma^{\nu}_{1} u(p_1) \Delta_{\nu \rho}(p_{2}-p_{4}-q) \overline{u}(p_4) \Gamma^{\rho}_{2} F(p_{2}-q)\Gamma^{\mu}_{2} u(p_2) \varepsilon^{*}_{\mu}(q),
\end{aligned}
\end{equation}
where the subscripts "$\chi_{j},\text{initial/final}$" label the external leg associated with the emission process. Here we define the interaction vertices as $\Gamma^{\mu}_{j} = -i g_{j} \gamma^{\mu}$, use the full propagator for a massive vector boson, given by $\Delta_{\nu\rho}(p) = {-i( g_{\nu\rho} - {p_\nu p_\rho}/{m_V^2} )}/({p^2 - m_V^2})$, as well as the fermionic propagator $F(p)={i(\slashed{p}+m_{\chi})}/{({p^2-m_{\chi}^2})}$, and denote by $\varepsilon^{\mu*}$ the polarization vector. 
The amplitudes~\eqref{ppt} can be adopted to describe the emission of a scalar from both  Majonara and Dirac DM candidates: one drops the Lorentz indices, removes the metric tensors and polarization vectors, and replaces the interaction vertices by $\Gamma_{j}=-iy_{j}$.

For scalar DM candidates, three diagram contributions should be taken into account: external leg, vertex and contact emission diagrams, Figs.~\ref{inifini}--\ref{Contacts}. First, the amplitudes of the external leg emission diagrams are obtained from~\eqref{ppt} by omitting the spinors, replacing the fermionic propagator with the scalar one $F(p)\to S(p)={i}({p^2-m^2})^{-1}$, and modifying the interaction vertices. The latter become, for vector emission, $\Gamma^{\mu}_{j} = -i g_{j} (p + p')^{\mu}$, where $p$ ($p'$) denote the incoming (outgoing) momenta of the DM candidate at the vertex; and for scalar emission, read $\Gamma_{j} = -i A_{j}$.
Second, the amplitudes from the vertex diagrams are given by
\begin{equation} \label{ptt2}
\begin{aligned}
    i\mathcal{M}_{\tilde{S}_{1}} &= \Gamma^{\prime\mu \rho }_{1} \Delta_{\rho \nu} \left(p_{2}-p_{4}\right) \Gamma^{\nu}_{2} \varepsilon^{*}_{\mu}(q), \\
i\mathcal{M}_{\tilde{S}_{2}} &= \Gamma_{1}^{\nu} \Delta_{\nu\rho}\left(p_{2}-p_{4}-q\right)\Gamma^{\prime \rho\mu}_{2} \varepsilon^{*}_{\mu}(q),
\end{aligned}
\end{equation}
with the subscripts "${\Tilde{S}_{j}}$" labeling from which particle vertex the emission occurs. For vector emission, the interaction vertices $\Gamma^{\prime\mu \rho }_{j} = 2i g_{j}^2 g_{\mu\rho}$ must be additionally introduced. For scalar emission, the Lorentz indices and the polarization vector are absent, and the additional interaction vertices are given by $\Gamma^{\prime}_{j} = -i \lambda_{j}$.
Finally, the dissipative channels corresponding to an emission from contact diagrams have the following amplitude
\begin{equation} 
\begin{aligned}
 i\mathcal{M}_{c}  =
 \begin{cases}
     \sum_{k=1,4} -\lambda_{S} g_{k} (2p_{k} + \eta_{k} q)^{\mu}\varepsilon_{\mu}^{*}(q) S(p_{k} +\eta_{k} q) &\text{for vector emission}, \\[10pt]
    \sum_{k=1,4} -\lambda_{S} A_{k}~ S(p_{k} +\eta_{k} q) &\text{for scalar emission}.
 \end{cases}
\end{aligned}
\end{equation}
The sum runs over external legs where $\eta_k=+1(-1)$ accounts for the emission from final(initial)-state leg $k$ with four-momentum $p_{k}$.

For identical particles scattering,~\eqref{ppt} and~\eqref{ptt2} describe the $t$-channel contributions. The $u$-channel contributions are obtained by performing the substitutions~$p_{3} \leftrightarrow p_{4}$ and, in the case of fermionic candidates, introducing an additional minus sign due to the odd permutation of spinors between the two types of channels.

\section{Contact diagrams and Virtual Internal Bremsstrahlung} \label{diagramcancellation}
In this Appendix, we discuss in detail the contributions to the emission amplitude of the contact (Fig.~\ref{Contacts}) and the VIB (Fig.~\ref{VIB}) diagrams. We show that, in addition to velocity enhancement, their contributions depend sensitively on the underlying parameter space.

\subsection{Contact diagram} \label{appendixContact}

When dealing with real (complex) scalar DM candidates, the quartic self-couplings $({\Tilde{\lambda}_{S}}/{4!})\Tilde{S}^4$ and $({\lambda}_{S}/{4})(S^{\dagger}S)^2$ yield four dissipative contact-diagrams (Fig.~\ref{Contacts}). While for short-range mediation, this dissipative channel should be considered in the derivation of the emission amplitude; see Sec.~\ref{shortrangeforce}, this is not the case for long-range mediation: the lowest-order contribution of these dissipative channels to the emission amplitude is velocity suppressed compared to the other channels. 

For $S_{1}S_{2} \to S_{1}S_{2} V$, in the non-relativistic limit, the sum of amplitudes of contact diagrams is reduced to 
\begin{equation} \label{contactamplitude}
\begin{aligned}
   \sum_{k=1,4} \mathcal{M}_{c,k}
   &\simeq \left(\frac{g_{1}}{m_{1}}- \frac{g_{2}}{m_{2}}\right) \lambda_{S} \frac{\mu \left(\Vec{v}_{f} - \Vec{v}_{i}\right)\cdot \Vec{e}^{*}}{\omega} + \mathcal{O}(v_{i}^{0}). 
   \end{aligned}
\end{equation}
Here, the polarization vectors are given in  Coulomb gauge, \( \varepsilon^{\mu*}(q) = (0, \vec{e}^{*}) \) where $\vec e^{*}$ is one of the two transverse polarization three-vectors satisfying $\vec q\cdot \vec e^{*} = 0$. The expression above is of order $\mathcal{O}(v_{i}^{-1})$, whereas dipole $\sim \mathcal{O}(v_{i}^{-3})$ and quadrupole $\sim \mathcal{O}(v_{i}^{-2})$ emission amplitudes are of lower order in velocity. 
We note that the associated squared amplitude, averaged (summed) over initial (final) spins, summed over polarizations, and angular averaged using~\eqref{averageqrelationsvector}, yields the energy-differential cross section~\eqref{dipoleshortrange} and the energy loss rate~\eqref{NRJ2}, setting $g'_{j}=0$.

While velocity-suppressed compared to the dipole and quadrupole emission amplitudes, these dissipative channels feature a weaker dependence on the coupling. It is important to note that~\eqref{contactamplitude} shares the same coupling-to-mass ratio prefactor as the dipole emission amplitude~\eqref{LOvectrans}. As a consequence, for a given process, both contributions are selected under the same conditions on the coupling structure. In the case of an equal mass-to-charge ratio between the two DM species, these leading-order contributions vanish, and the next-to-leading order of~\eqref{contactamplitude} becomes the dominant contact contribution to the emission amplitude. This otherwise subleading term shares the same coupling-to-mass-squared prefactor as the quadrupole emission amplitude~\eqref{massiveQEDT}. 

For~\eqref{contactamplitude} to compete with the dipole emission amplitude, and, similarly, for the subleading order denoted by $\mathcal{O}(v_i^0)$ in~\eqref{contactamplitude} to compete with the quadrupole emission amplitude, the quartic coupling must satisfy
\begin{equation} \label{conditioncontact}
    \lambda_{S} \gtrsim \frac{m_{1}m_{2}}{\mu^2}\frac{g_{1}g_{2}}{v_{i}^2}. 
\end{equation}
Moreover, the right-hand side of this condition is itself bounded by the validity of the perturbative expansion, which constrains the allowed values of the couplings. For instance, in the case of identical particle scattering with a coupling constant $g=\sqrt{4\pi \alpha_{\chi}}$, this condition can hold while remaining in the perturbative regime provided that $g \lesssim v_{i} $. For a typical relative velocity $v_{i}\sim 10^{-3} $, this imposes $\alpha_{\chi} \lesssim 10^{-6}$ and points 1) to a strong hierarchy of couplings and 2) renders the overall cross section for gauge emission minute, in conflict with the  SIDM paradigm.

For the processes $S_{1} S_{2} \to S_{1} S_{2} \phi$,\, $\Tilde{S}_{1} \Tilde{S}_{2}\to \Tilde{S}_{1} \Tilde{S} _{2}\phi$ similar conclusions on the relative size of quartic and trilinear couplings can be drawn with the replacements $g_{j}\to A_{j}/2m_{j}$, $\vec{e}^{*}\to \vec{\hat{q}}$, and adding an overall minus sign. The adopted condition~\eqref{conditioncontact} then again requires a hierarchy among the couplings of the dSIDM scenario that is hard to achieve in a perturbative regime, allowing us to neglect these contributions.

\subsection{Virtual Internal Bremsstrahlung} \label{VinternalB}

When the interaction is mediated by a long-range force, the inclusion of a trilinear self-coupling term $\frac{1}{3!}A_{\phi}\phi^3$ in the Lagrangian \eqref{angrangianSelfcoupling} gives rise to VIB (Fig.~\ref{VIB}) processes, where the emitted particle originates from the propagator. This occurs for the scenarios~$\chi_{1} \chi_{2}\to \chi_{1} \chi_{2} \phi\, , \Tilde{\chi}_{1} \Tilde{\chi}_{2}\to \Tilde{\chi}_{1} \Tilde{\chi}_{2} \phi\, , S_{1} S_{2} \to S_{1} S_{2} \phi$,~and~$\Tilde{S}_{1} \Tilde{S}_{2}\to \Tilde{S}_{1} \Tilde{S}_{2} \phi$. In contrast, for short-range mediation, VIB can also occur provided there exists a direct coupling between the light and the heavy mediator $\frac{1}{2}A_{\phi}'\phi^{\prime2}\phi$.
Finally, we note that for the scenarios~$\Tilde{S}_{1} \Tilde{S}_{2}\to \Tilde{S}_{1} \Tilde{S}_{2} \phi$,  emission from an $\tilde{S}$ mediator is also allowed. However, in the NR limit, and for both long and short-range-mediated forces, these contributions are not relevant compared to the other dissipative channels.

\subsubsection{Long-range mediated force}
For a long-range mediated force, the NR emission amplitude in VIB diagrams involve {\it two} almost massless scalar propagators, which, in the massless limit, induce a strong forward singularity. As a result, the corresponding angular integral shows a stronger degree of divergence than in the  2-to-2 elastic case.  In the following, we hence retain a finite propagator mass in the amplitude, which, in our setup, is equivalent to considering only the case of massive emission.\footnote{It was originally argued in \cite{Veneziano:1972rs}, that such trilinear couplings are theoretically inconsistent for massless scalar fields: both S-matrix and semi-classical vacuum stability considerations indicate that the coupling $A_{\phi}$ must vanish at least proportionally to the scalar mass $m_{\phi}$ as $m_{\phi} \rightarrow 0$.} 

Applying this prescription to the emission amplitude for distinguishable scattering, squared and angular averaged over the direction of the emitted particle, we obtain
\begin{equation} \label{VIBdistinguishable}
  \frac{1}{d_{1}d_{2}} \left\langle |\mathcal{M}_{\text{VIB}}|^2 \right\rangle_{\vec{\hat{q}}} = A_{\phi}^2 a^2_{1}a^2_{2} \frac{ 16 m_{1}^2m_{2}^2}{ \left(\mu^2|\vec{v}_{i}-\vec{v}_{f}|^2 + m_{\phi}^2\right)^4},
\end{equation}
with $a_{j}=y_{j}$ for fermionic DM and $a_{j}=A_{j}/2m_{j}$ for scalar DM. 
In this form, similarly to external leg emission, the corresponding differential cross section for the emission of a soft scalar can be expressed as, 
\begin{equation} \label{PartialVIBintro}
\begin{aligned}
\frac{1}{\omega}\frac{ d \sigma_{\text{VIB}}}{  d \omega d\cos{\theta}_{i,f}} &= \frac{1}{64 \pi^2} \frac{   A_{\phi}^2  }{\left( \mu^2|\vec{v}_{i}-\vec{v}_{f}|^2 + m_{\phi}^2 \right)^2} \, \frac{d\sigma_{2\to 2}}{d \cos{\theta}_{i,f} }
\end{aligned}
\end{equation}
To maintain a perturbative treatment despite the strong IR sensitivity of~\eqref{PartialVIBintro} for $m_\phi\to 0 $ in the forward direction, we require that the probability of single scalar emission remains small. 
Imposing $\int d\omega \, \left({ d \sigma_{\text{VIB}}}/d\omega\right)/{\sigma_{2-2}} \ll 1$ in the IR-dominated regime, yields\footnote{For emission of $n$ scalars from the $t$-channel propagators, the condition generalizes to
$m_{\phi} \gg A_{\phi}^{1/2}({\mu v_{i}^{2}}/{2n})^{1/2},$ so that considering a higher number of soft scalars emitted ($n>1$) yields a parametrically weaker bound.}
 \begin{equation} \label{boundVIB}
   m_{\phi} \gg \mu^{1/2} A_{\phi}^{1/2}  v_{i}.
\end{equation}

Moreover, the stability of the scalar potential requires the inclusion of a quartic self-interaction term $\lambda_{\phi}\phi^4$ with $\lambda_{\phi} >0$, limiting the trilinear coupling as $A_{\phi}\lesssim m_{\phi}\sqrt{\lambda_{\phi}}$~\cite{Veneziano:1972rs}. In our NR power counting scheme,  this translates to a coupling at most of order $A_{\phi}\sim \mathcal{O}(\mu v_{i}^2)$. However, saturating this upper limit, combined with~\eqref{boundVIB} violates the kinematic requirement $m_{\phi} < 1/2 \mu v_{i}^2$ for emission. We therefore require $A_{\phi} \ll \mu v_{i}^2$ which results, despite the velocity enhancement of~\eqref{VIBdistinguishable}, on processes that are tightly constrained by consistency arguments and, in conclusion, subleading to quadrupole emission. 
\paragraph{Distinguishable particle scattering} 
In the case of distinguishable particle scattering, the energy-differential cross section is obtained from~\eqref{VIBdistinguishable}, and is given by
\begin{equation} \label{VIBenergydiff}
    \omega \frac{d\sigma}{d\omega}= A_{\phi}^2 \frac{ a_{1}^2a_{2}^2 \mu^4 x^2 v_{i}^4 }{48\pi^3} \sqrt{1-\frac{4m_{\phi}^2}{\mu^2v_{i}^4x^2}} 
    \frac{(1-x)^{1/2}   \left[3 m_{\phi}^4 - 6 m_{\phi}^2v_{i}^2\mu^2(x-2) + \mu^4v_{i}^4(x-4)(3x-4) \right]}{\left(m_{\phi}^4 -2 m_{\phi}^2\mu^2v_{i}^2\left(x-2\right) + \mu^4v_{i}^4x^2\right)^3} .
\end{equation}
While the associated energy loss rate requires numerical evaluation, it can be conveniently expressed in terms of the dimensionless parameters~\cite{Brinkmann:1988vi},
\begin{equation}
\begin{aligned} \label{dimensionless}
\tilde{\omega} &= \frac{\omega}{T},\, 
u_{i} = \frac{|\Vec{p}_{i}|^2}{2 \mu T},\, 
u_{f} = \frac{|\Vec{p}_{f}|^2}{2 \mu T},\,
z = \frac{ \Vec{p}_{f} \cdot \Vec{p}_{i}}{|\Vec{p}_{i}||\Vec{p}_{f}|}= \cos{\theta_{if}},\, \tilde{m}_{\phi,V}= \frac{m_{\phi,V}}{T},
\end{aligned}    
\end{equation}
and reads,
\begin{equation} \label{VIBedotlongdistin}
\begin{aligned}
    \dot{\epsilon}=& A_{\phi}^2\frac{a_{1}^2a_{2}^2 n_{1}n_{2} \mu^{3/2}}{3 \sqrt{2} T^{1/2}\pi^{7/2} }  \int_{\tilde{m}_{\phi}}^{\infty} d u_{i}\, \sqrt{u_{i}}\,e^{-u_{i}} \\& \times \int_{\tilde{m}_{\phi}}^{u_{i}} d\tilde{\omega}\, \tilde{\omega}^2 \sqrt{1-\frac{\tilde{m}_{\phi}^2}{\tilde{\omega}^2}} \frac{(u_{i}-\tilde{\omega})^{1/2}\left[3\tilde{m}_{\phi}^4T^2 +12 \mu \tilde{m}_{\phi}^2 T \left(2u_{i}- \tilde{\omega}\right) + 4\mu^2\left(4 u_{i}-3\tilde{\omega}\right)(4u_{i}-\tilde{\omega}) \right]}{\left(\tilde{m}_{\phi}^4T^2+ 4T\tilde{m}_{\phi}^2\mu (2u_{i}-\tilde{\omega}) + 4 \mu^2 \tilde{\omega}^2 \right)^3} .
\end{aligned}
\end{equation}

\paragraph{Identical particle scattering}
In scenarios involving identical particle scattering, the $u$-channel of Fig.~\ref{VIB} must also be considered. The corresponding squared emission amplitude, averaged over the direction of the emitted particle, is given by
\begin{equation}
\begin{aligned}
   \frac{1}{d_{1}d_{2}}\left\langle |\mathcal{M}_{\text{VIB}}|^2 \right\rangle_{\vec{\hat{q}}} = & 16 A_{\phi}^2   a^4  m_{S,\chi}^4\left[ \frac{1}{\left(\mu^2|\vec{v}_{i}-\vec{v}_{f}|^2 + m_{\phi}^2\right)^4} + \frac{1}{\left(\mu^2|\vec{v}_{i}+\vec{v}_{f}|^2 + m_{\phi}^2\right)^4} \right. \\&+ \left. 2\frac{(-1)^{2s}}{(2s+1)} \frac{1}{\left(\mu^2|\vec{v}_{i}-\vec{v}_{f}|^2 + m_{\phi}^2\right)^2 \left(\mu^2|\vec{v}_{i}+\vec{v}_{f}|^2 + m_{\phi}^2\right)^2}\right],
\end{aligned}
\end{equation}
with $\{a,s\}=\{ A/2m_{S},0 \}$ for scalar DM candidates and $\{y,1/2 \}$ for fermionic DM candidates. The associated energy-differential cross sections are
\begin{subnumcases} 
{\omega\frac{d\sigma}{d\omega}=} 
\frac{A_{\phi}^2}{8\pi^3}
\left(\frac{A}{m_{S}}\right)^4 \,  \mathcal{I}\left(28,8,-2,3,3072\right) \quad & \text{for}\quad $SS\!\to\!SS\phi,\,
(S \leftrightarrow \tilde S )$,
\label{eq:VIBxs:scalar}
\\[10pt]
\frac{ A_{\phi}^2}{8\pi^3}
y^{4} \, \mathcal{I}\left(56,16,-4,-3,384\right)
\quad & \text{for} \quad $\chi\chi\!\to\!\chi\chi\phi,\,
(\chi \leftrightarrow \tilde \chi )$,
\label{eq:VIBxs:fermion}
\end{subnumcases} 
where to simplify the presentation, we have introduced a short-hand notation for the following $x$-dependent functions
\begin{equation}
\begin{aligned}
\mathcal{I}(a,b,c,d,e) &=
\frac{    m_{S,\chi}^2 v_{i}^{2}\,x^{2}
 } 
{  e \Bigl(m_{\phi}^{2} - \tfrac{1}{4} m_{S,\chi}^{2} v_{i}^{2}(x-2)\Bigr)^{3}\,
 } \sqrt{\,1 - \frac{16\,m_{\phi}^{2}}{m_{S,\chi}^{2} v_{i}^{4} x^{2}}} \\
&\times \Biggl[
\frac{m_{S,\chi}^{2} v_{i}^{2} \sqrt{1-x}\Bigl(m_{\phi}^{2} - \frac{1}{4} m_{S,\chi}^{2} v_{i}^{2}(x-2)\Bigr)\,
\Bigl(G(a) + \frac{1}{256} m_{S,\chi}^{8} v_{i}^{8} H(b,c,d)\Bigr)}{\Bigl(m_{\phi}^{4} - \frac{1}{2} m_{\phi}^{2} m_{S,\chi}^{2} v_{i}^{2}(x-2) + \frac{1}{16} m_{S,\chi}^{4} v_{i}^{4}x^{2}\Bigr)^{3}} \\
& +\, (-1)^{2s}\, 3 \ln\!\left(
\frac{m_{\phi}^{2} - \tfrac{1}{4} m_{S,\chi}^{2} v_{i}^{2}(x-2) + \tfrac{1}{2} m_{S,\chi}^{2} v_{i}^{2}\sqrt{1-x}}
     {m_{\phi}^{2} - \tfrac{1}{4} m_{S,\chi}^{2} v_{i}^{2}(x-2) - \tfrac{1}{2} m_{S,\chi}^{2} v_{i}^{2}\sqrt{1-x}}
\right)
\Biggr],
\end{aligned}
\end{equation}
and
\begin{equation}
\begin{aligned}
H(b,c,d)&= b(x-2)^{4} + c(2-x)^{2}x^{2} + dx^{4}, \\
G(a)&=
m_{\phi}^{2}\Bigl(m_{\phi}^{2} - \frac{1}{2}{m_{S,\chi}^{2} v_{i}^{2}}(x-2)\Bigr)\,
 \\ &\quad \times\left[
9 m_{\phi}^{4}
- \frac{9}{2} m_{\phi}^{2}\,{m_{S,\chi}^{2} v_{i}^{2}}(x-2)
+ \frac{1}{8}{m_{S,\chi}^{4} v_{i}^{4}}\,\bigl(a + x(-a + 9x)\bigr)
\right].
\end{aligned}
\end{equation}

The corresponding energy loss rates are then obtained by inserting~\eqref{eq:VIBxs:scalar} and~\eqref{eq:VIBxs:fermion} into~\eqref{Dot(e)} and evaluating the integrals numerically within the parameter bounds discussed above. Since this step is straightforward and leads to lengthy expressions, we do not reproduce them here. 

\subsubsection{Short-range mediated force}
A similar VIB process can also be studied in the heavy mediator limit. In that case, the interaction does not rely on the self-trilinear coupling $\frac{1}{3!}A_{\phi}\phi^3$, but rather on the mixed trilinear coupling between the light and heavy scalar mediators $\frac{1}{2}A^{\prime}_{\phi}\phi^{\prime 2}\phi$. In this setting, the coupling $A^{\prime}_{\phi}$ can be treated as a free parameter. This freedom allows one to study VIB for both massless and massive emission, since the emitted particle is now distinct from the heavy mediator.
\paragraph{Distinguishable particle scattering} For scattering processes involving distinguishable particles, the squared emission amplitude, averaged over the direction of the emitted particle, is given by
\begin{equation} \label{VIBamplitudeheavy}
  \frac{1}{d_{1}d_{2}} \left\langle |\mathcal{M}_{\text{int}}|^2 \right\rangle_{\vec{\hat{q}}} =  {A_{\phi}^{\prime}}^2 {a^{\prime}_{1}}^2{a^{\prime}_{2}}^2\frac{ 16 \, m_{1}^2 m_{2}^2 }{m_{\phi^{\prime}}^{8}},
\end{equation} 
with $a'_{j}=y'_{j}$ for fermionic DM and $a'_{j}=A'_{j}/2m_{j}$. The corresponding energy-differential cross section reads
\begin{equation} \label{VIBheabydcs}
    \omega \frac{d\sigma}{d\omega}= {A_{\phi}^{\prime}}^2 {a_{1}^{\prime }}^2{a_{2}^{\prime }}^2\frac{ x^2\mu^4v_{i}^4\sqrt{1-x}}{16\pi^3m_{\phi^{\prime}}^{8}}\sqrt{1-\frac{4m_{\phi}^2}{\mu^2v_{i}^4x^2}}.
\end{equation}
For massless emission, the associated energy loss rate is
\begin{equation} \label{eTri}
\dot{\epsilon} = n_{1}n_{2}{A_{\phi}^{\prime}}^2 {a_{1}^{\prime}}^2{a_{2}^{\prime}}^2 \frac{64 \sqrt{2}  \mu^{3/2}  T^{7/2}}{35\pi^{7/2}m_{\phi'}^8},
\end{equation}
while for massive emission,~\eqref{Dot(e)} must be evaluated numerically using~\eqref{VIBheabydcs}.

We note that~\eqref{eTri} has the same $T$-dependence as quadrupole emission---the second term in~\eqref{NRJ2}---and vertex emission~\eqref{shortvecterNRJ}.  However, it exhibits an additional~$1/m_{\phi'}^4$-scaling compared to the latter,\footnote{This $m_{\phi'}^{-8}$ behavior is typical for diagrams involving two propagators associated with a heavy mediator; see, e.g., electroweak $W/Z$-strahlung in \cite{Bell:2011if} and wino-like DM bremsstrahlung in \cite{Ciafaloni:2012gs}.} which, inside the kinematic window $\sqrt{s} \gg m_{\phi'} \gg \mu v_{i}$, drives a much stronger suppression of the cooling rate. We conclude that, within the perturbative regime, the VIB dissipative channel remains sub-dominant in emission processes even with arbitrarily large hierarchies among the couplings $y_{j}',A_{j}'/m_{j},\lambda_{j}'$ and~$A^{\prime}_{\phi}$.

\paragraph{Identical particle scattering}
For identical particle scattering involving fermionic DM candidates, i.e., for the processes \( \chi\chi \to \chi\chi\phi \) and \( \tilde\chi \tilde\chi \to \tilde\chi \tilde\chi\phi \), the emission amplitude is the same as in~\eqref{VIBamplitudeheavy}. The energy-differential cross section~\eqref{VIBheabydcs} must be multiplied by a factor \( S_f = 1/2 \), and the energy loss rate~\eqref{eTri} by \( S_i  S_f = 1/4 \).
However, for scalar DM candidates, i.e., for \( S S \to S S\phi \) and \( \tilde{S} \tilde{S} \to \tilde{S} \tilde{S}\phi \), the emission amplitude is larger by a factor of 4. This enhancement carries over to the energy-differential cross section that must be, in addition, multiplied by \( S_f = 1/2 \), and to the energy loss rate, cancelling the factor \( S_i  S_f = 1/4 \).

\section{Scalar DM candidates: scalar emission} \label{scalarscalarquad}
In contrast to the process \( S_{1}S_{2} \to S_{1}S_{2}\,V \), where gauge symmetry constrains the amplitude structure, the processes \(S_{1}S_{2} \to S_{1}S_{2}\,\phi\) and \(\tilde{S}_{1} \tilde{S}_{2} \to \tilde{S}_{1} \tilde{S}_{2}\,\phi\) are less restricted. As a consequence, these scenarios involve a larger set of independent couplings that enter the next-to-leading order term in the \(v_i\) or $r$-expansion of the emission amplitude: \((A_j/m_j,\, \lambda_j)\) for long-range mediation, and \((A_j/m_j,\, A'_j/m_j,\, \lambda'_j,\, \lambda_S)\) for short-range mediation.  The strength of interaction is being measured by the dimensionless coupling constant $A^{(\prime)}_j/m_j$. A vanishing ``coupling-to-mass'' ratio then refers to the  combination $A^{}_1/m_1^2 -A^{}_2/m_2^2$. For completeness, we provide below the corresponding full expressions for the squared emission amplitude and the soft factor~\eqref{softfactorscalarA}.

\subsection{Long-range mediation}
We begin by considering the processes mediated by a light mediator, corresponding to a long-range mediation.
\paragraph{Emission amplitude}
The emission amplitude of the processes $S_{1}S_{2}\to S_{1}S_{2} \phi$ and $\tilde{S}_{1}\tilde{S}_{2}\to \tilde{S}_{1}\tilde{S}_{2}\phi$ mediated by long-range interaction, contains contributions from the external leg and vertex emission diagram, Figs.~\ref{inifini}-\ref{vertex}. The relative importance of each process is determined by the magnitude of its corresponding coupling constant. As a result, at sub-leading order in the $v_{i}$-expansion, the squared emission amplitude, averaged over initial spins, summed over final spins, and angularly averaged over the direction of the emitted particle using~\eqref{averagingquadrupole2}, can be decomposed as follows,
\begin{equation} \label{fullqua}
\begin{aligned}
\left \langle |\mathcal{M}|^2 \right \rangle_{\vec{\hat{q}}} &=  \left \langle |\mathcal{M}_{\text{external leg}}|^2 \right \rangle_{\vec{\hat{q}}} +  \left \langle |\mathcal{M}_{\text{vertex}}|^2 \right \rangle_{\vec{\hat{q}}}  + \left \langle |\mathcal{M}_{\text{cross}}|^2 \right \rangle_{\hat{q}}.
\end{aligned}
\end{equation}

The first term corresponds to the contribution from the external leg emission diagram, which dominates the dissipative process when \( (A_1/m_1)^2,\, (A_2/m_2)^2 \gg \lambda_1,\, \lambda_2 \),

\begin{equation} \label{externallegscalarlong}
\begin{aligned}
\left \langle |\mathcal{M}_{\text{external leg}}|^2 \right \rangle_{\vec{\hat{q}}} &=   
  \frac{A_{{1}}^2} {m_{1}^2}\frac{A_{{2}}^2} {m_{2}^2}\left(\frac{A_{1}}{m_{1}^2}-\frac{A_{2}}{m_{2}^2} \right)  \frac{\Delta m_{12}}{ \mu^2 |\vec{v}_{i} - \vec{v}_{f}|^4}\left[ \frac{\Delta m_{12}}{4} \left(\frac{A_{1}}{m_{1}^2}-\frac{A_{2}}{m_{2}^2} \right)  \right. \\& \left.  + \left(\frac{A_{1 }}{m_{1}^3} + \frac{A_{2}}{m_{2}^3}   \right)\frac{2   m_{1} m_{2} }{3  } \right] + A_{1}^2 A_{2}^2 \left(\frac{A_{1 }}{m_{1}^3} + \frac{A_{2}}{m_{2}^3}   \right)^2 \frac{ T_{1} + 11 T_{2} }{60\omega^2|\Vec{v}_{i}-\Vec{v}_{f}|^4},
\end{aligned}
\end{equation}
where $\Delta m_{12}=m_{1}-m_{2}$ and $T_{1,2}$ are defined in~\eqref{T1} and~\eqref{T2}. The second term in~\eqref{fullqua} corresponds to the contribution from the vertex emission diagram, which becomes relevant when \( (A_1/m_1)^2,\, (A_2/m_2)^2 \ll \lambda_1,\, \lambda_2 \),
\begin{equation}
\left \langle |\mathcal{M}_{\text{vertex}}|^2 \right \rangle_{\vec{\hat{q}}} =  \left(\frac{A_{1}\lambda_{2} + A_{{2}}\lambda_{{1}}}{\mu^2|\Vec{v}_{i}- \Vec{v}_{f}|^2}\right)^2. 
\end{equation}
Finally, \(\langle |\mathcal{M}_{\text{cross}}|^2 \rangle_{\hat{q}}\) contains the cross terms involving both types of couplings, which become subdominant in the limits \( (A_1/m_1)^2,\, (A_2/m_2)^2 \gg \lambda_1,\, \lambda_2 \) or \( (A_1/m_1)^2,\, (A_2/m_2)^2 \ll \lambda_1,\, \lambda_2 \),
\begin{equation}
\begin{aligned}
 \left \langle |\mathcal{M_{\text{cross}}}|^2 \right \rangle_{\hat{q}}= 
    - \frac{A_{1} A_{2}\left(A_{1} \lambda_{2} + A_{2} \lambda_{1}\right)}{\mu^3 |\vec{v}_{i} - \vec{v}_{f}|^4} \left[ \frac{4}{3 }\left(\frac{A_{1 }}{m_{1}^3} + \frac{A_{2}}{m_{2}^3}   \right)    +  \left(\frac{A_{1}}{m_{1}^2} - \frac{A_{2}}{m_{2}^2}\right) \frac{\Delta m_{12}}{m_{1} m_{2} } \right].
\end{aligned}   
\end{equation}

We note that the emission amplitude squared~\eqref{externallegscalarlong} decomposes into two contributions. The first one vanishes in the case of equal coupling-to-mass ratios, a regime where quadrupole emissions dominate. We accordingly treat this contribution as a subleading, non-quadrupole correction and neglect it in the definition of the quadrupole emission amplitude squared. 
The second term, which survives in the case of equal coupling-to-mass ratios, is what we define as the quadrupole emission amplitude, 
\begin{equation} \label{quadscalarlongapp}
   \left \langle |\mathcal{M_{Q}}|^2 \right \rangle_{\vec{\hat{q}}} =  A_{1}^2 A_{2}^2 \left(\frac{A_{1 }}{m_{1}^3} + \frac{A_{2}}{m_{2}^3}   \right)^2 \frac{ \left[ 3\left(|\Vec{v}_{f}|^{4} + |\Vec{v}_{i}|^4\right) -|\Vec{v}_{f}|^2|\Vec{v}_{i}|^2\left( (\Hat{v}_{i} \cdot \Hat{v}_{f})^2 + 5\right) \right] }{15\omega^2|\Vec{v}_{i}-\Vec{v}_{f}|^4}.
\end{equation}

\paragraph{Eikonal factor}
The next-to-leading order term in the \(v_i\)-expansion of the eikonal factor~\eqref{softfactorscalarA}, multiplied by the correction factor~\eqref{correctionsoft}, squared, and angularly averaged over the direction of the emitted particle using~\eqref{averagingquadrupole2}, is given by
\begin{equation} \label{scalarscalarfactor}
 \begin{aligned}
  \left \langle |\mathcal{E}_{{S\phi}}|^2 \right \rangle_{\vec{\hat{q}}} &=  \left(\frac{A_{1}}{m_{1}^2}-\frac{A_{2}}{m_{2}^2} \right)  \frac{\Delta m_{12}\mu^2}{ m_{1}m_{2}}\left[ \frac{\Delta m_{12}}{4m_{1}m_{2}} \left(\frac{A_{1}}{m_{1}^2}-\frac{A_{2}}{m_{2}^2} \right)  \right. \\ 
  & \left.  + \frac{2    }{3  }\left(\frac{A_{1 }}{m_{1}^3} + \frac{A_{2}}{m_{2}^3}   \right) \right] +  \left( \frac{A_{{1}}}{m_{1}^3} +   \frac{A_{{2}}}{m_{2}^3}\right)^2 \frac{\mu^4 \left( T_{1} +11 T_{2} \right)}{60 \omega^2}.
\end{aligned}  
\end{equation}

\subsection{Short-range mediation}
We turn now to the case of a heavy mediator, leading to a short-range mediation.
\paragraph{Emission amplitude}
The emission amplitude for the processes \(S_{1}S_{2} \to S_{1}S_{2}\,\phi\) and \(\tilde{S}_{1}\tilde{S}_{2} \to \tilde{S}_{1}\tilde{S}_{2}\,\phi\), mediated by short-range interactions, receives contributions from the external leg, vertex, and contact emission diagrams, shown in Figs.~\ref{inifini}–\ref{Contacts}. As a result, at sub-leading order in the $r$-expansion, the squared emission amplitude, averaged over initial spins, summed over final spins, and angularly averaged over the direction of the emitted particle using~\eqref{averagingquadrupole2}, can be decomposed as~\eqref{fullqua}. 
The first term corresponds to the contribution from the external leg and contact emission diagrams, and is given by
\begin{equation}
\begin{aligned} \label{externallegscalarheavy}
\left \langle |\mathcal{M}_{\text{external leg}}|^2 \right \rangle_{\vec{\hat{q}}} &=    \frac{m_{1}m_{2}C_{A'_{j},\lambda_{S}}^2\mu^2\Delta m_{12}}{m_{\phi'}^4} \left(\frac{A_{1}}{m_{1}^2}-\frac{A_{2}}{m_{2}^2} \right)  \left[ \frac{1}{3}\left(\frac{A_{1 }}{m_{1}^3} + \frac{A_{2}}{m_{2}^3}   \right)\right. \\& \left. + \frac{\Delta m_{12} }{4m_{1}m_{2}}  \left(\frac{A_{1}}{m_{1}^2}-\frac{A_{2}}{m_{2}^2} \right)  \right] +  m_{1}^2m_{2}^2C_{A'_{j},\lambda_{S}}^2\left(\frac{A_{1 }}{m_{1}^3} + \frac{A_{2}}{m_{2}^3}   \right)^2 \frac{\mu^4\left(T_{1} +2 T_{2}\right) }{60\omega^2m_{\phi'}^4}, 
\end{aligned}
\end{equation}
where $C_{A'_{j},\lambda_{S}}=(A'_{1}/m_{1}) (A'_{2}/m_{2}) - \lambda_{S}m_{\phi'}^2/(m_{1}m_{2})$. The second term of~\eqref{fullqua} corresponds to the contribution from the vertex emission diagram, and reads
\begin{equation}
\left \langle |\mathcal{M}_{\text{vertex}}|^2 \right \rangle_{\vec{\hat{q}}} =  \left(\frac{A'_{1}\lambda'_{2} + A'_{{2}}\lambda'_{{1}}}{m_{\phi'}^2}\right)^2. 
\end{equation}
Finally, \( \left \langle |\mathcal{M}_{\text{cross}}|^2 \right \rangle_{\hat{q}}\) contains the cross terms involving all four couplings, and is defined by
\begin{equation} \label{crossheavy}
\begin{aligned}
\left  \langle |\mathcal{M_{\text{cross}}}|^2 \right \rangle_{\hat{q}}=&  
   - \frac{2 m_{1}m_{2}C_{A'_{j},\lambda_{S}}\mu}{3 m_{\phi'}^4} \left(A^{\prime}_{1}\lambda^{\prime}_{2} + A^{\prime}_{2}\lambda^{\prime}_{1}\right) \left[  \left(\frac{A_{1 }}{m_{1}^3} + \frac{A_{2}}{m_{2}^3}   \right)   +  \left(\frac{A_{1}}{m_{1}^2}-\frac{A_{2}}{m_{2}^2} \right) \frac{3\Delta m_{12}}{2 m_{1}m_{2}} \right].
\end{aligned}   
\end{equation}
In the regime \(\langle |\mathcal{M}_{\text{external leg}}|^2 \rangle_{\vec{\hat{q}}} \gg \langle |\mathcal{M}_{\text{vertex}}|^2 \rangle_{\vec{\hat{q}}}\), the dissipative process is dominated by external leg emission. In the opposite limit, \(\langle |\mathcal{M}_{\text{external leg}}|^2 \rangle_{\vec{\hat{q}}} \ll \langle |\mathcal{M}_{\text{vertex}}|^2 \rangle_{\vec{\hat{q}}}\), the vertex emission provides the dominant contribution. In both cases, the cross terms~\eqref{crossheavy} become subdominant and can be neglected. 

Similarly to the long-range mediation case, we note that the squared emission amplitude~\eqref{externallegscalarheavy} is composed of two terms. We identify the first piece, that vanishes in the case of equal coupling-to-mass ratios, as a subleading non-quadrupole contribution, and the second term, which stays finite in this regime, as the quadrupole piece of the emission amplitude squared,
\begin{equation}
\begin{aligned} \label{quadscalarheavyApp}
\left \langle |\mathcal{M}_{\mathcal{Q}}|^2 \right \rangle_{\vec{\hat{q}}} &=   m_{1}^2m_{2}^2C_{A'_{j},\lambda_{S}}^2\left(\frac{A_{1 }}{m_{1}^3} + \frac{A_{2}}{m_{2}^3}   \right)^2 \frac{\mu^4\left[-2|\Vec{v}_{f}|^2|\Vec{v}_{i}|^2\left( 2(\Hat{v}_{i} \cdot \Hat{v}_{f})^2 + 1\right) + 3\left(|\Vec{v}_{f}|^{4} + |\Vec{v}_{i}|^4\right) \right] }{60\omega^2m_{\phi'}^4}.  
\end{aligned}
\end{equation}

\paragraph{Eikonal factor}
The next-to-leading order in the $v_{i}$-expansion of the eikonal factor~\eqref{softfactorscalarA}, squared, and angular averaged over the direction of the emitted particle using~\eqref{averagingquadrupole2} is given by
\begin{equation} \label{scalarscalarfactorheavy}
 \begin{aligned}
\left   \langle |\mathcal{E}_{{S\phi}}|^2 \right \rangle_{\vec{\hat{q}}} &=  
    \frac{\mu^2\Delta m_{12}}{m_{1}m_{2}} \left(\frac{A_{1}}{m_{1}^2}-\frac{A_{2}}{m_{2}^2} \right)  \left[ \frac{1}{3}\left(\frac{A_{1 }}{m_{1}^3} + \frac{A_{2}}{m_{2}^3}   \right) \right.  \left. \right. \\
    & \left. + \frac{\Delta m_{12} }{4m_{1}m_{2}}  \left(\frac{A_{1}}{m_{1}^2}-\frac{A_{2}}{m_{2}^2} \right)  \right]  +\left(\frac{A_{1 }}{m_{1}^3} + \frac{A_{2}}{m_{2}^3}   \right)^2  \frac{\mu^4\left(T_{1}+2T_{2}\right) }{60\omega^2} .
\end{aligned}  
\end{equation}

\section{Details of scalar vs.~vector emission from fermionic DM candidates} \label{scalarvsgauge}
We outline in the following the main steps in deriving the dipole and quadrupole emission amplitude from fermionic DM candidates for long-range mediation following the $v_{i}$-expansion discussed in Sec.~\ref{NRhierarchyofscales}. In order to expand the four-component spinors, we apply the conventions defined in Sec.~\ref{4spinors}. We then carry out the analogous analysis for short-range mediators—presenting the key steps and results.

For scalar and vector emission from fermionic DM candidates, the dissipative processes happen via emission from an external leg, as illustrated in Fig.\ref{inifini} leading---for distinguishable particle scattering---to four diagrams. In the following, we consider only the two diagrams corresponding to an emission from the DM species 1. The two remaining diagrams, associated with an emission from the DM species 2, are simply obtained by applying the replacement $g_{1},y_{1} \leftrightarrow g_{2},y_{2}$, $m_{1} \leftrightarrow  m_{2}$ and $p_{1},p_{3}\leftrightarrow p_{2},p_{4}$ to the emission amplitude pieces defined bellow.

\subsection{Scalar emission from fermionic DM candidates} 
We start with the emission of massless scalars from fermionic DM candidates in the limit of a light mediator. The sum of the emission amplitudes associated with the $\chi_{1}$- or $\Tilde{\chi}_{1}$-leg can be decomposed as follows,
\begin{equation} \label{decomposition}
    \mathcal{M}|_{\chi_1\text{-leg}}= \mathcal{M}_{\slashed{q}=0} + \mathcal{M}_{\slashed{q}}.
\end{equation}
The first term is obtained by setting $\slashed{q}=0$ in the numerator of the emission amplitudes and reads,
\begin{equation} \label{M0yuka}
\begin{aligned}
\mathcal{M}_{\slashed{q}=0} &= -y_{1}^2 y_{2}\frac{2m_{1}\overline{u}(p_{3},s_{3})u(p_{1},s_{1}) \overline{u}(p_{4},s_{4})u(p_{2},s_{2})}{ \left(p_{2}-p_{4}\right)^2 - m_{\phi}^2  } \left[ \frac{1}{(p_{3}+q)^2 -m_{1}^2} +\frac{1}{(p_{1}-q)^2 -m_{1}^2 } \right],
\end{aligned}
\end{equation}
while the second term collects the remaining $\slashed{q}$-dependent terms, namely,
\begin{equation} \label{Mqyuka}
\begin{aligned}
\mathcal{M}_{\slashed{q}} &= -y_{1}^2 y_{2}\frac{\overline{u}(p_{3},s_{3})\slashed{q}u(p_{1},s_{1}) \overline{u}(p_{4},s_{4})u(p_{2},s_{2})}{ (p_{2}-p_{4})^2 - m_{\phi}^2  } \left[ \frac{1}{(p_{3}+q)^2 -m_{1}^2} -\frac{1}{(p_{1}-q)^2 -m_{1}^2 } \right].
\end{aligned}
\end{equation}

Applying to $\mathcal{M}_{\slashed{q}=0}$, which contributes both at dipole and quadrupole order, the expansion scheme discussed in Sec.~\ref{NRhierarchyofscales}, yields
\begin{equation} \label{withoutemissionYuka}
\begin{aligned}
\mathcal{M}_{\slashed{q}=0} 
   \overset{\text{CM}}{\simeq} & y_{1}^2 y_{2}\frac{m_{1} }{|\Vec{p}_{i}-\Vec{p}_{f}|^2}\left[\frac{1}{m_{1}\omega\left( 1- \frac{\Vec{p}_{f} \cdot \Vec{q}}{m_{1}\omega} +\mathcal{O}(v_{i}^2) \right)} - \frac{1}{m_{1}\omega\left( 1- \frac{\Vec{p}_{i} \cdot \Vec{q}}{m_{1}\omega} +\mathcal{O}(v_{i}^2) \right) } \right] \\ &\times \xi_{s_{3}}^{\dagger}\left(Av_{i}^{0} + Bv_{i}^2 + \mathcal{O}(v_{i}^4) \right)\xi_{s_{1}} \xi_{s_{4}}^{\dagger}\left(A'v_{i}^{0} + B'v_{i}^2 + \mathcal{O}(v_{i}^4) \right)\xi_{s_{2}}.  
\end{aligned}
\end{equation}
where the label $\text{CM}$ indicates that the $v_{i}$-expansion is carried out in the CM frame. 
Here, $A,A'$ have no dependence on Pauli matrices, whereas $B,B'$ do, and are obtained from the spinors' $v_{i}$-expansion. We find, similarly to the $\nu$-expansion in Sec.~\ref{dipolevsquadrupole}, that, in a first-order expansion, the term enclosed inside the square bracket vanishes. In order to obtain the dipole (quadrupole) emission contribution, a second-order (third-order) expansion of this term is necessary. In these cases, only the $A,A'$ terms are relevant, as the 
$B,B'$ terms are comparatively velocity-suppressed. Consequently, the spinors' terms from the $v_{i}$-expansion that contains Pauli matrices do not contribute to the dipole and quadrupole emission amplitude, meaning that spin does not enter at these orders of the velocity expansion in~\eqref{withoutemissionYuka}. 

The second term in~\eqref{decomposition}, which is $\slashed{q}$-dependent, contributes to the quadrupole emission amplitude only. Applying to the latter the $v_{i}$-expansion leads to
\begin{equation} \label{}
\begin{aligned}
\mathcal{M}_{\slashed{q}} & 
 \overset{\rm CM}{\simeq}  \frac{ y_{1}^2 y_{2}}{|\Vec{p}_{i}-\Vec{p}_{f}|^2}\left[\frac{1}{2m_{1}\omega\left( 1- \frac{\Vec{p}_{f} \cdot \Vec{q}}{m_{1}\omega} +\mathcal{O}(v_{i}^2) \right)} + \frac{1}{2m_{1}\omega\left( 1- \frac{\Vec{p}_{i} \cdot \Vec{q}}{m_{1}\omega} +\mathcal{O}(v_{i}^2) \right) } \right] \\
&\times \xi_{s_{3}}^{\dagger}\left(2m_{1} \omega + \mathcal{O}(v_{i}^4) \right)\xi_{s_{1}} \xi_{s_{4}}^{\dagger}\left(A'v_{i}^{0} + B'v_{i}^2 + \mathcal{O}(v_{i}^4) \right)\xi_{s_{2}}.
\end{aligned}
\end{equation}
As expected, only the scalar part of the emission current is relevant and does not lead to non-trivial spin-dependent expressions. We note that the contribution of this term to the quadrupole emission amplitude, denoted by a subscript~\(\mathcal{Q}\), reads
\begin{equation} \label{withemissionYukaq}
\begin{aligned}
\mathcal{M}_{\mathcal{Q},\slashed{q}} & \overset{}{=}  \frac{4 m_{2}  y_{1}^2 y_{2} }{\mu^2|\Vec{v}_{i}-\Vec{v}_{f}|^2} \delta_{s_{1}s_{3}} \delta_{s_{2}s_{4}},
\end{aligned}
\end{equation}
which is independent of the energy and momentum of the emitted particle. 

In the case of short-range mediation, the two contributions to the emission amplitude associated with the $\chi_{1}$- and $\tilde{\chi}_{1}$-leg, i.e.,~\eqref{M0yuka} and~\eqref{Mqyuka}, must be expanded at leading and sub-leading order in $r$ to isolate the dipole and quadrupole terms. 
Finally, the same arguments apply to the second set of diagrams, where the emission originates from $\chi_{2}$ and $\Tilde{\chi}_{2}$. 

\paragraph{Soft emission limit}
We now turn to the regime where the particle emitted is soft. Taking the limit $\omega\to 0$, the expression~\eqref{M0yuka} reduces to the elastic amplitude multiplied by
\begin{equation} \label{eyuka0}
\mathcal{E}_{0}|_{\chi_1\text{-leg}}= y_{1}\frac{m_{1}}{p_{3}\cdot q } - y_{1}\frac{m_{1}}{p_{1}\cdot q },
\end{equation}
which corresponds to the leading term in the soft factor~\eqref{softfactorscalar} restricted to one emission leg. In the NR limit,~\eqref{eyuka0} times the elastic amplitude precisely matches the dipole and quadrupole terms obtained from~\eqref{withoutemissionYuka}.

In addition to these leading terms, applying the relations
\begin{equation}
\begin{aligned}
&-\frac{\slashed{q}}{(p_{1} -q)^2-m_1^2}u(p_1)
= \frac{(\slashed{p_{1}} - \slashed{q} -m_{1})}{(p_{1} -q)^2-m_1^2}u(p_1)
= \frac{1}{(\slashed{p_{1}}-\slashed{q}+m_{1})}u(p_1),
\end{aligned}
\end{equation}
to~\eqref{Mqyuka}, together with the corresponding relations for an emission from the outgoing leg, one obtains an additional contribution multiplying the elastic amplitude,
\begin{equation} \label{sub}
\mathcal{E}_{\slashed{q}}|_{\chi_1\text{-leg}}= \frac{y_{1}}{m_{1} } + \mathcal{O}(q),
\end{equation}
namely, the subleading piece of the soft factor~\eqref{softfactorscalar}. Therefore, neglecting~\eqref{sub} during the soft limit computation corresponds to omitting the contribution of $\mathcal{M}_{\slashed{q}}$, as given in~\eqref{Mqyuka}, to the emission amplitude. However, for long-range mediation, from the full NR amplitude, we find that~\eqref{Mqyuka} gives rise to the quadrupole contribution~\eqref{withemissionYukaq}, which is enhanced as \( \omega \to 0 \) (i.e., in the limit \( \vec{v}_{i} \to \vec{v}_{f} \)). Consequently, this contribution to the emission amplitude cannot be neglected in the soft emission regime. For short-range mediation the quadrupole piece of $\mathcal{M}_{\slashed{q}}$ is obtained by the replacement \( \mu^2|\vec{v}_{i} - \vec{v}_{f}|^2 \to m_{\phi'}^2 \) in~\eqref{withemissionYukaq}. Unlike the long-range case, this contribution is not enhanced as \( \omega \to 0 \) and can be omitted without affecting the \( x \ll 1 \) behavior of the soft energy-differential cross section.
  
\subsection{Vector emission from fermionic DM candidates} For vector emission from fermionic DM candidates mediated by a long-range force, the amplitude can be decomposed similarly to the scalar emission case. Following~\eqref{decomposition}, the contribution from the $\chi_{1}$ legs can be written as
\begin{equation} \label{M0QED}
\begin{aligned}
\mathcal{M}_{\slashed{q}=0} &= \frac{g_{1}^2 g_{2}
\overline{u}(p_{3},s_{3})\gamma^{\nu}u(p_{1},s_{1})\overline{u}(p_{4},s_{4})\gamma_{\nu}u(p_{2},s_{2})}{\left(p_{2}-p_{4}\right)^2 - m_{V}^2} \left[ \frac{2 p_{3}\cdot \varepsilon^{*}(q) \ }{(p_{3}+q)^2 -m_{1}^2} +\frac{2 p_{1}\cdot \varepsilon^{*}(q)}{(p_{1}-q)^2 -m_{1}^2 } \right],
\end{aligned}
\end{equation}
together with a $\slashed{q}$-dependent part,
\begin{equation} \label{MqQED}
\begin{aligned}
\mathcal{M}_{\slashed{q}} &= g_{1}^2 g_{2}\frac{\overline{u}(p_{3},s_{3})\gamma^{\mu}\slashed{q}\gamma^{\nu}u(p_{1},s_{1})\overline{u}(p_{4},s_{4})\gamma_{\nu}u(p_{2},s_{2}) \varepsilon^{*}_{\mu}(q)}{\left[(p_{2}-p_{4})^2 - m_{V}^2\right] \left[(p_{3}+q)^2 -m_{1}^2\right]} \\
& - g_{1}^2 g_{2}\frac{\overline{u}(p_{3},s_{3})\gamma^{\nu}\slashed{q}\gamma^{\mu}u(p_{1},s_{1})\overline{u}(p_{4},s_{4})\gamma_{\nu}u(p_{2},s_{2}) \varepsilon^{*}_{\mu}(q)}{\left[(p_{2}-p_{4})^2 - m_{V}^2\right] \left[(p_{1}-q)^2 -m_{1}^2\right]} .
\end{aligned}
\end{equation}
Here, the inversion of the order of the gamma matrices between an emission in the final and initial state preserves the gauge invariance of the amplitude.

As for scalar emission, $\mathcal{M}_{\slashed{q}=0}$ contributes to both the dipole and quadrupole emission amplitudes. Applying to the latter the $v_{i}$-expansion yields
\begin{equation} \label{withoutemissionQED}
\begin{aligned}
\mathcal{M}_{\slashed{q}=0} 
   \overset{\text{CM}}{\simeq} & \frac{ g_{1}^2 g_{2}}{|\Vec{p}_{i}-\Vec{p}_{f}|^2}\left[  \frac{\Vec{p}_{f}\cdot \Vec{e}^{*}}{m_{1}\omega\left( 1- \frac{\Vec{p}_{f} \cdot \Vec{q}}{m_{1}\omega} +\mathcal{O}(v_{i}^2) \right)} - \frac{\Vec{p}_{i}\cdot \Vec{e}^{*}}{m_{1}\omega\left( 1- \frac{\Vec{p}_{i} \cdot \Vec{q}}{m_{1}\omega} +\mathcal{O}(v_{i}^2)\right)} \right]  \\
 & \times \xi_{s_{3}}^{\dagger}\left(Cv_{i}^{0} + Dv^2_{i} + \mathcal{O}(v_{i}^4) \right)\xi_{s_{1}} \xi_{s_{4}}^{\dagger}\left(C'v_{i}^{0} + D'v^2_{i} + \mathcal{O}(v_{i}^4) \right)\xi_{s_{2}},
\end{aligned}
\end{equation}
by assuming that, for a massless emitted particle, the polarization vectors are in the Coulomb gauge, \( \varepsilon^{\mu*}(q) = (0, \vec{e}^{*})  \), where $ \vec{e}^{*}$ denotes the transverse polarization vector in Cartesian components satisfying $\vec{e}^{*}\cdot \vec{q}=0$.  The spinor expansion separates into \( C^{(\prime)} \) terms, which are independent of Pauli matrices, and \( D^{(\prime)} \) terms, which explicitly depend on them. However, both for dipole and quadrupole emission, only the $C^{(\prime)}$-terms are relevant compared to the $D^{(\prime)}$-term. Consequently,~\eqref{withoutemissionQED} is spin-independent for both types of emission.

We apply the same procedure to the $\slashed{q}$-dependent part of the amplitude, given in~\eqref{MqQED}, and obtain
\begin{equation} \label{emissionQED}
\begin{aligned}
  \mathcal{M}_{\slashed{q}} 
 \overset{\text{CM}}{\simeq} & -\frac{ g_{1}^2 g_{2} }{|\Vec{p}_{i}-\Vec{p}_{f}|^2}\xi_{s_{3}}^{\dagger}\left[ \frac{(\Vec{e}^{*} \cdot \boldsymbol{\sigma}) (\Vec{q} \cdot \boldsymbol{\sigma}) + \mathcal{O}(v_{i}^{3}) }{\omega\left( 1- \frac{\Vec{p}_{f} \cdot \Vec{q}}{m_{1}\omega} +\mathcal{O}(v_{i}^2) \right)} + \frac{(\Vec{q} \cdot \boldsymbol{\sigma})( \Vec{e}^{*} \cdot \boldsymbol{\sigma}) + \mathcal{O}(v_{i}^{3}) }{\omega\left( 1- \frac{\Vec{p}_{i} \cdot \Vec{q}}{m_{1}\omega} +\mathcal{O}(v_{i}^2)\right)}\right]\xi_{s_{1}}  \\
   &\times  \xi_{s_{4}}^{\dagger}\left(C'v_{i}^{0} + D'v^2_{i} + \mathcal{O}(v_{i}^4) \right)\xi_{s_{2}},
\end{aligned}
\end{equation}
where the numerator structure arises from the vector-like nature of the interaction. Only the leading velocity term of this expression would, in principle, be relevant, contributing at quadrupole order. However, the $v_i^{-2}$ contribution~\eqref{emissionQED} vanishes identically:
\begin{equation} \label{withemissionQED}
\begin{aligned}
\mathcal{M}_{\mathcal{Q},\slashed{q}} &=  -\frac{ 2 g_{1}^2 g_{2}  m_{2}}{\omega |\Vec{p}_{i}-\Vec{p}_{f}|^2 }\xi_{s_{3}}^{\dagger}\left[ (\Vec{q} \cdot \boldsymbol{\sigma}) (\Vec{e}^{*} \cdot \boldsymbol{\sigma}) + (\Vec{e}^{*} \cdot \boldsymbol{\sigma})( \Vec{q} \cdot \boldsymbol{\sigma}) \right]\xi_{s_{1}}\delta_{s_{4}s_{2}} 
\\
& = \frac{-4 g_{1}^2 g_{2} m_{2}}{\omega |\Vec{p}_{i}-\Vec{p}_{f}|^2 }  \Vec{q}\cdot \Vec{e}^{*} \, \delta_{s_{1}s_{3}} \delta_{s_{2}s_{4}}=0 ,
\end{aligned}
\end{equation}
where we have applied the identity
\( \{\boldsymbol{\sigma}_{i}, \boldsymbol{\sigma}_{j}\} = 2 \delta_{ij} \).  Consequently, only $\mathcal{M}_{\slashed{q}=0}$ contributes to the dipole and quadrupole emission amplitudes, and $\mathcal{M}_{\slashed{q}}$ can be neglected in the derivation of the NR full amplitude. In contrast, for scalar emission the contributions arising from $\mathcal{M}_{\slashed{q}}$, i.e.~\eqref{Mqyuka}, do not vanish at quadrupole order and must also be taken into account.
These conclusions also apply to the process $\chi_{1}\chi_{2}\to\chi_{1}\chi_{1} V$ interacting via a heavy mediator, where the $r$-expansion should be applied.

Finally, we note that the scalar products in~\eqref{emissionQED} can be rewritten as
\begin{equation} \label{spineffects}
 (\Vec{e}^{*} \cdot \boldsymbol{\sigma}) (\Vec{q} \cdot  \boldsymbol{\sigma} )= i(\Vec{e}^{*} \times \Vec{q})\cdot \boldsymbol{\sigma},\quad (\Vec{q} \cdot \boldsymbol{\sigma}) (\Vec{e}^{*} \cdot  \boldsymbol{\sigma} )= i(\Vec{q} \times \Vec{e}^{*})\cdot \boldsymbol{\sigma},
\end{equation}
which, in QED, corresponds to magnetic-like components. Beyond the quadrupole order, the denominator expansion yields different kinematic prefactors for the two components defined in~\eqref{spineffects}, implying that spin-dependent effects only arise at higher orders. The subleading nature of these effects is expected in a QED-like model,  where spin-induced magnetic transitions are known to be suppressed compared to electric ones~\cite{Gould:1981an}. 
Accordingly, the processes \( \chi_{1} \chi_{2} \to \chi_{1} \chi_{2} V \) and \( S_{1} S_{2} \to S_{1} S_{2} V \) yield the same dipole and quadrupole emission amplitudes in the case of distinguishable particle self-scattering. For identical particle self-scattering, both the \( t \)- and \( u \)-channels must be included according to the underlying statistical properties of the DM candidates, leading to different emission amplitudes in the two dissipative scenarios.

\paragraph{Soft emission limit} For a soft emitted particle,~\eqref{M0QED} reduces to the standard soft factors~\eqref{softfactorgauge} multiplying the elastic amplitude. Including the quadrupole correction discussed in Sec.~\ref{softquadrupole}, the NR amplitude is fully recovered by this factorization, as the $\slashed{q}$-dependent part~\eqref{withemissionQED} vanishes identically.

\section{Derivation of $\vec q$-dependent eikonal emission factors} \label{Softderivations}

In this appendix, we sketch the derivation of the $v_{i}$-expansion of the soft emission factors~\eqref{softfactorgauge} and~\eqref{softfactorscalar}. A standard discussion of vector emission can be found in many textbooks such as in~\cite{Berestetskii:1982qgu}. However, in our derivations, we use the physical momenta of the emission process~\eqref{CMpj}, leading to a $\vec{q}$-dependent correction that is important to be carried along in the quadrupole amplitude.

Before explicitly showing the derivation of the soft factors, we briefly comment on the validity of expanding and squaring the different components of the soft amplitude individually as we did in Sec.~\ref{soft}. Indeed, to obtain the soft emission amplitudes, the soft factors must be multiplied by a piece \( \mathcal{M}_{\text{elastic}} \) that has the same structure as an elastic matrix element, but still originates from the dissipative processes. Accordingly, when applying the $v_{i}$-expansion, this piece should be expressed in terms of the physical momenta of the emission process and scales as,\footnote{The term $B$ arises from enforcing the full dissipative conservation of momentum in  \( \mathcal{M}_{\text{elastic}} \).} 
\begin{equation}
 \mathcal{M}_{\text{elastic}} \simeq A v_{i}^{-2} + Bv_{i}^{-1} + C v_{i}^{0} + \mathcal{O}(v_{i}) .   
\end{equation}
On the other hand, the soft factors~\eqref{softfactorgauge},~\eqref{softfactorscalarA} and~\eqref{softfactorscalar} scale as
\begin{equation}
\mathcal{E} \simeq \underbrace{\mathcal{E_{D}}}_{\mathcal{O}(v_{i}^{-1})} + \underbrace{\mathcal{E_{Q}}}_{\mathcal{O}(v_{i}^0)} + \mathcal{O}(v_{i}^1).    
\end{equation}
Taking the product \( \mathcal{M}_{\text{elastic}} \mathcal{E}_{D} \) prior to squaring it gives rise to a term $B\mathcal{E}_{D} \sim \mathcal{O}(v_{i}^{-2}) $, i.e., of the same order as the soft quadrupole emission amplitude.
However, the term $B$ cancels in the case of an equal coupling-to-mass ratio, making this contribution irrelevant when the quadrupole term dominates the emission. Therefore, we can safely square both \( \mathcal{M}_{\text{elastic}} \) and \( \mathcal{E} \) individually without losing any relevant terms for the dipole or quadrupole soft amplitudes.

\subsection{Vector soft emission}
The soft factor~\eqref{softfactorgauge} from the dissipative processes $S_{1}S_{2}\to S_{1}S_{2} V$ and $\chi_{1} \chi_{2} \to \chi_{1} \chi_{2} V$, reduces the well-known expression
\begin{equation} \label{eikovector}
    \begin{aligned}
   \mathcal{E}_{V}  =  g_{1} \frac{\varepsilon^{*}(q) \cdot p_{3}}{p_{3} \cdot q} + g_{2} \frac{\varepsilon^{*}(q) \cdot p_{4}}{p_{4} \cdot q} - g_{2} \frac{\varepsilon^{*}(q) \cdot p_{2}}{p_{2} \cdot q} - g_{1} \frac{\varepsilon^{*}(q) \cdot p_{1}}{p_{1} \cdot q} . 
    \end{aligned}
\end{equation}
We make the assumption of a massless emitted vector boson in Coulomb gauge $\varepsilon^{\mu*}(q)=(0,\vec{e}^{*})$, where $\vec{e}^{*}$ denotes the transverse polarization vector in Cartesian components,  and obtain for the $v_{i}$-leading order of~\eqref{eikovector}
\begin{equation} \label{dipolevectoreikode}
    \begin{aligned}
    \mathcal{E}_{V}  &\overset{CM}{\simeq} -  \frac{1}{\omega}\left[\frac{g_{1}}{m_{1}}(\Vec{p}_{f} - \Vec{p}_{i}) - \frac{g_{2}}{m_{2}}(\Vec{p}_{f} - \Vec{p}_{i}) \right] \cdot \textbf{e}^{*} + \mathcal{O}(v_{i}^{0}) \\
     &\simeq -  \frac{1}{\omega} \left(\frac{g_{1}}{m_{1}}- \frac{g_{2}}{m_{2}}\right) (\Vec{p}_{f} - \Vec{p}_{i})\cdot \textbf{e}^{*} + \mathcal{O}(v_{i}^{0}),
    \end{aligned}
\end{equation}
We identify this term as the dipole contribution, which vanishes in the case of equal coupling-to-mass ratios. As previously noted, this term can be directly squared, summed over polarizations, and angularly averaged using~\eqref{averageqrelationsvector}, yielding
   \begin{equation}
\begin{aligned}
  \langle  |\mathcal{E}_{\mathcal{D},V}|^2 \rangle_{\vec{\hat{q}}} =  \frac{2\mu^2}{3 \omega^2} \left(\frac{g_{1}}{m_{1}}- \frac{g_{2}}{m_{2}}\right)^2  |\Vec{v}_{f} - \Vec{v}_{i}|^2. 
    \end{aligned} 
\end{equation}

In the case of processes with an equal coupling-to-mass ratio, this term vanishes, and $v_{i}$-subleading terms of~\eqref{eikovector} become the dominant terms. The corresponding derivation is similar, but the inclusion of the correction factor~\eqref{correctionsoft} is necessary to obtain~\eqref{eq:equalityQuad}. The relevance of this factor in the quadrupole soft factor becomes apparent when considering, for example, an emission from the final external $\chi_{2}$-leg,
\begin{equation} \label{examplecorrectionsoft}
\begin{aligned}
 \mathcal{E}_{V}|_{\chi_2\text{ outgoing}}&=g_{2} \frac{p_{4}\cdot \varepsilon^{*}(q)}{p_{4}\cdot q} \frac{k_{\text{elastic}}^2}{k_4^2} 
\overset{\text{CM}}{\simeq}  g_{2}\frac{\vec{p}_{f}\cdot \vec{e}^{*}}{m_{2}\omega+\vec{p}_{f}\cdot \vec{q}  +\mathcal{O}(v_{i}^4) } \frac{|\Vec{p}_{i}-\Vec{p}_{f}|^2}{|\Vec{p}_{i}-\Vec{p}_{f}-{\Vec{q}\mu}/{m_{2}}|^2}. %
\end{aligned}
\end{equation}
During the $v_{i}$-expansion of~\eqref{examplecorrectionsoft}, the contribution from the correction factor~\eqref{correctionsoft} is trivial at leading order and only appears at subleading order. It introduces some~$\vec{q}$-dependent terms necessary to ensure consistency between the soft energy-differential cross section and the full energy-differential cross section over a significant energy range.

\subsection{Scalar soft emission}
We now turn to the case of a soft scalar emission. As the derivation is similar for both fermionic and scalar DM candidates, we restrict the following discussion to the fermionic case. 
The soft factor~\eqref{softfactorscalar} in the case of the dissipative processes $\chi_{1} \chi_{2} \to \chi_{1} \chi_{2} \phi$ and $\tilde\chi_{1} \tilde\chi_{2} \to \tilde\chi_{1} \tilde\chi_{2} \phi$ is
\begin{equation} \label{dipolescalareikode}
\begin{aligned}
\mathcal{E}_{\phi \chi}= \frac{y_{1} m_{1}}{p_{3} \cdot q} + \frac{y_{2} m_{2}}{p_{4} \cdot q} - \frac{y_{2} m_{2}}{p_{2} \cdot q} - \frac{y_{1} m_{1}}{p_{1} \cdot q} +\frac{y_{1}}{m_{1}} +\frac{y_{2}}{m_{2}}  .
\end{aligned}
\end{equation}
We apply to it the $v_{i}$-expansion and find at leading order
 \begin{equation}
     \begin{aligned}
       \mathcal{E}_{\phi \chi} & \overset{\text{CM}}{\simeq }\frac{y_{1}}{\omega}\left[ \left(1 + \frac{\Vec{p}_{f}\cdot \Vec{q}}{m_{1}\omega}\right) - \left(1 + \frac{\Vec{p}_{i}\cdot \Vec{q}}{m_{1}\omega}\right) \right] +   \frac{y_{2}}{\omega}\left[ \left(1 - \frac{\Vec{p}_{f}\cdot \Vec{q}}{m_{1}\omega}\right) - \left(1 - \frac{\Vec{p}_{i}\cdot \Vec{q}}{m_{1}\omega}\right) \right] + \mathcal{O}(v_{i}^{0})    \\
        &\simeq \frac{1}{\omega^2} \left(\frac{y_{1}}{m_{1}}- \frac{y_{2}}{m_{2}}\right) (\Vec{p}_{f} - \Vec{p}_{i}) \cdot \Vec{q}  + \mathcal{O}(v_{i}^{0}).  
     \end{aligned}
 \end{equation}
We identify this factor as the one corresponding to scalar dipole emission.
Squaring and applying the angular average using~\eqref{averageqrelations} leads to the dipole soft factor
\begin{equation}
\langle  |\mathcal{E}_{\mathcal{D},\phi \chi}|^2 \rangle_{\vec{\hat{q}}} = \frac{\mu^2}{3 \omega^2} \left(\frac{y_{1}}{m_{1}}- \frac{y_{2}}{m_{2}}\right)^2 |\Vec{v}_{f} - \Vec{v}_{i}|^2 . 
\end{equation}
We note that, as in the full QFT amplitude~\eqref{withoutemissionYuka} and the time-dependent perturbation theory matrix element~\eqref{Mscalar}, the leading-order term in the \( v_i \)-expansion of~\eqref{dipolescalareikode}—corresponding to the additive term in the numerator—vanishes. 

In the case an equal coupling-to-mass ratio, the next $v_{i}$-subleading term of~\eqref{dipolescalareikode} becomes the dominant term. The introduction of the correction factor~\eqref{correctionsoft} and the use of the full $\vec{q}$-dependent kinematics has to be applied, similarly to the vector emission case.

\section{Finite-mass corrections} \label{Massiveemission}

We consider in this App. the case where the emitted boson carries a non-zero mass, while remaining within the kinematically allowed regime, i.e., $m_{\phi,V} \lesssim \mu v_{i}^2$. In the case of scalar emission, a mass term can readily be introduced for the emitted particle. For vector gauge emission, the mass is introduced via the Stückelberg mechanism, see \cite{Ruegg:2003ps} for a review.

\subsection{Massive vector boson emission}

Considering a St\"uckelberg-type massive vector field $V^{\mu}$, the additional physical longitudinal polarization degree of freedom is given by the polarization vector,
\begin{equation} \label{longivector}
    \varepsilon^{\mu*}_{(L)} = \frac{1}{m_{V}} \left( |\vec{q}|,\, \omega \frac{\vec{q}}{|\vec{q}|} \right).
\end{equation}
In the limit $m_V \to 0$, corresponding to a vanishingly small but finite mass of the emitted particle, the longitudinal polarization vector approaches a singular limit,
\begin{equation}
    \varepsilon^{\mu*}_{(L)} \underset{m_{V}\rightarrow0}{=} \frac{q_{\mu}}{m_{V}} + \mathcal{O}\left(\frac{m_{V}}{\omega}\right),
\end{equation}
raising the concern that its contribution to the amplitude may lead to unphysical behavior in this regime.
In the dissipative scenarios we explore, the vector boson couples to a conserved current. Accordingly, the Ward identity, 
\begin{equation} \label{Ward}
    q_{\mu} \mathcal{M}^{\mu} = 0 \quad \Rightarrow \quad \omega\, \mathcal{M}^{0} = \, \oldhat{{q}}^{i}\mathcal{M}^{i}
\end{equation}
holds for $V^{\mu}$ \cite{Kribs:2022gri}, and once applied to the emission amplitude contracted with the longitudinal polarization vector
yields 
\begin{equation} \label{ward2}
 \mathcal{M}^{(L)} =  \varepsilon^{(L)*}_{\mu} \mathcal{M}^{\mu}
= -\frac{m_{V}}{\omega} \oldhat{{q}}^{i}\mathcal{M}^{i} = - \frac{m_{V}}{|\vec{q}|} \mathcal{M}^{0}.
\end{equation}
This identity is exact and remains valid in all kinematic regimes, including the limit $m_{V}\to 0$, where the potentially divergent term cancels and the longitudinal amplitude remains finite. In the limit $m_V \sim \omega$, the corrections, while remaining finite, become parametrically of order unity. We have explicitly checked that our emission amplitudes for $V$-emission satisfy~\eqref{ward2} at both dipole and quadrupole order.

\subsection{Emission amplitudes and energy-differential cross sections}
In the following, we present the corresponding $m_{\phi,V}$-corrections to the emission amplitudes and the energy-differential cross sections for distinguishable particle scattering.

The squared amplitudes are obtained, for the transverse modes of $V$-emission, by first summing over the transverse polarization states $\lambda$ of the polarization vector $\vec e^{(\lambda)*}$ and then averaging over the direction of the emitted particle, whereas for $\phi$-emission and the longitudinal $V$-emission only the angular averaging is required. Accordingly, we use the identities given in~\eqref{averagingquadrupole1},~\eqref{averagingquadrupole2} and~\eqref{averageqrelations}--\eqref{averageqrelationsvector}. 

\subsubsection{Long-range mediation}

\paragraph{Vector emission} We begin by treating vector emission. The dipole emission components are obtained by expanding the emission amplitude to leading order in $v_i$ using the appropriate polarization vectors. The longitudinal contribution, obtained from contraction with the polarization vector~\eqref{longivector}, is given by
\begin{equation} \label{LOveclong}
\mathcal{M}_{\mathcal D}^{(L)}=\varepsilon_{\nu}^{(L)*}\mathcal{M_{D}}^{\nu}= 4 \mathcal{F_{D}} \frac{m_{V}}{\omega} \frac{ m_{1}m_{2} (\Vec{v}_{f}-\Vec{v}_{i})\cdot \Vec{\hat{q}} }{\mu \omega|\Vec{v}_{i}-\Vec{v}_{f}|^2} \delta_{s_{1}s_{3}} \delta_{s_{2}s_{4}},
\end{equation}
where the factors $\delta_{s_{k}s_{l}}$ arise from the NR spinor expansion; see~\eqref{spinorchiral} for details. The associated transverse components coincide with those in the massless case; in particular, they exhibit no dependence on $m_V$,
\begin{equation} \label{LOvectrans}
\mathcal{M}_{\mathcal D}^{(T)} = \varepsilon_{\nu}^{(T)*}\mathcal{M_{D}}^{\nu}= 4 \mathcal{F_{D}}  \frac{ m_{1}m_{2} (\Vec{v}_{f}-\Vec{v}_{i})\cdot \Vec{e}^{*} }{\mu \omega|\Vec{v}_{i}-\Vec{v}_{f}|^2}\delta_{s_{1}s_{3}} \delta_{s_{2}s_{4}}.
\end{equation}
We emphasize that, at this order, the spinor expansion yields only trivial contributions. As a result, for the process $S_{1}S_{2} \to S_{1}S_{2}\, V$, the dipole components of the emission amplitude are also given by~\eqref{LOveclong} and~\eqref{LOvectrans}, with the spinor structures reducing to $\delta_{s_{1}s_{3}}\, \delta_{s_{2}s_{4}} \to 1$. Upon squaring these transverse and longitudinal components and performing the angular average individually using~\eqref{averageqrelations}--\eqref{averageqrelationsvector}, and subsequently summing these contributions, the resulting expression can be inserted into~\eqref{crosssection} to obtain the energy-differential cross section~\eqref{dipolelongrange}.

At quadrupole order, corresponding to the next-to-leading terms in the $v_{i}$-expansion, both the transverse and longitudinal components of the emission amplitude gain a dependence on~$m_V$. Using the same prescription as in the previous case, the longitudinal part reads
\begin{equation} \label{massiveQEDL}
\begin{aligned}
\mathcal{M}^{(L)}_{\mathcal{Q}} = & 8\mathcal{F_{Q}}\left(\frac{m_{V}}{\omega} \right) \frac{  m_{1}m_{2} |\Vec{q}| }{\omega^2 |\Vec{v}_{i}- \Vec{v}_{f}|^{4}} \oldhat{{q}}^{j} \oldhat{{q}}^{k}    \Bigg[  \left( \oldhat{{v}}_{f}^{k} \left( \oldhat{{v}}_{f}^{j} |\vec{v}_{f}|^3 \left(|\vec{v}_{f}| - |\vec{v}_{i}| \Hat{v}_{i}\cdot \Hat{{v}}_{f} \right) \right. \right.   \\& + \left. \left. \left. \frac{1}{2}\oldhat{{v}}_{i}^{j}|\vec{v}_{f}||\vec{v}_{i}|(|\vec{v}_{i}|^2- |\vec{v}_{f}|^2) \right) - (\vec{v}_{i} \leftrightarrow \vec{v}_{f} ) \right) \right.   \left.   + \frac{\delta^{jk}}{2 \mu}|\Vec{v}_{i}- \Vec{v}_{f}|^{2} \omega  \right]\delta_{s_{1}s_{3}} \delta_{s_{2}s_{4}},
\end{aligned}
\end{equation}
while the transverse part is given by
\begin{equation} \label{massiveQEDT}
\begin{aligned}
\mathcal{M}^{(T)}_{\mathcal{Q}} = & 4 \mathcal{F_{Q}}\frac{ m_{1}m_{2} |\vec{q}|}{\omega^2 |\Vec{v}_{i}- \Vec{v}_{f}|^{4}}\vec{\mathnormal{e}}^{*j} \oldhat{{q}}^{k} \left[   \oldhat{{v}}_{f}^{k} \left(2  \oldhat{{v}}_{f}^{j} |\vec{v}_{f}|^3 \left(|\vec{v}_{f}| - |\vec{v}_{i}| \Hat{v}_{i}\cdot \Hat{v}_{f} \right) \right. \right. \\ &  \left. \left. + \oldhat{{v}}_{i}^{j}|\vec{v}_{f}||\vec{v}_{i}|(|\vec{v}_{i}|^2- |\vec{v}_{f}|^2) \right)  - (\vec{v}_{i} \leftrightarrow \vec{v}_{f} ) \right] \delta_{s_{1}s_{3}} \delta_{s_{2}s_{4}}.
\end{aligned}
\end{equation}
We note that the $m_{V}$-dependence of~\eqref{massiveQEDT} appears from the on-shell relation of the emitted particle, i.e., $|\Vec{q}| = \sqrt{\omega^2 - m_{V}^2}$. Moreover, we have simplified the \( \Vec{e}^{*} \cdot \Vec{\Hat{q}} \) terms arising from the spinor expansion. Consequently, the components of the quadrupole emission amplitudes for scalar DM candidates are also given by~\eqref{massiveQEDL} and~\eqref{massiveQEDT}, treating the spinor factors as trivial. 

For the polarization modes given in~\eqref{massiveQEDL} and~\eqref{massiveQEDT} individually, we square the corresponding amplitudes, sum over the final spins, average over the initial ones, and perform the angular average using~\eqref{averagingquadrupole1} and~\eqref{averagingquadrupole2}. The sum over the three polarization states is then carried out by adding the two resulting contributions, and the combined expression is finally inserted into~\eqref{crosssection} to obtain the energy-differential cross section, 
\begin{equation} \label{QuadmassivelongV}
   \omega \frac{d\sigma}{d\omega} = (1-\kappa^2)^{3/2}  \frac{ \mathcal{F}_{\mathcal Q}^2 \mu^2   }{480 \pi^3 } \left[ 80\left( \frac{3}{8}\kappa^2  + \frac{1}{2} \right) \sqrt{1- x } + 
   12(2-x)\left(\frac{2}{3}\kappa^2 +1 \right)\ln{\left(\frac{ 1 + \sqrt{1-x}}{1-\sqrt{1-x}}\right)} \right],
\end{equation}
where we defined \(\kappa\equiv m_{\phi,V}/\omega\). 

\paragraph{Scalar emission}
We now turn to the case of scalar emission, beginning with dipole radiation. For fermionic DM candidates, the corresponding emission amplitude, obtained from the leading order $v_i$-terms, including spinor structures, reads
\begin{equation} \label{LOscalar}
\mathcal{M_{D}}=  4\mathcal{F_{D}} |\Vec{q}| \frac{ m_{1}m_{2} (\Vec{v}_{f}-\Vec{v}_{i})\cdot \Vec{\hat{q}} }{\mu \omega^2|\Vec{v}_{i}-\Vec{v}_{f}|^2} \delta_{s_{1}s_{3}} \delta_{s_{2}s_{4}},
\end{equation}
where the $m_{\phi}$-dependence sits in the on-shell relation for  $|\Vec{q}|$ of the emitted particle.

For scalar DM candidates, the dipole emission amplitude coincides with~\eqref{LOscalar}, up to the replacement $\delta_{s_{1}s_{3}} \delta_{s_{2}s_{4}} \to 1$. This emission amplitude, squared and angular-averaged using~\eqref{averageqrelations}, and subsequently inserted into~\eqref{crosssection}, yields the energy-differential cross section~\eqref{dipolelongrange}, which also applies to fermionic DM candidates.

We now turn to quadrupole emission, obtained from the subleading $v_{i}$-contributions to the emission amplitude. For both, fermionic and scalar DM candidates, the emission amplitude can be decomposed as 
\begin{equation} \label{massivescalardecomp}
    \mathcal{M_{Q}}= \mathcal{M_{Q}}|_{m_{\phi}\to 0} + \Delta\mathcal{M_Q}(m_\phi),
\end{equation}
with the first term corresponding to the emission amplitude describing massless emission and the second term arising in the case of massive emission.
In the case of fermionic DM candidates, the massless emission part of the emission amplitude reads 
\begin{equation}
\begin{aligned}
\mathcal{M_{Q}}|_{m_{\phi}\to 0} = & 8 \mathcal{F_{Q}} \frac{ m_{1}m_{2}}{\omega |\Vec{v}_{i}- \Vec{v}_{f}|^{4}}\oldhat{{q}}^{j} \oldhat{{q}}^{k}  \left. \Big[  \left( \oldhat{{v}}_{f}^{k} \left( \oldhat{{v}}_{f}^{j} |\vec{v}_{f}|^3 \left(|\vec{v}_{f}| - |\vec{v}_{i}| \Hat{v}_{i}\cdot \Hat{v}_{f} \right)  \right. \right. \right. \\ & \left. \left. \left. + \frac{1}{2}\oldhat{{v}}_{i}^{j}|\vec{v}_{f}||\vec{v}_{i}|(|\vec{v}_{i}|^2- |\vec{v}_{f}|^2) \right)   - (\vec{v}_{i} \leftrightarrow \vec{v}_{f} ) \right)  \right.  \left. + \frac{\delta^{jk}}{\mu}  |\Vec{v}_{i}- \Vec{v}_{f}|^{2} \omega  \right. \Big] \delta_{s_{1}s_{3}} \delta_{s_{2}s_{4}} ,
\end{aligned}
\end{equation}
and the massive emission part is given by 
\begin{equation} \label{massivecorrectionlong}
\begin{aligned}
\Delta\mathcal{M}_Q(m_\phi) = & -8 \mathcal{F_{Q}} \left(\frac{m_{\phi}}{\omega}\right)^2 \frac{ m_{1} m_{2}}{\omega |\Vec{v}_{i}- \Vec{v}_{f}|^{4}}\oldhat{{q}}^{j} \oldhat{{q}}^{k}   \left. \Big[ \left(   \oldhat{{v}}_{f}^{k} \left( \oldhat{{v}}_{f}^{j} |\vec{v}_{f}|^3 \left(|\vec{v}_{f}| - |\vec{v}_{i}| \Hat{v}_{i}\cdot \Hat{v}_{f} \right) \right. \right. \right.  \\ &  + \left. \left. \left. \frac{1}{2}\oldhat{{v}}_{i}^{j}|\vec{v}_{f}||\vec{v}_{i}|(|\vec{v}_{i}|^2- |\vec{v}_{f}|^2) \right)   - (\vec{v}_{i} \leftrightarrow \vec{v}_{f} ) \right) \right.  + \left. \frac{\delta^{jk}}{2\mu}|\Vec{v}_{i}- \Vec{v}_{f}|^{2} \omega  \right. \Big] \delta_{s_{1}s_{3}} \delta_{s_{2}s_{4}}.
\end{aligned}
\end{equation}

We sum these amplitude contributions according to~\eqref{massivescalardecomp}, square the resulting expression, sum over the initial spins, average over the final ones, and perform the angular average over the emission direction using~\eqref{averagingquadrupole2}. The energy-differential cross section is then obtained by inserting the resulting squared emission amplitude  into~\eqref{crosssection}, 
\begin{equation} \label{quadmassivefermionphi}
   \omega \frac{d\sigma}{d\omega} = \sqrt{1-\kappa^2} \frac{ \mathcal{F}_{\mathcal Q}^2 \mu^2   }{60 \pi^3 } \left[ 10\left( \frac{3}{8}\kappa^4 - \frac{1}{2}\kappa^2 + \frac{1}{2} \right) \sqrt{1- x } + 
   (2-x)\left(\kappa^2 -1 \right)^2\ln{\left(\frac{ 1 + \sqrt{1-x}}{1-\sqrt{1-x}}\right)} \right].
\end{equation}

For scalar DM candidates, we only keep the relevant quadrupole emission terms in the limit $(A_{j}/m_{j})^2 \gg 
\lambda_{j}$. We decompose the emission amplitude as in~\eqref{massivescalardecomp}, where the first term, which is the one that survives in the massless-emission limit, reads
\begin{equation}
\begin{aligned}
\mathcal{M_{Q}}|_{m_{\phi}\to0} = & -8 \mathcal{F_{Q}} \frac{ m_{1}m_{2} }{\omega |\Vec{v}_{i}- \Vec{v}_{f}|^{4}} \oldhat{q}^{j} \oldhat{q}^{k}  \biggl[ {\oldhat{v}}_{i}^{j}{\oldhat{v}}_{i}^{k}|\vec{v}_{i}|^3\left( |\vec{v}_{i}| - | \vec{v}_{f}| \hat{v}_{i}\cdot \hat{v}_{f} \right) - \left( \vec{v}_{i}\leftrightarrow \vec{v}_{f} \right) \\& \left. - \frac{2\omega}{\mu}\oldhat{v}_{i}^{j}\oldhat{v}_{f}^{k}|\vec{v}_{i}||\vec{v}_{f}| \right]   ,
\end{aligned}
\end{equation}
and the term \(\Delta\mathcal{M}_Q(m_\phi)\) is equal to one derived in the fermionic case, i.e.,~\eqref{massivecorrectionlong}, once the spinor structures are replaced by $\delta_{s_{1}s_{3}}\, \delta_{s_{2}s_{4}} \to 1$.

With these matrix elements at hand, we first sum them according to~\eqref{massivescalardecomp} and square the resulting expression, perform the angular average over the emission direction using~\eqref{averagingquadrupole2}, and finally insert the result into~\eqref{crosssection} to obtain the energy-differential cross section,
\begin{equation} \label{quadmassivescalarphi}
   \omega \frac{d\sigma}{d\omega} = \sqrt{1-\kappa^2}  \frac{ \mathcal{F}_{\mathcal Q}^2 \mu^2   }{60 \pi^3 } \left[ 10\left( \frac{3}{8}\kappa^4 - \kappa^2 + 1 \right) \sqrt{1- x } + 
   (2-x)\left(\kappa^2 -1 \right)^2\ln{\left(\frac{ 1 + \sqrt{1-x}}{1-\sqrt{1-x}}\right)} \right] .
\end{equation}

\subsubsection{Short-range mediated force} \label{short-rangemediationamplitude}
The leading order in $r$ of the emission amplitudes, corresponding to dipole emission, is obtained by performing the substitution $|\vec{v}_{i} - \vec{v}_{f}|^2 \rightarrow m_{\phi,V}^2/\mu^2$ in the denominators of~\eqref{LOveclong} and~\eqref{LOvectrans} for vector emission, and of~\eqref{LOscalar} for scalar emission.
 
\paragraph{Vector emission}
We start by treating massive vector emission. Expanding the emission amplitude for the scattering of $\chi_{1}\chi_{2}\to\chi_{1}\chi_{2} V$ to next-to-leading order in $r$, using the longitudinal polarization vector~\eqref{longivector}, gives the  longitudinal quadrupole part of the emission amplitude,
\begin{equation} \label{massiveQEDLshort}
\begin{aligned}
\mathcal{M}^{(L)}_{\mathcal{Q}}= &2 \mathcal{F^{\prime}_{Q}} \left(\frac{m_{V}}{\omega}\right)  \frac{  m_{1}m_{2} \mu^2 |\vec{q}| }{m_{V'}^2 \omega^2} \oldhat{q}^{j} \oldhat{q}^{k} \left[  |\vec{v}_{f}|^2 \left( 2  \oldhat{v}_{f}^{j} \oldhat{v}_{f}^{k}-\delta^{jk}\right) - |\vec{v}_{i}|^2 \left( 2  \oldhat{v}_{i}^{j} \oldhat{v}_{i}^{k}- \delta^{jk}\right)  \right] 
\delta_{s_{1}s_{3}}\delta_{s_{2}s_{4}}.
\end{aligned}
\end{equation}
Following the same steps, but using the transverse polarization vector, yields
\begin{equation} \label{massiveQEDTshort}
\mathcal{M}^{(T)}_{\mathcal{Q}}= 4 \mathcal{F^{\prime}_{Q}}   \frac{ m_{1}m_{2} \mu^2 |\Vec{q}| }{m_{V'}^2\omega^2}e^{*j} \oldhat{q}^{k}\left[{ |\vec{v}_{f}|^2\oldhat{v}_{f}^{j} \oldhat{v}_{f}^{k}  - |\vec{v}_{i}|^2\oldhat{v}_{i}^{j} \oldhat{v}_{i}^{k}  } \right]\delta_{s_{1}s_{3}} \delta_{s_{2}s_{4}}.
\end{equation}
Similarly to long-range mediation, the $m_{V}$-dependency of~\eqref{massiveQEDTshort} only comes from the on-shell mass relation of the emitted particle in~$|\vec q|$. Moreover, we have simplified the \( \Vec{e}^{*} \cdot \Vec{\Hat{q}} \) terms arising from the spinor expansion. Accordingly, the quadrupole emission amplitudes apply equally well to scalar DM candidates, with the spinor Kronecker deltas being replaced by unity.

Each polarization component is squared, summed over the final spins, averaged over the initial ones, and angularly averaged over the emission direction using~\eqref{averagingquadrupole1} and~\eqref{averagingquadrupole2}. The resulting contributions are then summed to obtain the squared amplitude summed over the three polarization states. Once the corresponding expression is inserted into~\eqref{crosssection}, we obtain the energy-differential cross section,
\begin{equation} \label{csshortmassiveQED}
   \omega \frac{d \sigma}{d \omega}=\frac{(1 - \kappa^2)^{3/2} \, \mathcal{F}^{\prime2}_{\mathcal{Q}} \mu^6 v_i^4}{30 \pi^3 m_{V'}^4 } 
\left( B(1, 2) + \frac{\kappa^2}{24} B(21, 32) \right).
\end{equation}

\paragraph{Scalar emission} For the scalar emission amplitude associated at sub-leading order in $r$, we follow the same decomposition as in~\eqref{massivescalardecomp}, analogously to the long-range mediation case.
In the case of fermionic DM candidates, the piece of the emission amplitude that arises from massless emission reads
\begin{equation} 
\begin{aligned}
\mathcal{M_{Q}}|_{m_{\phi}\to0} = & 4  \mathcal{F^{\prime}_{Q}} \frac{m_{1}m_{2}\mu^2   }{m_{\phi'}^2 \omega}\oldhat{q}^{j} \oldhat{q}^{k}\left[  |\vec{v}_{f}|^2 \left(  \oldhat{v}_{f}^{j} \oldhat{v}_{f}^{k}-\delta^{jk}\right) - |\vec{v}_{i}|^2 \left(  \oldhat{v}_{i}^{j} \oldhat{v}_{i}^{k}-\delta^{jk}\right)   \right] \delta_{s_{1}s_{3}} \delta_{s_{2}s_{4}},
\end{aligned}
\end{equation}
while the massive emission part is given by
\begin{equation} \label{massivecorrectionscalarheavy}
\begin{aligned}
\Delta\mathcal{M}_Q(m_\phi) = & -2  \mathcal{F^{\prime}_{Q}}  \left(\frac{m_{\phi}}{\omega} \right)^2 \frac{m_{1}m_{2}\mu^2   }{m_{\phi'}^2 \omega}\oldhat{q}^{j} \oldhat{q}^{k}\left[ |\vec{v}_{f}|^2 \left( 2  \oldhat{v}_{f}^{j} \oldhat{v}_{f}^{k}-\delta^{jk}\right) - |\vec{v}_{i}|^2 \left( 2  \oldhat{v}_{i}^{j} \oldhat{v}_{i}^{k}- \delta^{jk}\right)   \right] \\ & \times \delta_{s_{1}s_{3}} \delta_{s_{2}s_{4}}.
\end{aligned}
\end{equation}

From these matrix elements, by adding them according to~\eqref{massivescalardecomp} and squaring them, summing over the initial spins, averaging over the final ones, and averaging over the direction of the emitted particle using~\eqref{averagingquadrupole2}, we obtain the energy-differential cross section
\begin{equation} \label{csshortmassiveyuka}
\omega\frac{d\sigma}{d\omega}=\frac{2 \mathcal{F}_{\mathcal Q}'^2 \sqrt{1 - \kappa^2} \, \mu^6 v_i^4}{45 \pi^3 m_{\phi'}^4  } 
\left[ B(3,1) - \frac{\kappa^2}{4} B(9,8) +  \frac{\kappa^4}{32} B(21,32)  \right]. 
\end{equation}

For scalar DM candidates, we only keep the sub-leading terms in $r$ that are relevant for external leg-dominated emission, corresponding to the limit \(\langle |\mathcal{M}_{\text{external leg}}|^2 \rangle_{\vec{\hat{q}}} \gg \langle |\mathcal{M}_{\text{vertex}}|^2 \rangle_{\vec{\hat{q}}}\), and, following the decomposition of the emission amplitude~\eqref{massivescalardecomp}, obtain for the part arising from massless emission,
\begin{equation} \label{scalarscalarmatrixelement}
\mathcal{M_{Q}}|_{m_{\phi}\to0} =  4  \mathcal{F^{\prime}_{Q}} \frac{m_{1}m_{2}\mu^2   }{m_{\phi'}^2 \omega}\oldhat{q}^{j} \oldhat{q}^{k}\left[ |\vec{v}_{f}|^2 \oldhat{v}_{f}^{j} \oldhat{v}_{f}^{k} - |\vec{v}_{i}|^2  \oldhat{v}_{i}^{j} \oldhat{v}_{i}^{k}  \right].
\end{equation}
As in the case of long-range mediation, the massive correction part $\Delta\mathcal{M}_Q(m_\phi)$ is identical to the one obtained for fermionic DM candidates presented in~\eqref{massivecorrectionscalarheavy}, with the spinor factors set to unity.

The corresponding energy-differential cross section, obtained from the sum of these matrix elements according to~\eqref{massivescalardecomp}, squared, and averaged over the direction of the emitted particle using~\eqref{averagingquadrupole2}, reads
\begin{equation} \label{csshortmassiveyukascalar}
  \omega \frac{d \sigma}{d \omega}= \frac{\sqrt{1 - \kappa^2} \, \mathcal{F}_{\mathcal Q}'^2 \mu^6 v_i^4}{180 \pi^3m_{\phi'}^4} 
\left[ B(9, 8) - \kappa^2 B(3, 16) + \frac{\kappa^4}{4} B(21, 32) \right].
\end{equation}

\subsection{Universality of massive emission}

Based on the results of the previous section, it follows that, up to an overall prefactor, the kinematic and angular structures of the $m_{\phi,V}$-dependent terms are identical for both the dipole and quadrupole emission amplitudes across all dissipative models, except for the case of the transverse polarization mode in $V$-emission. In the following, we show that this universal behavior stems from the combined effect of the Ward identity obeyed by the longitudinal mode and the NR reduction of the relevant emission amplitudes. 

At dipole order, applying the identity~\eqref{ward2} to the longitudinal component of the $V$-emission amplitude given in~\eqref{LOveclong}, we obtain a relation with the amplitude for $\phi$-emission~\eqref{LOscalar},
\begin{equation} \label{longitudinalidentitydipole}
\varepsilon_{\nu}^{(L)*}\mathcal{M}_{\mathcal{D},V}^{\nu}= \frac{m_{V} }{|\vec{q}|} \mathcal{M}_{\mathcal{D,\phi}} \quad \overset{\eqref{ward2}}{\Rightarrow} \quad \mathcal{M}^{0}_{\mathcal{D},V} \;=\; -\,\mathcal{M}_{\mathcal{D},\phi}.
\end{equation}
This equality holds for both long-range and short-range mediation by applying
$|\vec{v}_{i} - \vec{v}_{f}|^2 \rightarrow m_{\phi,V}^2/\mu^2$ to~\eqref{LOveclong} and~\eqref{LOscalar} for the latter case.

At quadrupole order, two results are important to emphasize:  
1) for \( \phi \)-emission, \( \Delta \mathcal{M}(m_{\phi}) \) is identical for scalar and fermionic DM candidates, and  
2) the relations  
\begin{equation} \label{longitudinalidentityquadrupole}
\varepsilon_{\nu}^{(L)*} \mathcal{M}^{\nu} = -\frac{m_{V} |\vec{q}|}{m_{\phi}^{2}} \, \Delta \mathcal{M}_{Q}(m_{\phi})  \quad \overset{\eqref{ward2}}{\Rightarrow} \quad \mathcal{M}^{0}_{V,\mathcal{Q}} = \frac{|\vec{q}|^2}{m_{\phi}^2} \Delta\mathcal{M_Q}(m_\phi),
\end{equation}
can be inferred, from~\eqref{massiveQEDL} and~\eqref{massivecorrectionlong} for long-range mediation, and from~\eqref{massiveQEDLshort} and~\eqref{massivecorrectionscalarheavy} for short-range mediation, respectively. For the two mediation cases, they follow from applying~\eqref{ward2} to the contraction of the longitudinal polarization vector with the quadrupole emission amplitudes.

The equalities in~\eqref{longitudinalidentitydipole} and~\eqref{longitudinalidentityquadrupole} are non-trivial, as is the observation that 
\( \Delta\mathcal{M}_{Q}(m_{\phi}) \) coincides for scalar and fermionic DM candidate for $\phi$-emission. This can be understood from two considerations. 
First, for a conserved current, the scalar mode embedded in the definition of the longitudinal mode decouples~\cite{Kribs:2022gri}, 
and therefore cannot account for these equalities. Second, model dependence typically arises at quadrupole order; in particular, massless \( \phi \)-emission from scalar DM candidates exhibits a quadrupole amplitude that differs from its fermionic DM counterpart.

\paragraph{Non-relativistic amplitude} We now show that the relations to the right of~\eqref{longitudinalidentitydipole} and \eqref{longitudinalidentityquadrupole} can in fact be directly inferred from the NR expression of the emission amplitudes of the relevant processes. In the following, we consider only the case of distinguishable particle scattering with long-range mediation\footnote{The short-range mediation case follows the same arguments, with an $r$-expansion instead. It additionally includes the contact diagrams shown in Fig.\ref{Contacts}, whose emission amplitudes reduce to the same analytic expressions as the one from external-leg emissions.}, where the $m_{\phi,V}$-corrections arise exclusively from the four external leg emission diagrams, Fig.\ref{inifini}. We restrict ourselves to the two diagrams in which the DM species~1 emits the boson, since the remaining diagrams follow by the simple substitutions $a_{1}\leftrightarrow a_{2}$, $m_{1}\leftrightarrow m_{2}$, $E_{1},E_{3}\to E_{2},E_{4}$, and $p_{1},p_{3}\to p_{2},p_{4}$ applied to the amplitude pieces defined below. 
For a given scenario $k$, the emission amplitude contributions (or, for $V$-emission, its $0$-component) associated with an emission from the external legs of the DM species~1, read
\begin{equation} \label{amplitudedemonstrationmassive}
\begin{aligned}
\mathcal{M}^{(0)}_{k}&=\sum_{n\in\{1,3\}}\frac{J^{(0)}_{k,n}}{\left[ \left(p_{2}-p_{4}\right)^2 -m_{\phi,V}^2 \right]\left[2 \eta_{n}\left(E_{n}\omega -\vec{p}_{n}\cdot\vec{q}\right) + m_{\phi,V}^2 \right] },
\end{aligned}    
\end{equation}
where the sum runs over the two external legs of the DM species 1. The factor $\eta_n=+1(-1)$ accounts for the emission from final-(initial)-state leg~$n$ with energy $E_{n}$, three-momentum $\vec{p}_{n}$ and current $J^{(0)}_{k,n}$.  For the processes $S_{1}S_{2}\to S_{1}S_{2}\phi$ and $\tilde{S}_{1}\tilde{S}_{2}\to \tilde{S}_{1}\tilde{S}_{2}\phi$, the latter take the simple form, 
\begin{equation} \label{current1}
    \begin{aligned}
        J_{\phi S,1}&= J_{\phi S,3} = -A_{1}^2 A_{2}.
    \end{aligned}
\end{equation}
The remaining scenarios involve non-trivial currents, on which we perform the $v_{i}$-expansion discussed in Sec.~\ref{Powercounting}. In the case of the processes
\(\chi_{1}\chi_{2} \to \chi_{1}\chi_{2}\phi\) and
\(\tilde\chi_{1}\tilde\chi_{2} \to \tilde\chi_{1}\tilde\chi_{2}\phi\),
keeping only the terms relevant at dipole and quadrupole order, we obtain
\begin{equation} \label{current2}
    \begin{aligned}
        J_{\phi\chi,n} \overset{\text{CM}}{\simeq} -4 y_{1}^2y_{2}m_{1}m_{2}\left(2m_{1} +\eta_{n}\omega  \right) \delta_{s_{1},s_{3}}\delta_{s_{2},s_{4}} + \mathcal{O}(v_{i}^{3}). 
    \end{aligned}
\end{equation}
Applying the same prescription to $\chi_{1} \chi_{2}\to \chi_{1} \chi_{2} V$, the corresponding $0$-component of the currents reads
\begin{equation} \label{current3}
    \begin{aligned}
        J^{0}_{V\chi,n} \overset{\text{CM}}{\simeq} 4 g_{1}^2g_{2}m_{1}m_{2}\left[2m_{1} + 2(E_{n}-m_{1}) +\eta_{n}\omega\right]\delta_{s_{1},s_{3}}\delta_{s_{2},s_{4}} + \mathcal{O}(v_{i}^{3}). 
    \end{aligned}
\end{equation}
Here, $E_{3}=m_{1}+ {|\vec{p}_{f}|^2}/{2m_{1}} + \mathcal{O}(v_{i}^3)$ and $E_{1}=m_{1}+ {|\vec{p}_{i}|^2}/{2m_{1}} + \mathcal{O}(v_{i}^4)$. For $S_{1} S_{2} \to S_{1} S_{2} V$, the $0$-component of the currents are found by applying the substitution  $\delta_{s_{1},s_{3}}\delta_{s_{2},s_{4}}\to 1$ in~\eqref{current3}; with this replacement, the contributions from vertex emission are effectively accounted for.

The NR emission amplitudes associated with the DM species 1 are then obtained by applying the $v_{i}$-expansion to~\eqref{amplitudedemonstrationmassive}, and read
\begin{equation} \label{emissionamplitudeexpanded}
    \begin{aligned}
        \mathcal{M}^{(0)}_{k} \overset{\text{CM}}{\simeq} & -\frac{J^{(0)}_{k,3}}{|\Vec{v}_{i}-\Vec{v}_{f}|^2m_{1}\omega}\left(1 + Bv_{i}^{1} +C(m_{\phi,V}^2)v_{i}^{2} + \mathcal{O}(v_{i}^{3}) \right)  \\ & +\frac{J^{(0)}_{k,1}}{|\Vec{v}_{i}-\Vec{v}_{f}|^2m_{1}\omega}\left(1 + B^{\prime}v_{i}^{1} +C^{\prime}(m_{\phi,V}^2)v_{i}^{2}  + \mathcal{O}(v_{i}^{4}) \right),
    \end{aligned}
\end{equation}
where the coefficients $B^{(\prime)}$ and $C^{(\prime)}(m_{\phi,V}^{2})$ arising from the expansion of the propagators. The terms proportional to $m_{\phi,V}^2$ in $C^{(\prime)}(m_{\phi,V}^{2})$ originate both from the explicit term in~\eqref{amplitudedemonstrationmassive} and from the on-shell condition of the emitted particle. With these expressions at hand, the model-independent kinematic structure of the $m_{\phi,V}$-correction terms at dipole and quadrupole order follows directly.

At dipole order, only the $\mathcal{O}(1)$ pieces of the currents~\eqref{current1}–\eqref{current3}  multiplied by the common factors $B^{(\prime)}$ survive. Each of these terms reduces, up to an
overall sign, to the same universal structure, thereby leading to~\eqref{longitudinalidentitydipole}.
At quadrupole order, 
the second equality in~\eqref{longitudinalidentityquadrupole} takes the form,
\begin{equation} \label{M0VvsdeltaM}
    \mathcal{M}^{0}_{V,\mathcal{Q}} = -\left(1- \frac{\omega^2}{m_{\phi}^2} \right) \Delta\mathcal{M_Q}(m_\phi).
\end{equation}
The first term arises by the same argument as the dipole term, namely from the product of the universal $\mathcal{O}(1)$ piece of the currents~\eqref{current1}–\eqref{current3} with the common factors $C^{(\prime)}(m_{\phi,V}^2)$. 
Accordingly, it follows that the massive correction term \( \Delta \mathcal{M}(m_{\phi}) \) is identical for $\phi$-emission from scalar and fermionic DM candidates. 
In contrast, the second term in~\eqref{M0VvsdeltaM} being independent of \( m_{\phi} \), originates from two sources: the product of the \( \mathcal{O}(v_i^2) \) pieces of currents~\eqref{current3} with the leading \( \mathcal{O}(1) \) propagator term, and the \( \mathcal{O}(1) \) part of the currents multiplied by \( C^{(\prime)}(0) \).

Finally, the absence of a relation analogous to~\eqref{ward2} for the  transverse component of the $V$-emission amplitude, accounts for the model-dependent behavior of the $m_{V}$-corrections arising in~\eqref{LOvectrans},~\eqref{massiveQEDT} (long-range mediation) and in~\eqref{LOvectrans} with $|\vec{v}_{i} - \vec{v}_{f}|^2 \to m_{\phi,V}^2/\mu^2$,~\eqref{massiveQEDTshort} (short-range mediation).

\subsection{Scattering of identical particles} \label{EqualmassAmplitudes}

In the following, we present the emission amplitudes for long-range mediated processes in the case of identical particle scattering, that include the $t$-channel and $u$-channel contributions. As already discussed, for short-range mediation, the latter exhibit the same amplitudes. Consequently, the amplitudes defined in Sec.~\ref{short-rangemediationamplitude} remain valid for identical particle scattering of fermionic DM candidates. For scalar DM candidates, only the overall prefactor differs; see Sec.~\ref{identicalparticles} for details. 

When considering the scattering of indistinguishable particles, it is convenient to introduce the variables \( k = p_2 - p_4 \) and \( l = p_2 - p_3 \). In the NR framework, and using the power counting scheme introduced in Sec.~\ref{NRhierarchyofscales}, the square of these four momenta reduce to \( k^2 \simeq -|\vec{k}|^2 \) and \( l^2 \simeq -|\vec{l}|^2 \), with \( |\vec{k}|/m_{\chi,S} \sim \mathcal{O}(v_i) \) and \( |\vec{l}|/m_{\chi,S} \sim \mathcal{O}(v_i) \).

\paragraph{Vector emission}
For the scattering of identical particles of spin $s=0$ $(S)$ and $s=1/2$ $(\chi)$ with a massive vector emission, the squared emission amplitude summed over the transverse and longitudinal polarizations of the vector boson, averaged over the initial and summed over the final spins, and subsequently angular averaged, is given by
\small \begin{equation} \label{EMQEDampli}
\begin{aligned}
\left\langle |\mathcal{M_{Q}}|^2 \right\rangle_{\vec{\hat{q}}}= &\frac{32 g^{6}}{15\omega^2} |\vec{k}|^2 |\vec{l}|^2 \left(1-\kappa^2\right) \left\{ \frac{\left( 3 + 13 (\hat{k}\cdot \hat{l})^2 \right) + \frac{\kappa^2}{2}\left( 4 + 19 (\hat{k}\cdot \hat{l})^2 \right)}{|\vec{l}|^4} + (\vec{k} \leftrightarrow \vec{l})  \right.
\\ &  + 2\frac{(-1)^{2s}}{(2s+1) |\vec{k}|^2 |\vec{l}|^2 }\left[ \left( 3 + 7 (\hat{k}\cdot \hat{l})^2 + 6 (\hat{k}\cdot \hat{l})^4 \right)  \left. \left. \right.+ \frac{\kappa^2}{2}\left( 4 + 11 (\hat{k}\cdot \hat{l})^2 + 8 (\hat{k}\cdot \hat{l})^4 \right) \right] \right\}. 
\end{aligned}
\end{equation}
In the case of a massless emission $(\kappa=0)$ our results agree with \cite{Pradler:2020znn}. 

\paragraph{Scalar emission}
The massive dissipative processes $\chi \chi \to \chi \chi \phi$ and $\tilde\chi \tilde\chi \to \tilde\chi \tilde\chi \phi$ have as squared emission amplitude, averaged over the initial and summed over the final spins, and angularly averaged,
 \begin{equation} \label{EMyuka}
\begin{aligned}
\left \langle |\mathcal{M_{Q}}|^2 \right \rangle_{\vec{\hat{q}}} = &\frac{64 y^{6}}{15\omega^2}|\Vec{k}|^2|\Vec{l}|^2 \left\{ \frac{ \left(1 + 6 (\hat{k} \cdot \hat{l})^2 \right) - \kappa^2 \left( 2 + 7 (\hat{k} \cdot \hat{l})^2 \right) + \frac{\kappa^4}{4}\left( 4 + 19 (\hat{k} \cdot \hat{l})^2 \right) }{|\Vec{k}|^4} + (\vec{k} \leftrightarrow \vec{l}) \right. \\
& - \frac{1}{|\Vec{k}|^2|\Vec{l}|^2} \Bigg[ \left(1 + 4 (\hat{k} \cdot \hat{l})^2 + 2 (\hat{k} \cdot \hat{l})^4 \right)- \kappa^2 \left(2 + 3 (\hat{k} \cdot \hat{l})^2 + 4 (\hat{k} \cdot \hat{l})^4 \right)   \\
& \left. \left. + \frac{\kappa^4}{4}\left(4 + 11 (\hat{k} \cdot \hat{l})^2 + 8 (\hat{k} \cdot \hat{l})^4 \right) \right] \right\}.
\end{aligned}
\end{equation}

For the processes  $S S \to S S \phi$ and $\tilde S \tilde S \to \tilde S \tilde S \phi$ involving massive emission, the squared amplitude—averaged over initial spins, summed over final spins, and angularly averaged—is given in the limit $ ({A}/{m_{S}})^2 \gg \lambda$ by
 \begin{equation} \label{EMscalar}
\begin{aligned}
\left \langle |\mathcal{M_{Q}}|^2 \right \rangle_{\vec{\hat{q}}}= &\left(\frac{A}{m_{S}}\right)^6\frac{|\Vec{k}|^2|\Vec{l}|^2}{15\omega^2} \left\{ \frac{ \left(1+11 (\hat{k}\cdot\hat{l})^2 \right) - 2\kappa^2\left( 1+6 (\hat{k}\cdot\hat{l})^2 \right) + \frac{\kappa^4}{4}\left( 4 +19 (\hat{k}\cdot\hat{l})^2 \right)  }{|\Vec{k}|^4} + (\vec{k} \leftrightarrow \vec{l}) \right. \\ &
 \left. + 2\frac{1}{{|\Vec{k}|^2|\Vec{l}|^2}} \Bigg[ \left(1 + 9 (\hat{k}\cdot\hat{l})^2 + 2 (\hat{k}\cdot\hat{l})^4 \right) - 2\kappa^2 \left(1 + 4 (\hat{k}\cdot\hat{l})^2 + 2 (\hat{k}\cdot\hat{l})^4 \right) \right. 
\\& \left. \left. + \frac{\kappa^4}{4} \left(4 + 11 (\hat{k}\cdot\hat{l})^2 + 8 (\hat{k}\cdot\hat{l})^4 \right)  \right] \right\},
\end{aligned}
\end{equation}  
while in the opposite limit, $ ({A}/{m_{S}})^2 \ll \lambda$, it is given by
\begin{equation}
  \left \langle |\mathcal{M_{Q}}|^2 \right \rangle_{\vec{\hat{q}}} = 4 \left(\frac{A}{m_{S}}\right)^2\lambda^2 \frac{|\Vec{k}|^2|\Vec{l}|^2  (\hat{k}\cdot \hat{l})^2 } {\omega^2}\left( \frac{1}{|\vec{k}|^4} + \frac{1}{|\vec{l}|^4} + \frac{2}{|\vec{k}|^2|\vec{l}|^2} \right).
\end{equation}
We note that in this limit, the emission amplitude does not gain any correction from the mass of the emitted particle.

\newpage
\bibliographystyle{JHEP}
\input{output.bbl}

\end{document}

%% file: output.bbl
\providecommand{\href}[2]{#2}\begingroup\raggedright\endgroup